\UseRawInputEncoding
\documentclass[journal]{IEEEtran}
\usepackage{textalpha}
\usepackage{amsmath,amsfonts}
\usepackage[T1]{fontenc}
\usepackage{algorithmic}
\usepackage{algorithm}
\usepackage[utf8]{inputenc}
\usepackage{textgreek}
\usepackage{array}
\usepackage[caption=false,font=normalsize,labelfont=sf,textfont=sf]{subfig}
\usepackage{textcomp}
\usepackage{stfloats}
\usepackage{url}
\usepackage{verbatim}
\usepackage{graphicx}
\usepackage{bm}
\usepackage{hyperref}
\usepackage{cite}
\usepackage{multirow}
\usepackage{pifont}
\usepackage{xcolor} 
\renewcommand{\arraystretch}{1.5}
\newcommand{\triangleq}{\stackrel{\triangle}{=}}

\begin{document}

\title{Reconfigurable Intelligent Surfaces for 6G and Beyond:  \\
A Comprehensive Survey from Theory to Deployment}

\author{Prasetyo Putranto,~\IEEEmembership{Student Member,~IEEE}, 
Anis Amazigh Hamza,~\IEEEmembership{Member,~IEEE},  
Sameh Mabrouki,~\IEEEmembership{Student Member,~IEEE},  
Nasrullah Armi,~\IEEEmembership{Member,~IEEE}, 
Iyad Dayoub,~\IEEEmembership{Senior Member,~IEEE} 
\thanks{Prasetyo Putranto is with  Institut d’Électronique de Microelectronique et de Nanotechnologie (IEMN) UMR CNRS 8520, Université Polytechnique Hauts-de-France, 59313 Valenciennes,
France, and Research Center for Telecommunications, National Research and Innovation Agency, Bandung, Jawa Barat 40135, Indonesia (email: pras009@brin.go.id).}
\thanks{Anis Amazigh Hamza is with XLIM, UMR CNRS 7252, SRI (Smart Systems and Network), University of Limoges, 16 Rue Atlantis, 87068 Limoges Cedex, France (email: anis-amazigh.hamza@unilim.fr).}
\thanks{Sameh Mabrouki and Iyad Dayoub are with Institut d’Électronique de Microelectronique et de Nanotechnologie (IEMN) UMR CNRS 8520, Université Polytechnique Hauts-de-France, 59313 Valenciennes,
France (email: \{sameh.mabrouki, iyad.dayoub\}@uphf.fr).}
\thanks{Nasrullah Armi is with Research Center for Telecommunications, National Research and Innovation Agency, Bandung, Jawa Barat 40135, Indonesia (email: nasr004@brin.go.id).}

}

\markboth{Journal of \LaTeX\ Class Files,~Vol.~14, No.~8, August~2021}%
{Shell \MakeLowercase{\textit{et al.}}: A Sample Article Using IEEEtran.cls for IEEE Journals}

\IEEEpubid{0000--0000/00\$00.00~\copyright~2021 IEEE}

\maketitle

\begin{abstract}
As the wireless research community moves toward shaping the vision of sixth-generation (6G) networks, reconfigurable intelligent surfaces (RIS) have emerged as a promising technology for controlling the propagation environment. Although RIS has not yet been standardized, its versatile applications and enabling capabilities have attracted growing attention in both academia and industry. This survey presents a comprehensive review of RIS technology spanning theoretical foundations, design aspects, and practical deployment considerations. In contrast to existing surveys that focus on isolated aspects, this work offers an integrated view covering use cases, control mechanisms, channel sounding methodologies, and channel estimation strategies. Each of these topics is reviewed through the lens of recent literature, synthesizing the latest advancements to provide updated insights for both academic researchers and industry practitioners. It further addresses emerging topics such as standardization activities and industrial perspectives, which are often overlooked in prior literature. By bridging theoretical insights with practical challenges, this survey aims to provide a holistic understanding of RIS and support its evolution from a research concept toward real-world implementation.
\end{abstract}


\section{Introduction}
\label{sec:introduction}
\IEEEPARstart{W}{ireless} communication technology continues to advance and evolve. With the commercialization of fifth-generation (5G) technology worldwide, scientists and engineers have been collaborating over the last few years to envision and develop the sixth-generation (6G) technology \cite{wang_road_2023,vaezi_cellular_2022,saad_vision_2020, jiang_road_2021, ishteyaq_unleashing_2024}. A crucial milestone is now being approached, with the official 6G standardization phase expected to be started by the 3rd Generation Partnership Project (3GPP) in June 2025 \cite{jain_3gpp_2024, zhao_6g_2025}. This evolution toward 6G is driven by the need for faster data rates, lower latency, increased reliability, and greater capacity to support advanced applications such as tactile/haptic internet, extended reality (XR), fully automated driving, wireless brain-machine interfaces, and the Internet of Everything (IoE), which cannot be fully realized with 5G. As exemplified by these applications, the transition in telecommunication services from basic connectivity to enhanced interaction is the key factor driving these advancements forward\cite{zhang_6g_2019}.

A major challenge to the advancement of wireless networks is the uncontrollable channel condition, which potentially distorts the propagating signal through various physical phenomena, such as reflection, fading, scattering, and diffraction. Reflection occurs when a signal bounces off surfaces like buildings or walls, often causing multipath propagation, where multiple copies of the signal arrive at the receiver at different times, potentially leading to interference. Fading refers to the fluctuation in signal strength due to destructive interference from multipath signals or environmental changes. Scattering happens when the signal encounters small obstacles, such as foliage or rough surfaces, dispersing the signal energy in different directions and weakening the intended transmission. Diffraction occurs when the signal bends around obstacles, such as corners or edges, which can lead to signal attenuation and distortion. These phenomena, individually or collectively, degrade the signal quality and the reliability of wireless communications. 

\begin{figure}[tb]
\centering
\includegraphics[width=\linewidth]{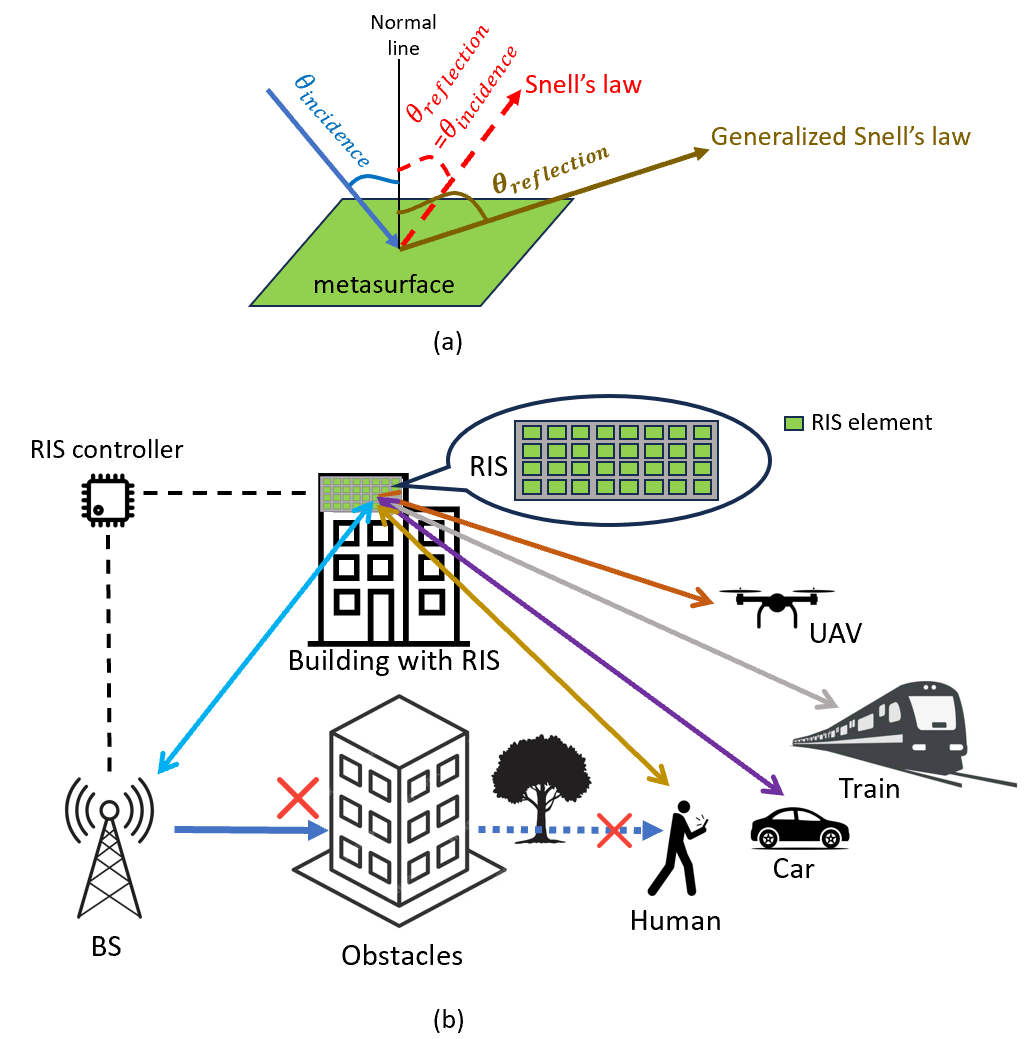}
\caption{Illustrations of (a) the generalized Snell's law applied to a metasurface, and (b) implementation of RIS for various user situation.}
\label{fig:ris1}
\end{figure}

\IEEEpubidadjcol 
Since the challenge stems from a lack of control over the propagation environment, an effective solution is to introduce mechanisms that enable the manipulation of the wireless channel. A currently popular solution is the reconfigurable intelligent surfaces (RIS), which is also called intelligent reflecting surfaces (IRS) \cite{wu_towards_2020}, smart reflect-arrays \cite{tan_enabling_2018}, software-controlled metasurfaces \cite{liaskos_new_2018}, or intelligent beamforming metasurfaces \cite{shafi_6g_2024}. Throughout this paper, the term RIS will be used exclusively. The term "reconfigurable" refers to their ability to adapt their properties over time in response to network conditions.
The "intelligent" aspect refers to their capability for real-time programmability, enabling algorithm-driven reconfiguration. The term "surface" highlights that RIS is a two-dimensional, planar, non-volumetric structure. RIS is capable of manipulating signals by employing the generalized Snell’s law \cite{cai_hierarchical_2021}. Unlike Snell’s law, which states that the angle of incidence equals the angle of reflection when the refractive indices of the incoming and outgoing media are identical, the generalized Snell’s law allows these angles to differ through phase shift \cite{yu_light_2011}. Fig. \ref{fig:ris1}a illustrates the comparison between the conventional Snell's law and generalized Snell's law as well as how this principle enables RIS to reach various user locations. Fig. \ref{fig:ris1}b shows how an RIS can establish a virtual line-of-sight (LoS)  connection in scenarios where obstacles like buildings or trees block the LoS path between the base station (BS) and user equipments (UEs). By intelligently controlling the phase of incoming signals, the RIS reflects them along unobstructed paths toward the UE, effectively bypassing physical barriers and fostering a smart radio environment \cite{di_renzo_smart_2020,bjornson_reconfigurable_2022, wu_towards_2020}.

RIS comprises multiple reflecting elements, often called unit cells. Each unit cell is composed of a metasurface, a thin-layer implementation of metamaterials, which are artificially engineered materials made from composite substances such as metals and plastics \cite{etsi01}.  The electromagnetic properties of a metasurface can be adjusted by incorporating semiconductor components (e.g. PIN and varactor diodes) or tunable materials (e.g., graphene and liquid crystals) into the surface \cite{etsi01, fu_fundamentals_2025}. 
For instance, each diode embedded within an RIS unit cell can modify the unit cell’s impedance, and by orchestrating varying impedance patterns across the unit cells, the surface can effectively alter its wavefront shaping characteristics, creating the appearance that the surface shape is changing. Instead of a single large design, the modular structure of RIS facilitates scalability, allowing the surface to be expanded or reduced based on application requirements. It simplifies maintenance as individual unit cells can be replaced or repaired without affecting the entire surface. A large number of unit cells enables the RIS to perform beamforming, a technique that focuses the reflected signal toward a specific direction, which is particularly crucial for high-frequency communications such as millimeter wave (mmWave) and terahertz (THz), where signals are highly directional and suffer from high path loss.

With its basic features, RIS offers low energy consumption and hardware cost \cite{di_renzo_smart_2020}, better coverage \cite{bjornson_reconfigurable_2022,pan_reconfigurable_2021}, and improved spectral efficiency \cite{di_renzo_reconfigurable_2020}. RIS can also be combined with other emerging technologies, such as massive multiple-input multiple-output (MIMO) \cite{an_low-complexity_2022, wang_ris-aided_2023, yang_spatially_2024}, mmWave communication \cite{alsenwi_intelligent_2022,yu_channel_2023,lyu_crb_2024}, THz communication \cite{dash_active_2022,zarini_resource_2023,le_performance_2024,su_channel_2024}, visible light communication (VLC)\cite{aboagye_ris-assisted_2023,maraqa_optimized_2023,abdeljabar_sum_2024}, non-orthogonal multiple access (NOMA) \cite{arslan_reconfigurable_2023,hamza_error_2022,dogukan_reconfigurable_2024,allouis_maximum_2023,liu_road_2024}, vehicle-to-everything (V2X)\cite{gu_socially_2022,singh_visible_2022,saikia_proximal_2024}, and machine learning (ML) \cite{zhong_ai_2022,xu_deep_2022,li_double_2023,abdallah_multi-agent_2024,zhou_heuristic_2024,qi_deep-reinforcement-learning-based_2025}.
\begin{figure}[tb]
\centering
\includegraphics[width=\linewidth]{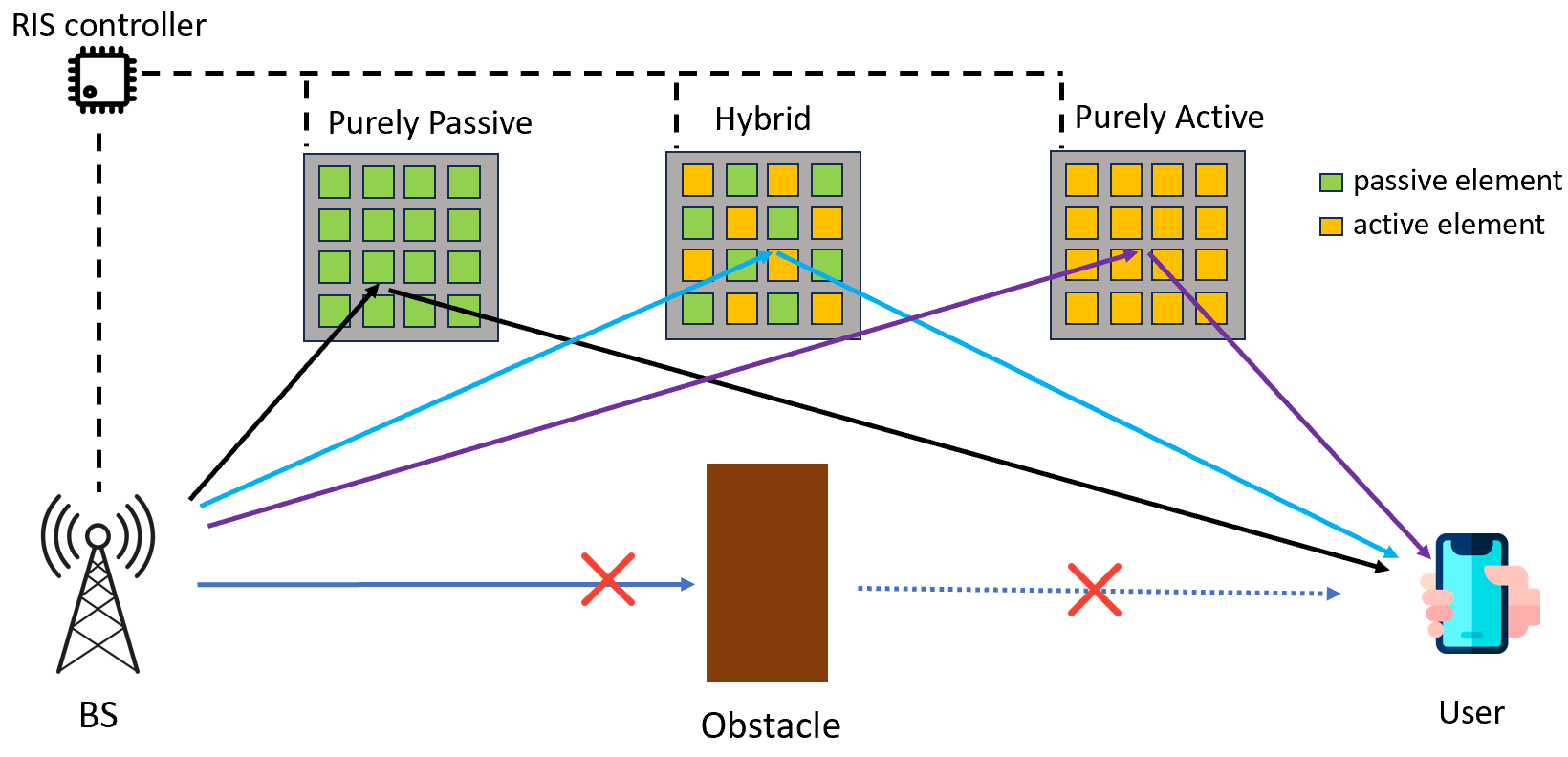}
\caption{Different types of RIS.}
\label{fig:ristype}
\end{figure}

There are three different types of hardware architecture of RIS: passive, active, and hybrid, as illustrated in Fig. \ref{fig:ristype}. Passive RIS is the basic type of RIS, which uses low-power tunable electronic components such as PIN diodes or varactors to adjust the phase, resulting in almost negligible power consumption \cite{etsi01}. It requires neither a dedicated power source nor a radio frequency (RF) chain. This feature makes passive RIS interesting from the energy consumption perspective. However, the passive RIS has limitations, as the signal reflected by the RIS must traverse two paths: BS-RIS and RIS-UE. Without amplification, the received signal is subject to double-path attenuation, causing significant power loss \cite{zhi_active_2022}. Addressing this challenge, researchers have introduced the concept of active RIS \cite{zhang_active_2023}. Unlike passive RIS, active RIS  can amplify incoming signals. Consequently, it includes an integrated amplifier, which requires more power consumption. Moreover, this amplifier introduces thermal noise \cite{chen_channel_active_2023}. Another difference is about the deployment. As discussed in \cite{you_wireless_2021}, active RIS has to be installed closer to the receiver so that the amplification factors are sufficient to compensate for the amplification noise, while passive RIS should be properly deployed closer to either the transmitter or receiver to minimize the cascaded path loss. Beyond purely passive RIS and purely active RIS, there exists a combination of both types, known as hybrid RIS or semi-passive RIS, which has a small number of active elements among mostly passive ones. Hence, it empowers metasurfaces to reflect incoming signals in a precisely controllable manner while concurrently sensing a fraction of the signal \cite{zhang_channel_2023, alexandropoulos_hybrid_2024}. 

With RIS emerging as a paradigm shift in wireless communication technology, examining its fundamental characteristics, recent advancements, and practical deployment considerations becomes necessary.
This survey provides a broad overview of RIS technology and its enabling techniques towards seamless integration into next-generation wireless communication systems.

\subsection{Motivation and Prior Works}

\begin{table*}[htbp]
\centering
\small
\caption{Comparison of Related Surveys and Reviews}
\label{tab:comparison}
\resizebox{\textwidth}{!}{ 
\begin{tabular}{|>{\centering\arraybackslash}m{0.6cm}|>{\centering\arraybackslash}m{0.6cm}|>{\centering\arraybackslash}m{5cm}|>{\centering\arraybackslash}m{0.9cm}|>{\centering\arraybackslash}m{1.6cm}|>{\centering\arraybackslash}m{1.5cm}|>{\centering\arraybackslash}m{1.5cm}|>{\centering\arraybackslash}m{2.1cm}|>{\centering\arraybackslash}m{1.7cm}|} 
\hline
\textbf{Year} & \textbf{Ref} & \textbf{Summary} & \textbf{Use Cases} & \textbf{Control Mechanism} &\textbf{Channel Sounding} & \textbf{Channel Estimation} & \textbf{Standardization Effort} & \textbf{Industrial Perspective} \\
\hline
2020 & \shortstack{\cite{gong_toward_2020}} & A survey on performance metrics, optimization techniques, emerging use cases, and the associated challenges of RIS for future wireless networks.  & \ding{51} & \ding{51} & \ding{55} & \ding{55} & \ding{55} & \ding{55}  \\
\hline
2021 & \shortstack{\cite{kisseleff_reconfigurable_2021}} & A review of potential use cases, deployment strategies, and design aspects for RIS devices in underwater IoT, underground IoT, Industry 4.0, and emergency networks.  & \ding{51} & \ding{55} & \ding{55} & \ding{55} & \ding{55} & \ding{55}  \\
\hline
2022 & \shortstack{\cite{huang_reconfigurable_2022}} & An overview of RIS wireless channels, including measurements, characterization, and modeling. & \ding{55} & \ding{55} & \ding{51} & \ding{55} & \ding{55} & \ding{55}  \\
\hline
2022 & \shortstack{\cite{swindlehurst_channel_2022}} & An overview of channel estimation in wireless systems using RIS, analyzing various channel models and training setups. & \ding{55} & \ding{55} & \ding{55} & \ding{51} & \ding{55} & \ding{55}  \\
\hline
2022 & \shortstack{\cite{pan_overview_2022}} & An overview of signal processing techniques for RIS-assisted wireless systems, with a focus on channel estimation, transmission design, and localization & \ding{55} & \ding{55} & \ding{55} & \ding{51} & \ding{55} & \ding{55}  \\
\hline
2022 & \shortstack{\cite{zheng_survey_2022}}  & A survey on channel estimation and passive beamforming in RIS-assisted systems, considering practical constraints and architectures. & \ding{55} & \ding{55} & \ding{55}  & \ding{51}  & \ding{55} & \ding{55}  \\
\hline
2024 & \shortstack{\cite{shafique_going_2024}} & A review of AI-driven RIS optimizations and various RIS tuning mechanisms.  & \ding{55} & \ding{51} & \ding{55} & \ding{55} & \ding{55} & \ding{55}  \\
\hline
2024 & \shortstack{\cite{zhou_survey_2024}} & A survey about optimization techniques
for RIS-assisted wireless communication by comparing model-based, heuristic, and machine learning techniques.
& \ding{55} & \ding{51} & \ding{55} & \ding{55} & \ding{55} & \ding{55}  \\
\hline
\textbf{2025} & \multicolumn{2}{c|}{\textbf{This survey}} & \scalebox{1.45}{\ding{51}}  & \scalebox{1.45}{\ding{51}} & \scalebox{1.45}{\ding{51}} & \scalebox{1.45}{\ding{51}} & \scalebox{1.45}{\ding{51}} & \scalebox{1.45}{\ding{51}} \\
\hline
\end{tabular}
}
\end{table*}

While RIS has demonstrated the potential to enhance wireless communication, its practical deployment faces several open questions:
\begin{itemize}
    \item What are the relevant use cases of RIS in different wireless environments?
    \item How can RIS be controlled to dynamically adjust its parameters?
    \item What methodologies have been explored for RIS-assisted channel sounding and how do they compare across different setups?
    \item What are the available channel estimation techniques suitable for various propagation environments and system configurations?
    \item  How are ongoing standardization efforts and industrial initiatives shaping the real-world adoption of RIS technology?
\end{itemize}

Addressing these questions requires a comprehensive investigation that unifies existing research efforts. Although several works have examined individual aspects of RIS, a holistic survey covering use cases, control mechanisms, channel sounding, and channel estimation has yet to be fully explored. Furthermore, standardization efforts and industrial perspectives play a critical role in transitioning RIS from theoretical models to practical deployment, yet they remain underexplored. 

Prior related surveys and reviews \cite{kisseleff_reconfigurable_2021, gong_toward_2020, shafique_going_2024, huang_reconfigurable_2022, swindlehurst_channel_2022, pan_overview_2022, zheng_survey_2022, zhou_survey_2024} have provided valuable insights into RIS applications but remain fragmented across different research domains, as summarized in Table \ref{tab:comparison}. For instance, the authors in \cite{gong_toward_2020} and \cite{ kisseleff_reconfigurable_2021} adopt different approaches in categorizing RIS use cases. The authors in \cite{gong_toward_2020} classify RIS applications based on implementation strategies, including deep learning-assisted RIS systems, wireless power transfer, uncrewed aerial vehicle (UAV) communications, and mobile edge computing. Conversely, the authors in \cite{kisseleff_reconfigurable_2021} categorize use cases based on extreme environmental scenarios, such as underwater, underground, industrial, and disaster response settings. While implementation-based and scenario-based categorizations offer practical insights for specific deployment contexts, they may have limited generalizability across diverse RIS applications. Implementation-based classifications focusing on technological integration may become obsolete as technology evolves. Similarly, scenario-based categorizations are useful for understanding RIS operation in specialized environments but may not comprehensively capture trade-offs in system performance. In contrast, a classification grounded in fundamental performance metrics, such as spectral efficiency, security, and energy efficiency, offers a more adaptable and scalable framework. These metrics are intrinsic to RIS operation across various deployment scenarios and technological implementations, ensuring that the categorization remains relevant as RIS technology evolves.

The authors in \cite{gong_toward_2020,zhou_survey_2024, shafique_going_2024} provide discussions about the control mechanisms of RIS. In \cite{gong_toward_2020}, the authors examine RIS control key components, including tunable materials, a central controller, inter-unit-cell communication, and phase tuning. On the other hand, \cite{shafique_going_2024} categorizes RIS control mechanisms based on the type of tunable material, including lumped elements, graphene, phase-change materials, and liquid crystals. The recent survey in \cite{zhou_survey_2024} provides a comprehensive discussion on optimization techniques in RIS-assisted wireless communication, including control aspects. Optimization techniques are important in RIS-assisted systems. However, given the recency and depth of the discussion in \cite{zhou_survey_2024}, we exclude RIS optimization from the scope of our survey to avoid redundancy. Nonetheless, the discussion about distribution of the control process within the system architecture remains unaddressed. This oversight is critical as the real-time adaptability of RIS systems depends not only on component selection but also on how control functions are orchestrated across network entities. Understanding the integration of RIS control within the overall network architecture is essential to ensure seamless coordination in practical deployments.

Channel sounding techniques are reviewed in \cite{huang_reconfigurable_2022}, which examines existing RIS-based approaches across various frequency bands, system configurations, and channel observations. The study emphasizes on the importance of validating theoretical and simulation-based analyses through practical experiments. While it offers an insightful review, its primary discussion focuses on indoor environments, with very limited coverage of outdoor scenarios. Given the unique challenges of outdoor deployments, including larger propagation distances, environmental variability, and mobility-induced Doppler shifts, further investigations are needed to extend RIS-based channel sounding to practical outdoor settings.

The overview in \cite{swindlehurst_channel_2022} provides a valuable discussion on channel estimation in RIS-assisted systems with various setups. However, the aspect of high mobility is not included in the scope of the discussion. The survey in \cite{pan_overview_2022} offers a comprehensive overview of signal processing techniques for RIS-assisted wireless systems, including channel estimation. While the review is thorough, aspects such as high-mobility scenarios and multi-RIS deployments are only identified as future work. Meanwhile, \cite{zheng_survey_2022} presents an in-depth analysis of RIS-assisted channel estimation across various setups and techniques, with a strong emphasis on passive RIS. However, discussions on hybrid and active RIS remain limited in scope. Additionally, although high-mobility channel estimation is mentioned, it is not explored in detail. Furthermore, the discussion on channel estimation in multi-RIS deployments primarily reflects studies from 2020 to 2021, leaving a gap in the literature regarding more recent advancements, which this survey aims to address.

Beyond these technical aspects, a significant gap remains in discussing standardization efforts and industrial perspectives, which are critical for RIS's real-world feasibility. This survey aims to bridge these  gaps by providing a comprehensive analysis that integrates use cases, control mechanisms, channel sounding, channel estimation, and the evolving landscape of standardization and industrial initiatives. The synthesis of these aspects establishes a holistic framework for understanding RIS across theoretical and applied domains.

\subsection{Scope of the Survey}
\label{sec:scope}

First, this survey provides an overview of the use cases and control mechanisms of RIS. Additionally, it presents the latest developments of channel sounding and channel estimation techniques for RIS-assisted systems. The survey also explores standardization efforts and industrial perspectives on RIS, offering insights into ongoing developments and future directions. These aspects differentiate our work from previous surveys. Table \ref{tab:comparison} highlights the positioning of this paper in comparison to existing literature.

\subsection{Organization of the Survey}
\label{sec:org}

\begin{figure}[tb!]
\centering
\includegraphics[width=\linewidth]{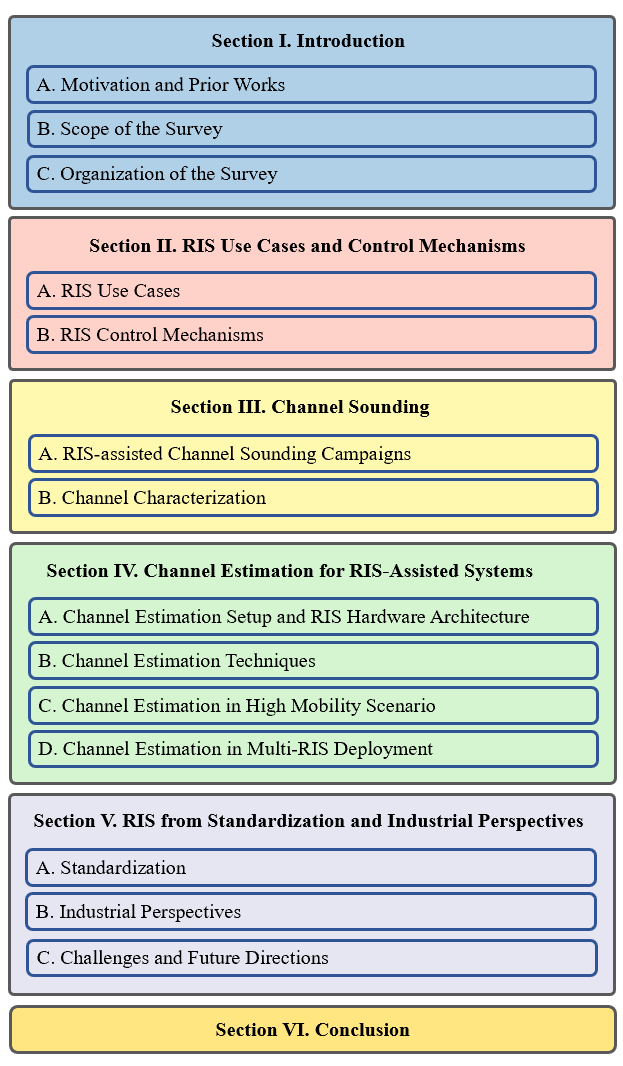}
\caption{Organization of the survey.}
\label{fig:organization}
\end{figure}

This paper is organized as illustrated in Fig. \ref{fig:organization}, and its outline is summarized as follows:

\begin{itemize}
\item Section \ref{sec:sec2} presents the use cases of RIS for contextualizing the utility of RIS in a practical scenario, followed by an overview of RIS control mechanisms.
\item Section \ref{sec:sec3} explores the channel sounding techniques in RIS-assisted systems. 
\item Section \ref{sec:sec4} discusses various channel estimation methods applicable to RIS.
\item Section \ref{sec:sec5} assesses RIS from the perspectives of standardization efforts and the industrial point of view.
\item Section \ref{sec:sec7} summarizes our key findings and insights drawn from the reviewed literature.
\end{itemize}

\section{RIS Use Cases and Control Mechanisms}
\label{sec:sec2}
RIS, as an emerging technology in telecommunications, requires sufficient investigation to understand its impact on network architecture. Therefore, it is essential to examine the use cases of RIS and its control mechanisms within such systems.
\subsection{RIS Use Cases}
As previously discussed, implementation-based and scenario-specific categorizations offer limited generalizability across diverse RIS potential applications. To address this limitation, we present a more general perspective on RIS use cases by reviewing two complementary viewpoints: the standardization-driven approach of the European Telecommunications Standards
Institute (ETSI) and the research-oriented vision of the RISE-6G\footnote{RISE-6G stands for Reconfigurable Intelligent Sustainable Environments for 6G Wireless Networks, a European Union-funded project from  January 1st, 2021, to December 31st,  2023 (36 months).  It aimed to explore the latest advancements in RIS that will actively contribute to standardization and promote its adoption in industry.}project. Together, they provide a comprehensive overview of how RIS can be effectively integrated into future wireless systems.

In 2023, ETSI published the first report (GR RIS-001 \cite{etsi01}) which specifies 11 key use cases of RIS which can be summarized as follows: 
\subsubsection{Coverage Enhancement} RIS can mitigate coverage gaps caused by obstacles, such as walls and buildings, by establishing a virtual LoS path. It provides a low-cost and energy-efficient solution compared to deploying additional BSs or relay nodes. More advanced architecture, such as multi-RIS systems, further enhances coverage robustness and spatial flexibility. Several studies have investigated the use of RIS for this purpose \cite{kayraklik_indoor_2024,qin_indoor_2022,nemati_ris-assisted_2020,ma_multi-ris_2024}.
\subsubsection{Spectral Efficiency} In wireless networks, high channel correlation between transmit and receive antennas reduces the number of available eigenchannels for parallel data transmission. Deploying an RIS will reshape the wireless environment which might mitigate this high correlation issue. As a result, networks benefit from increased spatial multiplexing gains and spectral efficiency. Studies \cite{eldeeb_energy_2023} and \cite{lu_ris-assisted_2025} validate this functionality.
\subsubsection{Beam Management} RIS supports dynamic beamforming by adjusting its elements to reflect signals in desired directions. This allows efficient steering of beams toward users, reducing the need for multiple transmission points. Compared to massive MIMO, RIS achieves beam gains with lower energy consumption, as demonstrated in \cite{wu_intelligent_2019} and \cite{huang_reconfigurable_2019}.
\subsubsection{Physical Layer Security} By customizing the wireless propagation environment, RIS prevents signal leakage to unintended receivers. It can guide reflections into “trusted zones,” reducing vulnerability to eavesdropping. RIS can also serve as a defense mechanism against malicious RIS deployments. The results in \cite{zhang_enhancing_2024} and \cite{zhang_physical_2021} align with this objective.
\subsubsection{Localization Accuracy} RIS enhances positioning accuracy by introducing additional, controllable signal paths. It can replace or complement BS in localization, especially in indoor environments where traditional methods fail. This enables finer spatial resolution and more reliable tracking. This functionality is demonstrated by  \cite{emenonye_ris-aided_2023,wu_multiple_2024,luan_phase_2022}.
\subsubsection{Sensing Capabilities} RIS introduces artificial reflections that support sensing in environments where LoS paths are unavailable or unreliable. It plays a key role in Integrated Sensing and Communication (ISAC) by extending sensing range and improving angular resolution. Depending on the application, either passive or hybrid RIS architectures can be deployed to optimize sensing performance \cite{li_user_2025, chen_simultaneous_2024}.
\subsubsection{Wireless Power Transfer} RIS helps deliver energy wirelessly to battery-powered devices in scenarios where direct energy transmission paths are blocked. It focuses electromagnetic energy onto target receivers, enhancing power transfer efficiency and thus extending battery life. \cite{zhao_wireless_2021} and \cite{zargari_multiuser_2022} exploit RIS for this purpose.
\subsubsection{Energy Harvesting} Some RIS elements can be dedicated to harvesting energy from environmental RF signals while others assist in communication. This strategy enables self-sustained or battery-less operation for IoT and M2M devices. RIS enhances both energy collection and data transmission simultaneously \cite{tavana_energy_2024,ntontin_wireless_2023}.
\subsubsection{Power Saving} RIS reduces the need for additional BSs and lowers the transmit power of both BSs and UEs. It enables green communication by improving network efficiency with minimal power consumption \cite{hashempour_power-efficient_2025,zhou_traffic_2023}.
\subsubsection{EMF Exposure Minimization} RIS enables lower electromagnetic field (EMF) exposure by focusing energy only where needed (BSs and UEs), avoiding excess radiation in other areas. This principle supports regulatory compliance and public health considerations in dense environments. So, RIS is able to enhance coverage and transmission quality, while keeping the network’s transmission power nearly unchanged. Several studies \cite{ibraiwish_emf-aware_2022,aghashahi_emf-aware_2024,zappone_energy_2022,yin_ris-aided_2023} investigated this use case.
\subsubsection{
Link Management} RIS facilitates programmable and reconfigurable wireless interconnects, particularly in settings like data centers where cabling can be cumbersome. It enables dynamic THz or mmWave links between servers or racks, increasing flexibility, as presented in \cite{aziz_use_2023}. Intelligent RIS control, possibly via machine learning, ensures robust and adaptable link performance even under changing network demands. 

\begin{figure}[tb]
\centering
\includegraphics[width=0.85\linewidth]{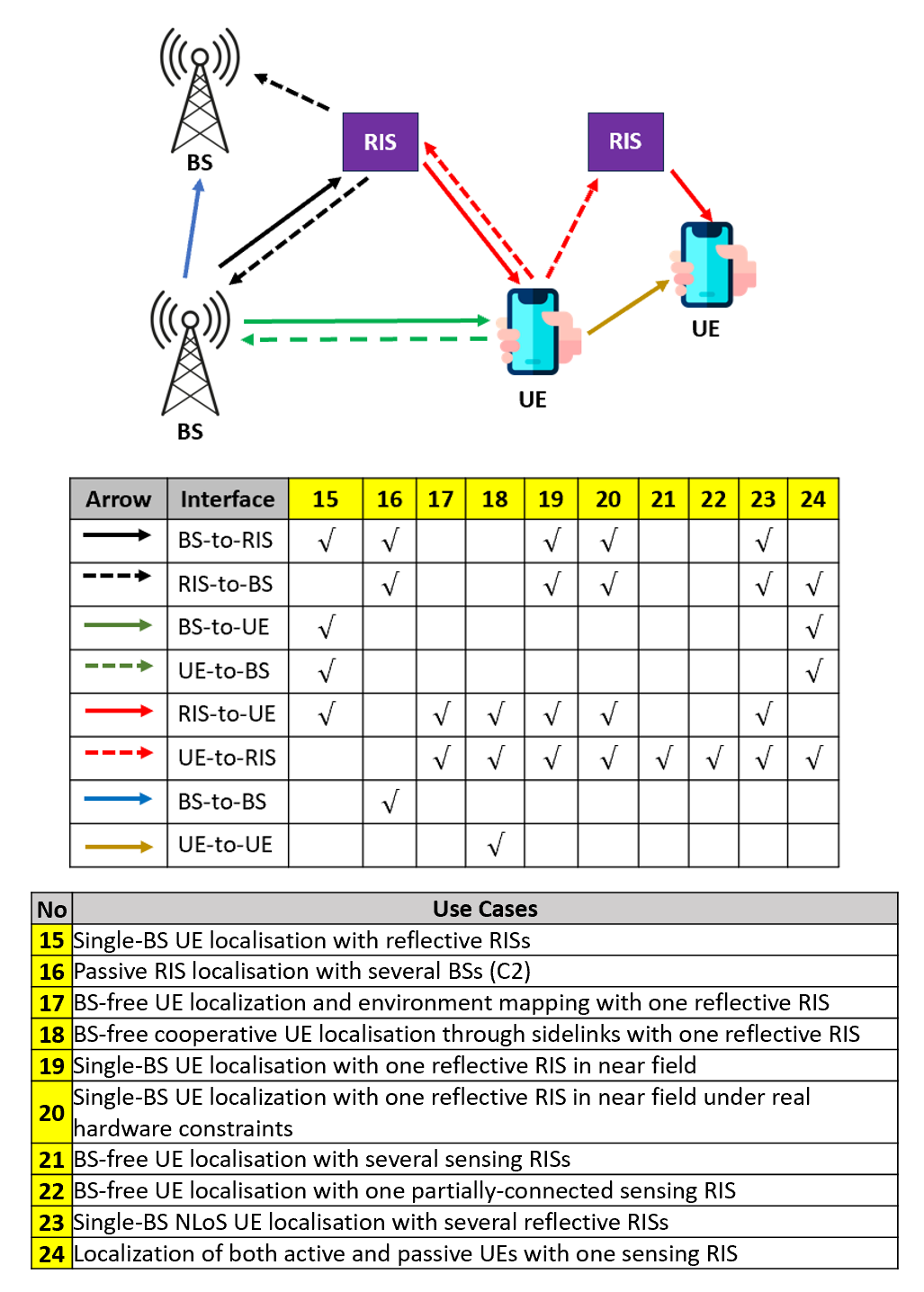}
\caption{Summary of category 2 of RIS use cases: enhanced localization and sensing, including the block diagram of the involved network entities, a summary table of all considered interfaces with corresponding use case, and the list of the use cases. \cite{wymeersch_control_2023}.}
\label{fig:usecase2}
\end{figure}

\begin{figure}[tb]
\centering
\includegraphics[width=\linewidth]{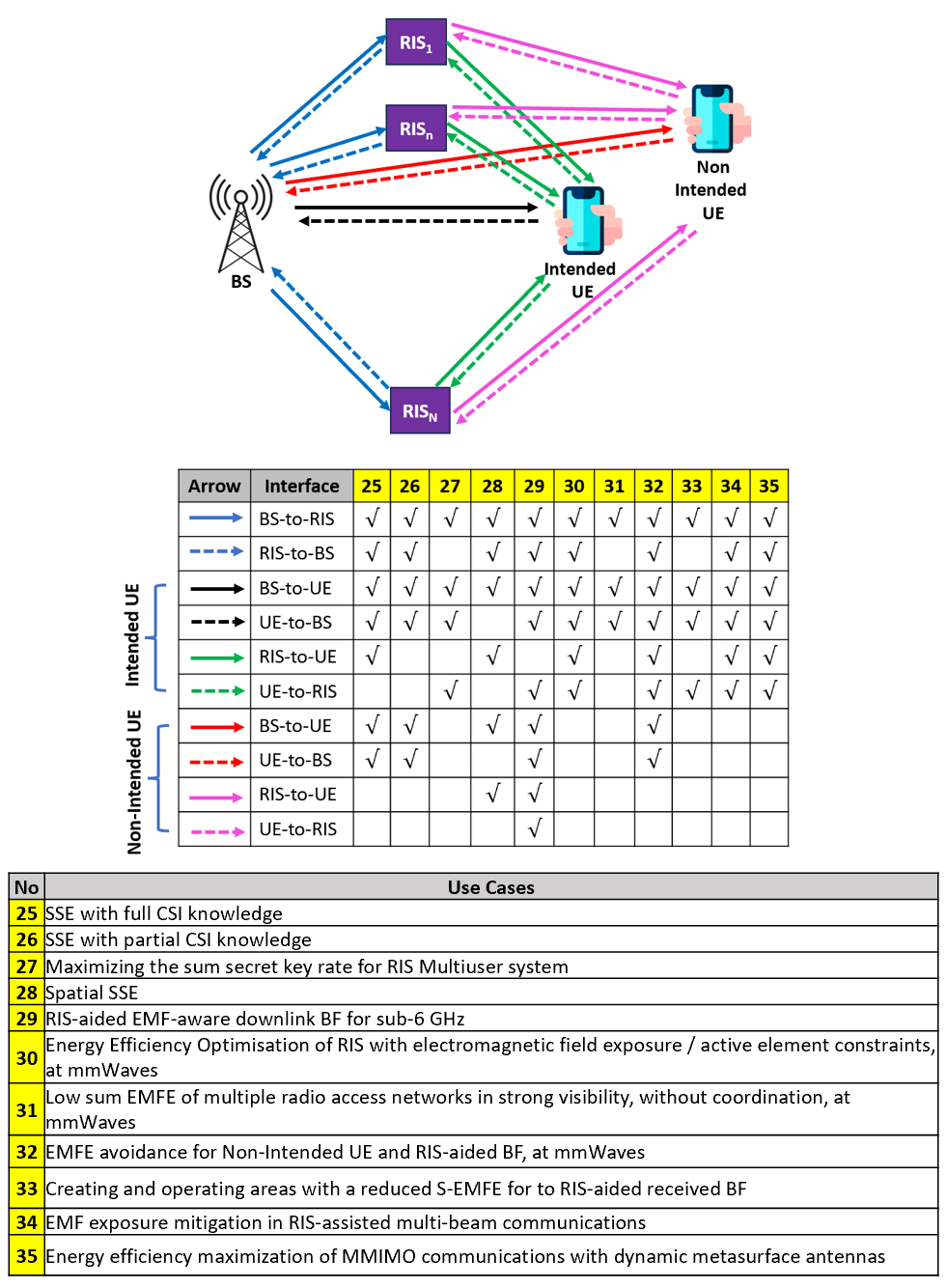}
\caption{Summary of category 3 of RIS use cases: enhanced
sustainability and security, including the block diagram of the involved network entities, a summary table of all considered interfaces with corresponding use case, and the list of the use cases. \cite{katsanos_network_2023}.}
\label{fig:usecase3}
\end{figure}

In the work of the RISE-6G project\cite{ mursia_deployment_2023, wymeersch_control_2023,katsanos_network_2023}, a total of 35 RIS use cases are classified into 3 main categories: (i) enhanced connectivity and reliability, (ii) enhanced
localization and sensing, and (iii) enhanced sustainability
and security. These 3 categories of use cases are illustrated in Figs. \ref{fig:usecase1}, \ref{fig:usecase2}, and \ref{fig:usecase3}, respectively. In each figure, an arrow represents an interface between two nodes with the head pointing toward the direction of the signal. The numbers in the columns correspond to a specific use case. These interfaces are essential to  ensuring proper system functionality for each use case. 

\setcounter{subsubsection}{0}
\subsubsection{Enhanced connectivity and reliability}

\begin{figure}[tb]
\centering
\includegraphics[width=\linewidth]{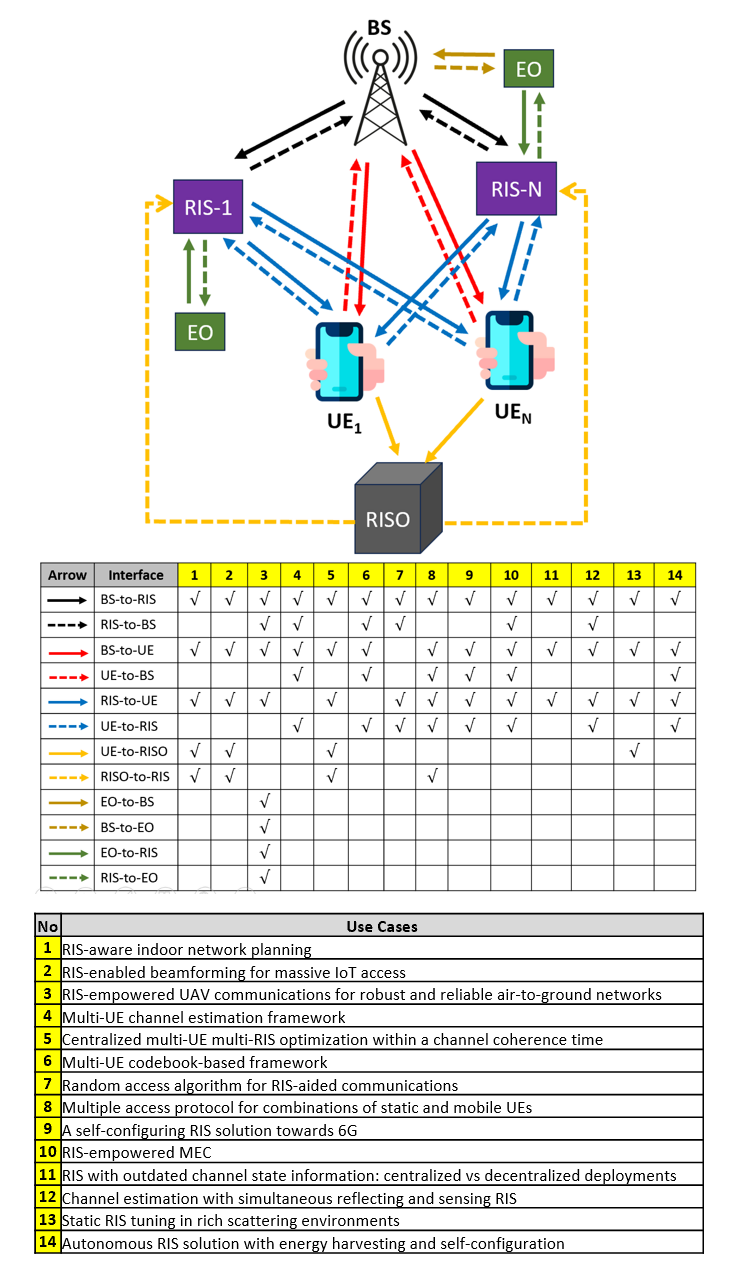}
\caption{Summary of category 1 of RIS use cases: enhanced connectivity and reliability, including the block diagram of the involved network entities, a summary table of all considered interfaces with corresponding use case, and the list of the use cases. \cite{mursia_deployment_2023}.}
\label{fig:usecase1}
\end{figure}

This category of use cases highlights the ability of RIS to facilitate the establishment of wireless communication by extending coverage while maintaining stable, high-quality connections. The details of this category can be found in \cite{mursia_deployment_2023}. This category encompasses 14 different use cases. The document \cite{mursia_deployment_2023} outlines each use case, detailing the involved network entities and their interfaces. In this category, the involved network entities include a BS, multiple UEs, multiple RISs, external operators (EOs), and an RIS orchestrator (RISO). An EO typically refers to a third-party entity that owns, deploys, manages, or controls the RIS infrastructure, independently of the mobile network operator, such as a human or a machine responsible for controlling the maneuvering of an RIS installed on a UAV. On the other hand, RISO operates at a higher layer coordinating
multiple RIS controllers. RISO is responsible for gathering all available information and solving the optimization problems that entail RIS configuration selection and potentially adjacent problems such as resource
allocation. A detailed summary of this category is provided in Fig. \ref{fig:usecase1}, including the involved network entities, the corresponding communication interfaces, and the associated use cases. In all use cases within this category, a set of RISs is exploited to enhance the communication between the BS and UEs by dynamically modifying the propagation environment to support various wireless tasks such as channel estimation, beamforming, and mobile edge computing (MEC).  

\subsubsection{Enhanced localization and sensing}
This category refers to RIS's ability to improve the precision of object localization and environmental sensing by controlling how signals reflect or scatter to improve navigation and monitoring in GPS-limited areas. \cite{wymeersch_control_2023} outlines each use case, detailing the involved network entities and their interfaces. In this category, the involved network entities includes multiple BSs, multiple UEs and multiple RISs.
This category encompasses the 10 use cases summarized in Fig. \ref{fig:usecase2}.

The technology that is relevant to this category is ISAC. In the past, communication and sensing functions were performed independently, requiring separate transceivers, additional bandwidth, and other devices. However, recent studies have highlighted the benefits of sharing the same resources for both communication and radar sensing, resulting in more efficient resource utilization \cite{ma_reconfigurable_2023,rihan_passive_2023, magbool_survey_2024}. ISAC provides a unified framework for communication and sensing, enabling both functions to operate within the same spectrum. This approach enables shared use of hardware, beamformers, and waveform designs, enhancing the overall system efficiency. As RIS is known for its capability in controlling the propagation environment, its integration with ISAC will further improve the resource efficiency, such as energy consumption and spectrum usage \cite{ma_reconfigurable_2023,rihan_passive_2023, magbool_survey_2024}.

\subsubsection{Enhanced sustainability and security}
This category highlights the role of RIS in optimizing energy efficiency by reducing power consumption and infrastructure requirements while enhancing data security through controlled signal propagation. By mitigating energy waste and minimizing the risk of eavesdropping, RIS contributes to a more sustainable and secure communication environment. To classify the use cases within this category, the term 'intended' is introduced. An intended UE refers to the legitimate recipient of the signal, whereas a non-intended UE denotes an entity that is not the designated destination. \cite{katsanos_network_2023} outlines each use case, detailing the involved network entities and their interfaces. In this category, the involved network entities includes a BS, multiple RISs, an intended UE, and a non-intended UE. This category encompasses 11 different use cases. Figure \ref{fig:usecase3} provides a detailed summary of this category.

\subsection{RIS Control Mechanisms} 
While the potential use cases of RIS are vast and transformative, their effective operation relies heavily on the underlying control mechanisms. In this section, the role of the control mechanism in managing and optimizing RIS functionalities is presented.

RIS is a metasurface that requires control to fully utilize its reconfigurability. For optimal operation, several key components must be in place: tunable chips, an RIS controller, inter-unit-cell communication, and a phase tuning mechanism \cite{gong_toward_2020}. These components work together to regulate and control the RIS, ensuring its intended functionality. RIS reconfigurability depends on a physical control mechanism that adjusts the impedance of its unit cells. This is achieved through tunable chips, typically PIN diodes, which are integrated into the metasurface. These chips connect the metasurface to the RIS controller, which governs their behavior to manipulate the electromagnetic response of the RIS. The RIS controller determines the state of the tunable chips by switching PIN diodes ON or OFF to modify the phase response of the metasurface. To enable flexible and adaptive operation, the RIS controller can be programmed via software, allowing real-time adjustments to dynamically change the system's behavior based on network conditions.

The reconfigurability of the RIS also relies on inter-unit-cell communication between the tunable chips, which collectively control the scattering elements of the metasurface, enabling the desired tunable behaviors. This communication can be implemented via wired or wireless means. Wired communication may offer advantages, such as facilitating integration with the controllers on the same chip. However, wireless communication has become a more feasible option for large-scale or densely packed metasurfaces. The design of inter-unit-cell communication protocols must address certain requirements to ensure optimal performance. 

Tuning the phase of the unit cell is another important feature for the reconfigurability of the RIS. Modifying the triggering parameter (electric, magnetic, or thermal) will change the electromagnetic properties of the metasurface.

The choice of the material to be layered with the RIS element influences the control performance, as different materials exhibit unique behaviors in processing amplitude, frequency, phase, and polarization of the incoming signal, as discussed in \cite{shafique_going_2024}. Lumped-based elements with varactors and PIN diodes operate below 100 GHz and can modify amplitude, phase, frequency, and polarization. In contrast, graphene-based elements, phase-changing materials, and liquid crystal-based elements can handle THz signals. However, graphene-based elements cannot vary in frequency, and phase-changing materials are highly sensitive to temperature variations. On the other hand, liquid crystal-based elements offer the advantage of negligible power consumption, making them particularly attractive from an energy-saving perspective.

\begin{figure}[tb]
\centering
\includegraphics[width=\linewidth]{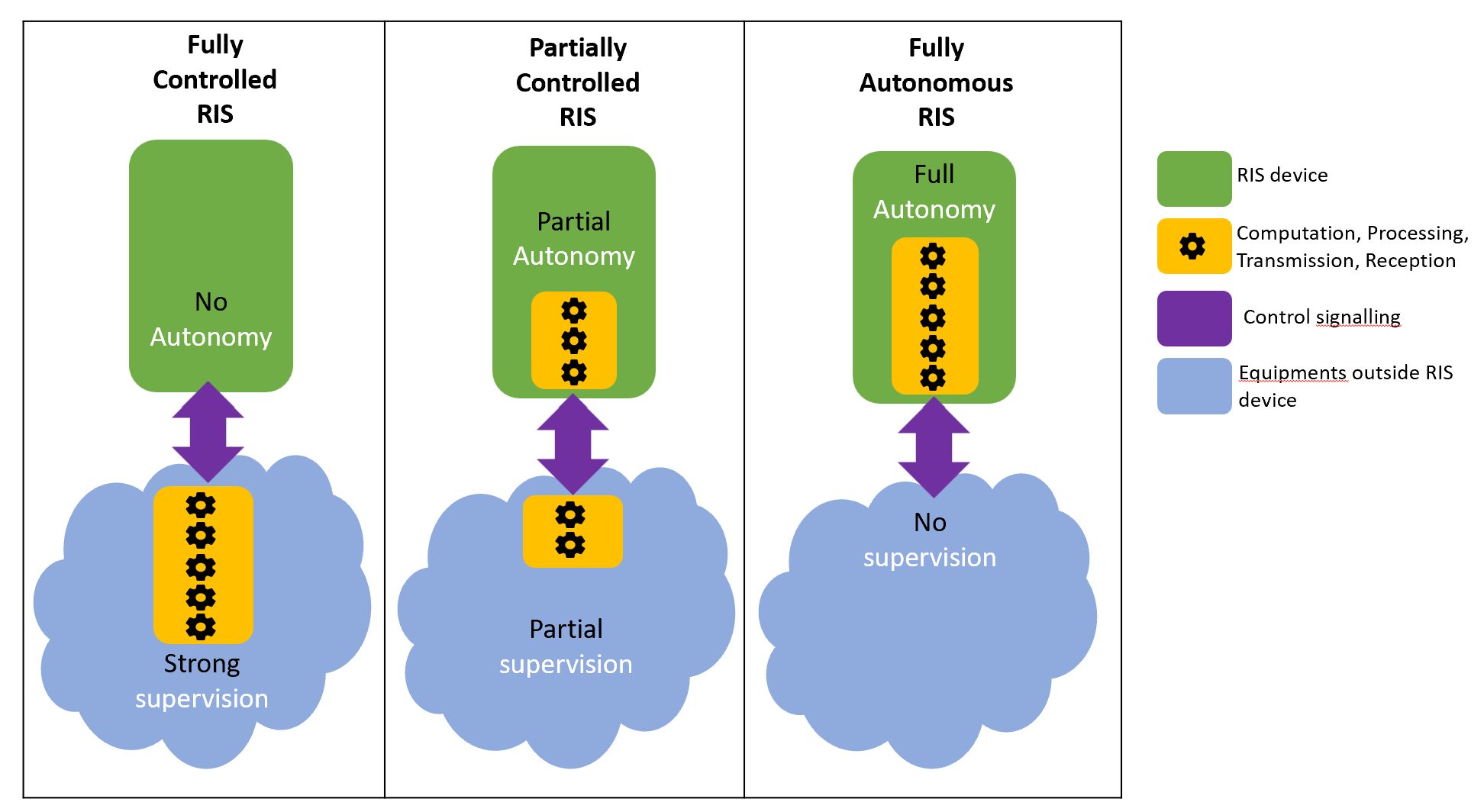}
\caption{RIS operation taxonomy \cite{dinh_orange}.}
\label{fig:riscontrol}
\end{figure}

The concept of the control mechanism of RIS extends beyond just the RIS controller, encompassing the system's overall functionality, including BS. In terms of operational control, RIS systems are typically categorized into three types: fully controlled, partially controlled, and fully autonomous \cite{dinh_orange, mursia_deployment_2023}. This categorization is illustrated in Fig. \ref{fig:riscontrol} \cite{dinh_orange}. 

In the fully controlled RIS configuration, the device lacks autonomy and cannot independently adjust its settings, necessitating strong supervision from external entities. In this configuration, the RIS performs only a limited set of functions, such as surface operation (e.g., reflection regulation via PIN diodes) and actuation (e.g., configuration by the microcontroller). However, the RIS does not make decisions regarding beamforming and reconfiguration computations. External entities, such as the BS, manage these functions. Consequently, the RIS relies on physical and MAC layer control signaling, which is facilitated by the RF modem to connect with the BS. The BS, in turn, handles control signaling to receive requests. This process is referred to as supervision, which is considered strong because the BS is responsible for various tasks, including pilot channel sensing (BS-RIS-UE and UE-RIS-BS), beamforming computation, RIS reconfiguration, scheduling, and sending reconfiguration commands \cite{dinh_orange}. In short, fully controlled RIS configuration's operations are controlled by
an external entity providing the
main computational processing.

In the partially controlled configuration, the control functions are distributed between the RIS and external entities. The RIS can adjust its settings based on information it independently acquires, but it also depends on additional signaling from other network components to complete its control processes. In this configuration, the RIS is capable of sensing and computing, including functions such as UE-RIS and BS-RIS channel sensing, RIS configuration computation, and applying reconfiguration along with scheduling. As a result, this configuration involves greater complexity in both functionality and hardware requirements, requiring a processing unit to handle these tasks. However, the RIS still operates under the supervision of the BS, as the BS manages the scheduling of reconfiguration.

Finally, the autonomous configuration, though relatively rare at present, allows the RIS to tune itself independently without any supervision from external entities. Hence, the RIS controller operates on its own to tune the RIS elements. However, the signaling to the BS is 
still needed for communication,
synchronization, or feedback
information.

To transmit a control signal of RIS and BS, a control channel is needed, and its category is shown in Table \ref{tab:ris_control_channel}. An explicit channel refers to a dedicated channel for transmitting control signals to the RIS, whereas an implicit channel has no dedicated one. In the case of an explicit channel, there are two types: out-of-band and in-band. The out-of-band type transmits control signals in a frequency band separated from the primary communication channel. This approach is cost-efficient but may not be highly spectrum-efficient. On the other hand, the in-band type allows control signals to overlap with the primary communication frequency band, offering improved spectrum efficiency but potentially increasing the complexity of the system design.

\begin{table}[t]
\centering
\caption{RIS Control Channel Taxonomy \cite{mursia_deployment_2023}}
\label{tab:ris_control_channel}
\begin{tabular}{|p{1.5cm}|p{5.5cm}|}
\hline
\textbf{Control Channel} & \textbf{Description} \\
\hline
\multirow{2}{*}{Explicit} & \textbf{Out-of-band}: Away from RIS main communication channel (by using different frequency), simpler design, but possibly lower spectral efficiency. \\
\cline{2-2}
 & \textbf{In-band}: Overlapping RIS operational spectrum resources, more complex design, but with possibly higher spectral efficiency. \\
\hline
Implicit & No dedicated channel. \\
\hline
\end{tabular}
\end{table}

\section{RIS-Assisted Channel Sounding}
\label{sec:sec3}

\begin{figure}[tb]
\centering
\includegraphics[width=0.6\linewidth]{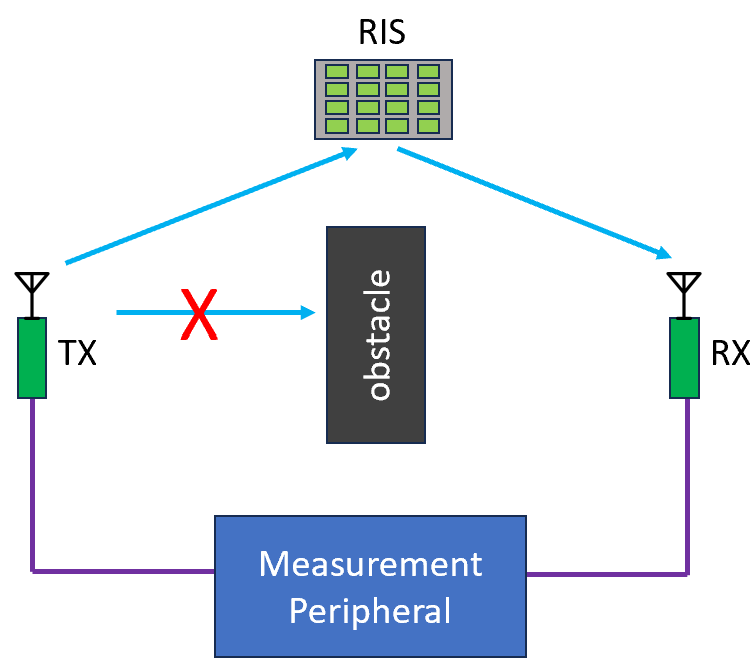}
\caption{A block diagram of an RIS-assisted channel sounding setup.}
\label{fig:sounding}
\end{figure}

While effective control mechanisms enable RIS to dynamically adapt to wireless environments, its performance is fundamentally constrained by the ability to accurately characterize the propagation channels in which they operate. Therefore, an empirical method to measure the actual wireless channel characteristics in a specific environment is needed. This method is known as channel sounding.  It involves transmitting signals to characterize the radio channel by analyzing the received impulse response, which helps in understanding how signals propagate through various paths. It provides real-world data on parameters such as delay spread, Doppler spread, path loss, and multipath effects. RIS-assisted systems may require rapid channel sounding operations to estimate channels between the RIS and UEs \cite{etsi01}. Fig. \ref{fig:sounding} shows a block diagram for channel sounding setup, using a transmitter, a receiver, RIS, and a measurement peripheral. One popular example of a measurement peripheral for channel sounding is the vector network analyzer (VNA) \cite{eckhardt_uniform_2023}. Multiple soundings (at least two) with different RIS phase shifts are necessary to resolve ambiguity problem in estimating the UE-RIS channel, specifically the complex path gain and the angles of arrival/departure.\cite{deepak_channel_2020}. 

\begin{table*}[!t]
\caption{Summary of Channel Sounding Experiments in RIS-Enabled Wireless Communications (2022-2024)
\label{tab:sounding}}
\centering
\begin{tabular}{|m{0.4cm}|>{\centering\arraybackslash}m{1cm}|>{\centering\arraybackslash}m{1cm}|>{\centering\arraybackslash}m{2.1cm}|>{\centering\arraybackslash}m{1.3cm}|>{\centering\arraybackslash}m{1.2cm}|>{\centering\arraybackslash}m{1.15cm}|>{\centering\arraybackslash}m{1.3cm}|>{\centering\arraybackslash}m{1.43cm}|>{\centering\arraybackslash}m{1.05cm}|>{\centering\arraybackslash}m{1.25cm}|}
\hline
    \textbf{Ref} & \textbf{Setup} & \textbf{Center Freq (GHz)} & \textbf{RIS Dimension} & \textbf{Sounding Method} & \textbf{Bandwidth} & \textbf{RIS Gain} & \textbf{Phase Resolution} & \textbf{Polarization} & \textbf{Tuning Element} & \textbf{Substrate} \\
\hline
\shortstack{\cite{rao_novel_2022}} & Indoor & 2.4  & 
  20 x 20 unit cells
 & VNA & 100 MHz &N/A & 4-bit digital & single & PIN diode & Rogers
4003C\\ \hline
\shortstack{\cite{liang_dual-band_2023}} & Indoor & 2.595, 3.45   & 
  8×11 unit cells, 
  248×429 mm$^2$
 & VNA & 150 MHz, 100 MHz &N/A & 3-bit digital & single & varactor diode & F4B\\ \hline
\shortstack{\cite{ren_time-domain_2024}} & Indoor & 2.6 & 
  32×16 unit cells,  
  1.6 × 0.8 m$^2$
 & NI USRP & 75 MHz & N/A& 1-bit digital & single & PIN diode & metal patch\\ \hline
\shortstack{\cite{liang_angle-insensitive_2022}} & Indoor & 3.15 & 254 × 272 mm$^2$ & VNA &450 MHz &N/A & 3-bit digital & single & varactor diode & F4B, metallic vias \\ \hline
\shortstack{\cite{araghi_reconfigurable_2022}} & Indoor & 3.5 & 
  2430 unit cells,  
  1.14 × 1.12 m$^2$
 & VNA &N/A & 15 dB & analog & single & varactor diode & F4BT450 \\ \hline
\shortstack{\cite{rains_high-resolution_2023}} & Indoor & 3.75 & 
  48×48 unit cells,  
  1.02 x 0.72 m$^2$
  & VNA & 105–240 MHz & N/A& 3-bit digital&single & PIN diode& F4BM-2 \\ \hline
\shortstack{\cite{tewes_comprehensive_2023}} & Indoor & 5.5 & 
  16 × 16 unit cells, 
  400 x 320 mm$^2$ 
  & VNA & 1 GHz& N/A&1-bit digital &single & binary switch & FR4\\ \hline
\shortstack{\cite{she_intelligent_2022}} & Indoor & 7.4 & 
  30 × 30 unit cells,  
  15 x 15 cm$^2$
  & VNA &5.2 GHz & N/A&1-bit digital & single& PIN diode & F4B, FR4\\ \hline
\shortstack{\cite{ma_wideband_2023}} & Indoor &  10.55 & 24 × 24 unit cells, 440 × 400 mm$^2$ & N/A & 2.4 GHz & 23.3 dBi &1-bit digital &single &PIN diode &GF265, RO4350B  \\ \hline
\shortstack{\cite{rahal_ris-aided_2023}} & Indoor & 28 & 20×20 unit cells & VNA & 2 GHz & 27.9 dBi &1-bit digital &dual &PIN diode &Meteorwave 8000, Rogers Duroid RT6002 \\ \hline
\shortstack{\cite{wang_wideband_2024-1}} & Indoor & 27.5 & 20×20 unit cells & VNA & 7 GHz & 24.7 dB &1-bit digital &single & PIN diode &Rogers 5880, FR4 \\ \hline
\shortstack{\cite{kim_independently_2023}} & Indoor & 28 & 50 × 50 mm$^2$ & VNA & N/A & 17.9 dBi &analog &single &liquid crystal (LC) &Taconic TLY-5, FR4 \\ \hline
\shortstack{\cite{jeong_improved_2022}} & Indoor & 29 & 24 x 24 unit cells & VNA & N/A & 15.46 dBi & 2-bit digital & single & PIN diode & IsolaI-TeraMT \\ \hline
\shortstack{\cite{jeong_mechatronic_2022}} & Indoor & 32 & 112 × 112 mm$^2$ & VNA & 4 GHz & 3 dBi & 1-bit digital & single & motor and gear structure & resin \\ \hline
\shortstack{\cite{zhou_wideband_2024}} & Indoor & 102 & 31 × 31 unit cells & VNA &  27.95 & 25.1 dBi & 1-bit digital & single & RF switch & RO3003, RO4450F,  FR4  \\ \hline
\shortstack{\cite{alexandropoulos_characterization_2024}} & Indoor & 304 & 
  100×100 unit cells,  
  50 x 50 mm$^2$
  & Sounder based on \cite{eckhardt_hybrid_2024} & 8 GHz & N/A& {1,2,3}-bit digital&N/A & N/A& dielectric \\ \hline
\shortstack{\cite{lan_measurement_2023}} & Outdoor & 2.6 & 
  32×16 unit cells,  
  1.6 x 0.8 m$^2$
  & VNA & 190 MHz & N/A& 1-bit digital&single & binary switch& dielectric \\ \hline
\shortstack{\cite{gao_multi-frequency_2024}} & Outdoor & 2.75, 35 & 
  32×16 (2.75 GHz),  
  60×60  (35 GHz) 
 & vector signal analyzer & 100 MHz, 300 MHz & 3.4 dBm, 22.4 dBm&1-bit digital &single &N/A & N/A\\ \hline
\shortstack{\cite{zheng_field_2023}} & Indoor, Outdoor & 2.6 & 
  32×16 unit cells, 
  1.6x0.8 m$^2$
  & NI USRP &20 MHz &N/A &1-bit digital &single &PIN diode &dielectric \\ \hline
\shortstack{\cite{pei_two-stage_2024}} & Indoor, Outdoor & 2.6, 5.8 & 
  N/A 
  & spectrum analyzer & 20 MHz & 4–22 dB & N/A &N/A & N/A & N/A\\ \hline
\shortstack{\cite{youn_liquid-crystal-driven_2023}} & Indoor, Outdoor & 29.5 & 60 x 60 mm$^2$
  & SDR & 800 MHz & 19.3–21.2 dBi & 1-bit digital & single & LC & corning glass\\ \hline
\shortstack{\cite{sang_multi-scenario_2024}} & Indoor, Outdoor, O2I & 2.6 & 
  32×16 unit cells,
  5 cm × 5 cm 
 & VNA &190 MHz & 30 dB&1-bit digital &single &PIN diode & N/A\\ \hline
\shortstack{\cite{yuan_field_2024}} & Indoor, Outdoor, O2I & 2.6 & 
  16×16 unit cells, 
  2 m × 2 m 
 & commercial 5G test equipment &200 MHz& N/A&2-bit digital & N/A& N/A& N/A\\ \hline
\end{tabular}
\end{table*}

\subsection{RIS-assisted Channel Sounding Campaigns}
Channel sounding campaigns are generally categorized based on the environment, into two types: indoor and outdoor. Additionally, they can be further classified according to the operating frequency range, including sub-6 GHz and mmWave bands. This survey focuses on papers published between 2022 and 2024, as \cite{huang_reconfigurable_2022} has already covered earlier works in this domain. Table \ref{tab:sounding} summarizes  the surveyed channel sounding experiments detailing various location setups, working frequencies, RIS dimensions, RIS substrates, tuning elements, and sounding methods. In most of the works in Table \ref{tab:sounding}, horn antennas are employed for both signal transmission and reception. Additionally, VNAs are the most commonly used measurement instruments, while a smaller subset of works utilizes National Instrument universal software radio peripheral (NI-USRP), software defined radio (SDR) platforms, vector signal analyzers, spectrum analyzers, and commercial 5G test equipment.

Indoor channel sounding has been investigated in \cite{rao_novel_2022,liang_dual-band_2023,ren_time-domain_2024,liang_angle-insensitive_2022,araghi_reconfigurable_2022,tewes_comprehensive_2023,she_intelligent_2022,ma_wideband_2023,rahal_ris-aided_2023,wang_wideband_2024-1,kim_independently_2023,jeong_improved_2022,jeong_mechatronic_2022,zhou_wideband_2024,rains_high-resolution_2023}.
Specifically, the experiments in \cite{liang_dual-band_2023,liang_angle-insensitive_2022,tewes_comprehensive_2023,she_intelligent_2022,ma_wideband_2023,kim_independently_2023,jeong_improved_2022} were conducted inside standard
microwave anechoic chambers, whereas those in \cite{rao_novel_2022,ren_time-domain_2024,araghi_reconfigurable_2022,rahal_ris-aided_2023,wang_wideband_2024-1,jeong_mechatronic_2022,zhou_wideband_2024,rains_high-resolution_2023,alexandropoulos_characterization_2024} were performed in general indoor environments without absorber treatment. Anechoic chamber measurements offer a highly controlled and isolated setting where multipath components and delay spread are minimized due to the presence of electromagnetic absorbers on surrounding surfaces. In contrast, measurements in typical indoor environments capture more realistic propagation characteristics, thereby providing insight into the practical performance of RIS in multipath-rich scenarios.
In most of those studies, the measurement setup follows an L-shaped configuration comprising the transmitter, RIS, and receiver. However, in \cite{wang_wideband_2024-1,araghi_reconfigurable_2022,she_intelligent_2022,alexandropoulos_characterization_2024}, the transmitter and receiver antennas are facing the RIS, which is placed at the same distance from both of them. The L-shaped transmitter–RIS–receiver configuration is typically employed to evaluate the angular reflection and beam steering capability of RIS, particularly under non-line-of-sight (NLoS) conditions. In contrast, the co-located transmitter–receiver setup with a distant RIS is designed to assess path loss, phase stability, and the distance-dependent behavior of RIS-assisted links.

Outdoor channel sounding has been studied in \cite{lan_measurement_2023,gao_multi-frequency_2024}. The experiment in \cite{lan_measurement_2023} considers the transmitter, receiver, and RIS to be positioned outside a building in an L-shaped position where the RIS is situated at the corner between the transmitter and receiver. The channel measurement is conducted using sub-6 GHz using VNA. On the other hand, the work in \cite{gao_multi-frequency_2024}, investigates sub-6 GHz and mmWave frequencies under LoS, NLoS, and obstructed-LoS (OLoS) conditions, considering different RIS phase configurations. The transmitter, RIS, and receiver are placed outside a building near a park in which the angle is varied from nearly $0^\circ$ until it forms an L-shape. The results demonstrate that RIS can enhance received power in both sub-6 GHz and mmWave frequency bands, with more pronounced gains observed at mmWave frequencies due to the inherently weaker multipath propagation in that band.

Research in \cite{zheng_field_2023,pei_two-stage_2024,youn_liquid-crystal-driven_2023} investigates both indoor and outdoor channel sounding scenarios. For indoor measurements, an L-shaped transmitter–RIS–receiver configuration is employed, whereas the outdoor setups utilize oblique incidence angles.

In \cite{sang_multi-scenario_2024} and \cite{yuan_field_2024}, the measurement includes indoor, outdoor, and outdoor-to-indoor (O2I) scenarios.
The experiments in \cite{sang_multi-scenario_2024} 
implemented an L-shaped transmitter–RIS–receiver configuration for all indoor, outdoor, and outdoor-to-indoor (O2I) channel soundings, whereas \cite{yuan_field_2024} used oblique incident angles. For the O2I scenario, the transmitter is located outside a building, and the receiver is located inside a building. 
 
Indoor settings typically involve shorter distances, higher multipath richness, and a more controlled environment, making channel sounding simpler but requiring careful consideration of material effects (such as furniture) and spatial configuration. In contrast, outdoor settings are characterized by larger distances, dynamic environments, and more complex propagation conditions, demanding more sophisticated channel sounding to optimize RIS performance. From the frequency perspective, free-space path loss is more pronounced in high-frequency signals than lower-frequency signals. Additionally, outdoor settings experience more frequent blockages and environmental changes, making outdoor RIS channel sounding more challenging and susceptible to error. It seems to be the reason why the successful experiments on outdoor channel soundings are relatively more limited compared to the indoor ones.

In addition, while RIS is anticipated to enhance UAV communication in Beyond-5G (B5G) systems, there is currently a lack of studies addressing channel sounding in RIS-assisted UAV communication, as noted in \cite{mao_survey_2024}. This highlights a potential area for further exploration.

\subsection{Channel Characterization}
Channel sounding techniques provide empirical data on the wireless propagation environment, enabling the estimation of key channel parameters. However, raw measurements alone are insufficient to fully understand the impact of RIS on wireless communication. Channel characterization aims to analyze these measurements (by extracting key channel properties such as path loss and delay spread) and relate them to theoretical models. In the context of cellular systems, this process is essential for understanding how scattering environments influence the power and frequency behavior of transmitted signals \cite{zhang_channel_2015}.

\subsubsection{Path Loss}
The RIS-assisted channel exhibits a cascaded structure involving the BS–RIS and RIS–UE links. According to \cite{yuan_white_2025,li_path_2022}, the path loss can be expressed as: 
\begin{align}
PL(d_1, d_2) &= PL_{\text{1}}(d_1) + PL_{\text{2}}(d_2) \notag \\
&\quad + 10 \log_{10} \left( \frac{\lambda^2}{4\pi} \right) - 10 \log_{10} (\sigma_{\text{RIS}}),
\end{align}
where $d_1$, $d_2$, $PL_{\text{1}}$, and $PL_{\text{2}}$ are the BS–RIS and RIS–UE channel distances and pathlosses, respectively. $\lambda$ represents the wavelength and $\sigma_{\text{RIS}}$ denotes the RIS average radiation gain. 

More accurate models that incorporate the angular dependence are the empirical modified Floating Intercept (FI) and Close-In (CI) models, given by \cite{sang_multi-scenario_2024} as
\begin{subequations}\label{eq:PL_models}
\begin{align}
PL_{\text{FI}}(d_1, d_2, \theta_t, \theta_r) &= \alpha + 10 \beta_1 \log_{10} d_1 + 10 \beta_2 \log_{10} d_2 \notag \\
&\quad - 10 \lambda_1 \log_{10} \cos \theta_t - 10 \lambda_2 \log_{10} \cos \theta_r \notag \\
&\quad + X^\text{FI}_\sigma ,\label{eq:FI}
\end{align}
\begin{align}
PL_{\text{CI}}(d_1, d_2, \theta_t, \theta_r) &= PL_{FS}(d_0^1,d_0^2,\theta_0^t, \theta_0^r) + 10 n_1 \log_{10} \frac{d_1}{d^1_0} \notag \\
&\quad + 10 n_2 \log_{10} \frac{d_2}{d^2_0}  - 10 \mu_1 \log_{10} \frac{\cos \theta_t}{\cos \theta_0^t} \notag \\
&\quad - 10 \mu_2 \log_{10} \frac{\cos \theta_r}{\cos \theta_0^r} + X^\text{CI}_\sigma, \label{eq:CI}
\end{align}
\end{subequations}
where $\theta_t$ and $\theta_r$ denote the angle of arrival (AoA) from BS to RIS and angle of departure (AoD) from RIS to BS, respectively. $\alpha$ is the intercept parameter of the PL. $\beta_1$ and $\beta_2$ represent the path-loss exponents (PLEs) on $d_\text{1}$ and $d_\text{2}$. $\lambda_1$ and $\lambda_2$ are the PLEs on $\theta_t$ and $\theta_r$. $X^\text{FI}_\sigma$ denotes the shadow factor (SF) that follows a Gaussian distribution. $PL_{FS}(d_0^1,d_0^2,\theta_0^t, \theta_0^r)$ corresponds to the free space path loss under these variables, with $d_0^1$, $d_0^2$, $\theta_0^t$, and $\theta_0^r$ denoting the reference distances of $d_1$ and $d_2$ as well as the reference angles of $\theta_t$ and $\theta_r$, respectively. $n_1$ and $n_2$ represent the PLEs on $d_1$ and $d_2$. $\mu_1$ and $\mu_2$ denote the PLEs on $\theta_t$ and $\theta_r$. $X^\text{CI}_\sigma$ is the SF with Gaussian distribution. 
The effects of the RIS physical size, number of unit cells, reflection magnitude, and reflection phase on the PL have been included in the terms $\alpha$ and $PL_{FS}(d_0^1,d_0^2,\theta_0^t, \theta_0^r)$ in the modified FI and CI models, respectively. Authors of \cite{yuan_white_2025} proposes a power control factor (PCF) parameter to model the coupling between the cascaded BS–RIS–UE channel and the direct BS–UE channel, as follows:
\begin{subequations}\label{eq:PL_coupled}
\begin{align}
PL_{\text{cou}} &= PL_{\text{cas}} + PL_{\text{dir}} \label{eq:PLcou_a} \\
PL_{\text{dir}} &= O_{\text{RIS}} \cdot PL_{\text{env}} \label{eq:PLcou_b}
\end{align}
\end{subequations}
where $PL_{\text{cou}}$ denotes the overall coupling PL, 
$PL_{\text{cas}}$ represents the cascaded BS-RIS-UE channel PL, and 
$PL_{\text{dir}}$ corresponds to the BS–UE direct channel PL. 
$PL_{\text{env}}$ is the background environmental PL without RIS, generally taken from standard channel models like 3GPP TR 38.901 \cite{3gpp_study_2022}.
$O_{\text{RIS}}$ indicates a PCF that offsets the environmental PL when RIS is introduced into the scenario.

\subsubsection{Scattering Gain Loss}
The study in \cite{stefanini_characterization_2023} characterizes scattering gain loss, considering practical factors such as manufacturing inaccuracies, physical deformations, temperature fluctuations, and dust accumulation, which can impact the scattering efficiency of RIS. This phenomenon is called electromagnetic random roughness. 

The RIS is characterized by a phase profile $\varphi_{M}(x)$ along the
x-direction, so the reflection
coefficient is given by
\begin{equation}
    \Gamma(x) = |\Gamma_0| \exp[j \varphi_{M}(x)],
\end{equation}
where $\varphi_{M}(x)$ denotes the phase profile that controls the tangential momentum for the incident plane wave, ensuring that it is reflected in the desired direction. As the RIS is assumed to be an impenetrable lossless reflective
surface, $|\Gamma_0|=1$. The reflected field $E_q(x, z)$ is connected to incident wave $E_p(x, z)$ at $z=0$, thus
\begin{equation}
    E_q(x, z = 0) = \Gamma(x) E_p(x, z = 0) \prod_D(x),
\end{equation}
where \( \prod_D(x) \) represents the rectangular function modeling the finite dimension of the RIS. The phase roughness is assumed to be random noise, in addition to $\Gamma(x)$. The randomness of the 
the phase error on the RIS is assumed to follow a Gaussian probability
density function (pdf)
\begin{equation}
\text{pdf}(\hat{\phi}) = \frac{1}{\sqrt{2 \pi \Delta \phi^2}} \exp\left( -\frac{1}{2} \left( \frac{\hat{\phi} - \bar{\phi}}{\Delta \phi} \right)^2 \right),
\end{equation}
where \( \hat{\phi} \) is the random variable of the additional phase, \( \bar{\phi} \) is the average deviation given by the additional phase, and \( \Delta \phi \) is the range of that deviation.
So, the renewed model of the RIS is a random complex variable given by
\begin{equation}
    \hat{\Gamma}_{\phi}(x) = \Gamma(x) \exp[j \hat{\phi}],
\end{equation}

The range of deviation \( \Delta \phi \) indicates the expected distance from the ideal profile, thus influencing the radiation performance. The expected scattering gain loss is given by
\begin{equation}
G_{\text{sc-loss, dB}} = -20 \log_{10} \left( \exp(-0.5 \Delta \phi^2) \right),
\label{eq:gloss}
\end{equation}

\subsubsection{Delay Spread} Delay spread is characterized as the maximum time difference between signals arriving directly from the BS and those reflected through an RIS. These differences arise because reflected signals travel longer paths and can be delayed further depending on how each RIS element modifies the signal phase. To reduce this delay spread, the work in \cite{wang_how_2022} optimizes the RIS phase settings to align the arrival times of the reflected signals as closely as possible with the direct signal.

The authors of \cite{wang_how_2022} conclude that, to effectively mitigate delay spread in high speed train (HST) communication, deploying the RIS on the train side is preferable to the railroad side. This recommendation aligns with the observation that reducing the distance of the cascaded link can alleviate channel variations. In the case of railroad-side RIS deployment, the relative distance between the static RIS and the high-mobility user varies rapidly, resulting in significant phase fluctuations and misalignment between direct and RIS-assisted links. Conversely, when the RIS is mounted on the train, the RIS-user distance becomes small and nearly constant, leading to reduced path loss and a quasi-static channel.

\subsubsection{Small-scale Fading}
Small-scale fading is characterized by the power delay profile (PDP) decay, which obeys either an exponential law or a power law. The PDP power-law decay satisfies $p_k = a / \tau_k^n$ where $p_k$ is the average power corresponding to each discrete delay $\tau_k$, $k$ is the discrete tap, and $a$ is a constant. In logarithmic form, the power-law decay model is expressed as

\begin{equation}
10 \log_{10} p_k = \eta_0 - 10 n_{\mathrm{PDP}} \log_{10} \tau_k + A_{\mathrm{PDP}},
\end{equation}
where $n_{\mathrm{PDP}}$ is the decay factor, $\eta_0 = 10 \log_{10} a$ is a constant, and $A_{\mathrm{PDP}}$ is a normally distributed random variable in dB, i.e., $A_{\mathrm{PDP}} \sim \mathcal{N}(0, \sigma^2)$. 
For a PDP exponential decay, $p_k = e^{-\tau_k / \gamma}$, which its logarithmic form is described as
\begin{equation}
10 \log_{10} p_k = \frac{-10 \tau_k}{\gamma_{\mathrm{PDP}} \ln 10} + A_{\mathrm{PDP}},
\end{equation}
where $\gamma_{\mathrm{PDP}}$ is the decay factor.

The measurements in \cite{ren_time-domain_2024} were performed using a USRP at 2.6 GHz in two indoor NLoS scenarios with an L-shaped configuration: a
corridor and a laboratory. The corridor setup included 30 grid points (10 rows × 3
columns), while the laboratory setup consisted of 22 points arranged in horizontal and vertical
directions. For each measurement point, multiple PDPs were recorded to ensure
accuracy. Specifically, 10 PDPs were averaged at each location to smooth out noise,
reliably representing the channel characteristics. The transmitter used a horn antenna, while the receiver employed an omnidirectional
antenna, both positioned at a height of 1.2 m. The transmitter was fixed and the receiver was moved to
various measurement points to capture the PDP at different angles of reflection. The angle of incidence on the RIS was
consistently set at 45°. Three channel setups were evaluated: without RIS (representing a standard NLoS channel), with
RIS specular reflection (passive metal-like reflection), and with RIS intelligent reflection
(phase-coded to focus the beam towards the receiver). The results indicated that the
PDPs under RIS intelligent reflection better fit the power-law decay model than the exponential one. The RIS intelligent reflection enhances energy focusing along the virtual LoS path, leading to reduced signal delay spread (up to 13 ns in the laboratory and 4.2 ns in the corridor) and diminished multipath fading. The authors also performed clustering analysis using an
improved Saleh-Valenzuela (S-V) model, revealing that the cluster characteristics are highly
scenario-dependent. In the corridor scenario, multiple clusters are observed due to repeated reflections from parallel walls, windows, and other reflective surfaces. In contrast, the laboratory scenario under RIS intelligent reflection typically exhibits a single dominant cluster associated with the strong virtual LoS path, characterized by a smoothly decaying envelope and minimal secondary clusters.

\subsubsection{Fast Fading}

The fast fading component of the RIS-assisted channel captures the cascaded nature of the BS–RIS and RIS–UE channels, as well as the influence of the RIS radiation pattern. A Geometry-Based Stochastic Model (GBSM) can be employed to characterize this behavior, with its parameters derived following the guidelines specified in 3GPP TR 38.901\cite{3gpp_study_2022}. GBSM convolves multipath components from the BS-RIS and RIS-UE channels \cite{yuan_white_2025}. With this model, the channel impulse response (CIR) of the BS–RIS–UE link is expressed by (\ref{eq:gbsm}), as proposed in \cite{gong_how_2024} and experimentally validated in \cite{zhang_cascaded_2024}. Equation (\ref{eq:gbsm}) considers a single-user scenario, where the BS is equipped with $M_{\text{BS}}$ transmit antennas indexed by $b = 1, 2, \ldots, M_{\text{BS}}$, and the UE is equipped with $M_{\text{UE}}$ receive antennas indexed by $u = 1, 2, \ldots, M_{\text{UE}}$. The equation describes the channel response coefficient from the $b$-th BS antenna to the $u$-th UE antenna through the RIS at time $t$ with delay $\tau$. $n_i$ and $m_i$, $i \in \{1, 2\}$ denote the cluster and path indices in the BS–RIS and RIS–UE sub-channels, respectively. $P_{n_1,m_1}$ and $P_{n_2,m_2}$ represent the power of the corresponding propagation paths. $p_1, p_2 \in \{v, h\}$ indicate the vertical and horizontal polarization components, respectively. $F^{v}_{\text{UE},u}, F^{h}_{\text{UE},u}, F^{v}_{\text{BS},b}$ and $F^{h}_{\text{BS},b}$ are the radiation pattern of the $u$-th UE antenna and the $b$-th BS antenna in the $v$ and $h$ polarization, respectively. $\phi^{\text{in}}_{n_1,m_1}, \theta^{\text{in}}_{n_1,m_1}, \theta^{\text{BS}}_{n_1,m_1}, \phi^{\text{BS}}_{n_1,m_1}$ denote the angle of the $(n_1, m_1)^{\text{th}}$ path in the BS–RIS sub-channel, corresponding to the azimuth angle of arrival, the zenith angle of arrival, the azimuth angle of departure, and the zenith angle of departure, respectively. Similarly, $\phi^{\text{UE}}_{n_2,m_2}, \theta^{\text{UE}}_{n_2,m_2}, \theta^{\text{out}}_{n_2,m_2}, \phi^{\text{out}}_{n_2,m_2}$ represent the azimuth angle of arrival, the zenith angle of arrival, azimuth angle of departure, and the zenith angle of departure of the $(n_2, m_2)^{\text{th}}$ path in the RIS-UE subchannel. $\Phi^{p_1 p_2}_{n_i, m_i},\ i \in \{1, 2\}$ indicates the random phase associated with the $(n_i, m_i)^{\text{th}}$ path, which departs with polarization $p_1$ and arrives with polarization $p_2$. $\kappa_{n_i, m_i},\ i \in \{1, 2\}$ is the cross polarization power ratio (XPR) of the corresponding path. $F^{p_1 p_2}_{\text{RIS}}$ denotes the radiation pattern of the RIS. $p_1p_2$ indicates that the RIS has different effects on the path incident in the $p_1$ polarization direction and emitted in the $p_2$ polarization direction. $\mathbf{r}^{\text{UE}}_{n_2,m_2}$ and $\mathbf{r}^{\text{BS}}_{n_1,m_1}$ are the unit direction vectors of the corresponding path at UE and BS. $\mathbf{d}^{\text{UE}}_u$ and $\mathbf{d}^{\text{BS}}_s$ represent the location vectors of the $u$-th UE antenna and the $b$-th BS antenna. $\lambda$ is the signal wavelength. $f_{n_2,m_2}$ denotes the Doppler shift of the $(n_2, m_2)^{\text{th}}$ path. $\tau_{n_i, m_i},\ i \in \{1, 2\}$ indicates the delay of the corresponding path.

\begin{figure*}[!b]
\hrulefill
\begin{equation}
\resizebox{\textwidth}{!}{$
\begin{aligned}
h_{u,b}^{\text{RIS}}(t, \tau) &= \sum_{n_1, m_1}^{N_1, M_1} \sum_{n_2, m_2}^{N_2, M_2} 
\sqrt{P_{n_1,m_1} P_{n_2,m_2}} 
\left[
\begin{array}{c}
F_{\text{UE},u}^{v}\left(\theta^{\text{UE}}_{n_2,m_2}, \phi^{\text{UE}}_{n_2,m_2}\right) \\
F_{\text{UE},u}^{h}\left(\theta^{\text{UE}}_{n_2,m_2}, \phi^{\text{UE}}_{n_2,m_2}\right)
\end{array}
\right]^T \cdot
\left[
\begin{array}{cc}
\exp(j \Phi_{n_2,m_2}^{vv}) & \sqrt{\kappa_{n_2,m_2}^{-1}} \exp(j \Phi_{n_2,m_2}^{vh}) \\
\sqrt{\kappa_{n_2,m_2}^{-1}} \exp(j \Phi_{n_2,m_2}^{hv}) & \exp(j \Phi_{n_2,m_2}^{hh})
\end{array}
\right] \\
&\quad \cdot
\left[
\begin{array}{cc}
F_{\text{RIS}}^{vv}\left(\phi^{\text{in}}_{n_1,m_1}, \theta^{\text{in}}_{n_1,m_1}, \phi^{\text{out}}_{n_2,m_2}, \theta^{\text{out}}_{n_2,m_2}\right) & 
F_{\text{RIS}}^{vh}\left(\phi^{\text{in}}_{n_1,m_1}, \theta^{\text{in}}_{n_1,m_1}, \phi^{\text{out}}_{n_2,m_2}, \theta^{\text{out}}_{n_2,m_2}\right) \\
F_{\text{RIS}}^{hv}\left(\phi^{\text{in}}_{n_1,m_1}, \theta^{\text{in}}_{n_1,m_1}, \phi^{\text{out}}_{n_2,m_2}, \theta^{\text{out}}_{n_2,m_2}\right) & 
F_{\text{RIS}}^{hh}\left(\phi^{\text{in}}_{n_1,m_1}, \theta^{\text{in}}_{n_1,m_1}, \phi^{\text{out}}_{n_2,m_2}, \theta^{\text{out}}_{n_2,m_2}\right)
\end{array}
\right] \\
&\quad \cdot
\left[
\begin{array}{cc}
\exp(j \Phi_{n_1,m_1}^{vv}) & \sqrt{\kappa_{n_1,m_1}^{-1}} \exp(j \Phi_{n_1,m_1}^{vh}) \\
\sqrt{\kappa_{n_1,m_1}^{-1}} \exp(j \Phi_{n_1,m_1}^{hv}) & \exp(j \Phi_{n_1,m_1}^{hh})
\end{array}
\right]
\left[
\begin{array}{c}
F_{\text{BS},b}^{v}\left(\theta^{\text{BS}}_{n_1,m_1}, \phi^{\text{BS}}_{n_1,m_1}\right) \\
F_{\text{BS},b}^{h}\left(\theta^{\text{BS}}_{n_1,m_1}, \phi^{\text{BS}}_{n_1,m_1}\right)
\end{array}
\right] \\
&\quad \cdot \exp\left( j \frac{2\pi}{\lambda} \left( \mathbf{r}^{\text{UE}}_{n_2,m_2} \cdot \mathbf{d}_u^{\text{UE}} + \mathbf{r}^{\text{BS}}_{n_1,m_1} \cdot \mathbf{d}_b^{\text{BS}} \right) \right) 
\cdot \exp(j 2\pi f_{n_2,m_2} t) \cdot \delta\left( \tau - \tau_{n_1,m_1} - \tau_{n_2,m_2} \right)
\end{aligned}
$}
\label{eq:gbsm}
\end{equation}
\end{figure*}

\section{Channel Estimation for RIS-assisted Systems}
\label{sec:sec4}

\begin{table*}[t]
\centering
\caption{Resume of the contribution papers about channel estimation in RIS. A red asterisk (\textcolor{red}{*}) indicates scenarios with high mobility.}
\label{tab:resume}
\resizebox{\linewidth}{!}{
\begin{tabular}{|>{\centering \arraybackslash}c|c|c|c|c|c|c|c|}
\hline
\multirow{3}{*}{\shortstack{Deployment \\Strategy}} & \multirow{3}{*}{User Setting} & \multirow{3}{*}{\shortstack{Antenna \\ Configuration \\ (Downlink)}}  & \multicolumn{5}{c|}{Channel Estimation Technique} \\ \cline{4-8} 
 &  &  & \multicolumn{3}{c|}{Conventional} & \multicolumn{2}{c|}{Machine-learning based} \\ \cline{4-8} 
 &  &  & With Passive RIS & With Active RIS & With Hybrid RIS & With Passive RIS & With Hybrid RIS \\ \hline
\multirow{8}{*}{Single RIS} & \multirow{4}{*}{Single User} & SISO & \shortstack{\cite{azizzadeh_pilot-aided_2023},\cite{xu_reconfigurable_2022}\textcolor{red}{*}, \cite{xu_channel_2023}\textcolor{red}{*},\cite{zheng_progressive_2023},\cite{li_variational_2024}} & \cite{kang_active-irs-aided_2024} &  & \shortstack{\cite{kim_deep_2023},\cite{zhang_deep_2021},\cite{yang_joint_2024}, \cite{wang_channel_2021},\cite{xu_deep_2022}\textcolor{red}{*}} & \cite{taha_enabling_2021} \\ \cline{3-8} 
 &  & SIMO &  &  &  & \shortstack{\cite{feng_mmwave_2023},\cite{tseng_untrained_2024}, \cite{dampahalage_supervised_2022}, \cite{fredj_variational_2023},\cite{xiao_multi-scale_2023},\cite{chen_attention-based_2024},\cite{tong_diffusion_2024}} &  \\ \cline{3-8} 
 &  & MISO & \shortstack{\cite{kundu_channel_2021},\cite{syed_design_2023}, \cite{sun_channel_2021},\cite{oyerinde_iterative_2024}} &  & \cite{iren_simple_2023} & \shortstack{\cite{kundu_channel_2021},\cite{xu_ordinary_2021},\cite{liu_deep_2020},\cite{feng_deep_2024}} & \cite{haider_channel_2024} \\ \cline{3-8}  
 &  & MIMO & \shortstack{\cite{naamani_three-stage_2023},\cite{li_parametric_2024}} &  &  & \shortstack{\cite{li_channel_2020},\cite{gao_deep_2021}} & \shortstack{\cite{qi_channel_2024}\textcolor{red}{*},\cite{gao_two-stage_2023}} \\ \cline{2-8} 
 & \multirow{5}{*}{Multi User} & SISO & \cite{bjornson_maximum_2022} &  &  &  &  \\ \cline{3-8} 
 &  & SIMO &  & \cite{chen_channel_active_2023} &  &  &  \\ \cline{3-8} 
 &  & MISO & \shortstack{\cite{yang_novel_2022},\cite{yang_regularized_2024},\cite{zhou_channel_2022},\cite{wei_parallel_2020},\cite{li_channel_2024}} &  & \cite{yang_separate_2023} & \shortstack{\cite{rahman_deep_2023},\cite{zhang_self-supervised_2023},\cite{seo_dbpn-based_2023}} &  \\ \cline{3-8}
  &  & MIMO & \shortstack{
\cite{jeong_joint_2023},\cite{su_channel_2024},\cite{eddine_zegrar_reconfigurable_2021},\cite{sahoo_mobility-aware_2022},\cite{beldi_parafac_2023},\\
\cite{gomes_channel_2023},\cite{jin_near-field_2023},\cite{zhou_individual_2024},\cite{zhang_low-complexity_2023},
\cite{zhang_sparse_2023},\cite{fang_channel_2022},\\\cite{sahoo_mobility-aware_2022},\cite{abdallah_ris-aided_2023},
\cite{benicio_tensor-based_2024},\cite{haghshenas_parametric_2024},\cite{shi_triple-structured_2022}} &  & \shortstack{\cite{schroeder_two-stage_2022},\cite{haider_sparse_2022},\cite{zhu_channel_2024},\\
\cite{guerra_channel_2024},\cite{schroeder_channel_2022},\cite{lin_tensor-based_2021},\\
\cite{ xiong_separate_2024}} & \shortstack{\cite{elbir_deep_2020},\cite{liu_deep_2022},\cite{ahmadinejad_performance_2024},\\\cite{shen_deep_2023},\cite{xiao_u-mlp-based_2023},\cite{elbir_federated_2022},\\\cite{shen_federated_2022},\cite{nguyen_channel_2023}, \cite{abdallah_ris-aided_2023} } & \shortstack{\cite{jin_channel_2021},\cite{marques_guerra_ris-aided_2022},\cite{zhai_cnn_2023}}  
 \\ \hline
\multirow{6}{*}{Multi RIS} & \multirow{3}{*}{Single User} & SISO & \cite{you_double} &  &  &  &  \\ \cline{3-8} 
 &  & MISO & \shortstack{\cite{huang_roadside_2023}\textcolor{red}{*},\cite{huang_intelligent_2024}\textcolor{red}{*}} &  &  &  &  \\ \cline{3-8} 
 &  & MIMO & \shortstack{\cite{zheng_intelligent_2022}\textcolor{red}{*},\cite{ardah_double-ris_2022}} &  &  &  &  \\ \cline{2-8} 
 & \multirow{3}{*}{Multi User} & SISO & \cite{gurgunoglu_impact_2023} &  &  &  &  \\ \cline{3-8} 
 &  & MISO & \cite{wu_common_2023} &  &  &  &  \\ \cline{3-8} 
 &  & MIMO & \shortstack{\cite{ding_two-phase_2024},\cite{zheng_efficient_2021},\cite{zheng_uplink_2021},\cite{gherekhloo_nested_2023}, \cite{li_double_2022}\textcolor{red}{*}} & \cite{yang_active_2023} &  & \cite{liu_deep_2023} &  \\ \hline
\end{tabular}
}
\end{table*}

The insights gained from channel sounding are important for developing accurate channel estimation algorithms, as they provide a foundational understanding of the propagation environment. This section reviews the literature of channel estimation (CE) for RIS-assisted communications. The reviewed works are discussed based on the following key aspects: CE setup and hardware type, CE techniques, and CE in High Mobility, and CE in Multi-RIS deployment. A summary of the representative works on channel estimation in RIS-assisted systems is presented in Table \ref{tab:resume}.

\subsection{Channel Estimation Setup and RIS Hardware Architecture}
\begin{figure}[tb]
\centering
\includegraphics[width=\linewidth]{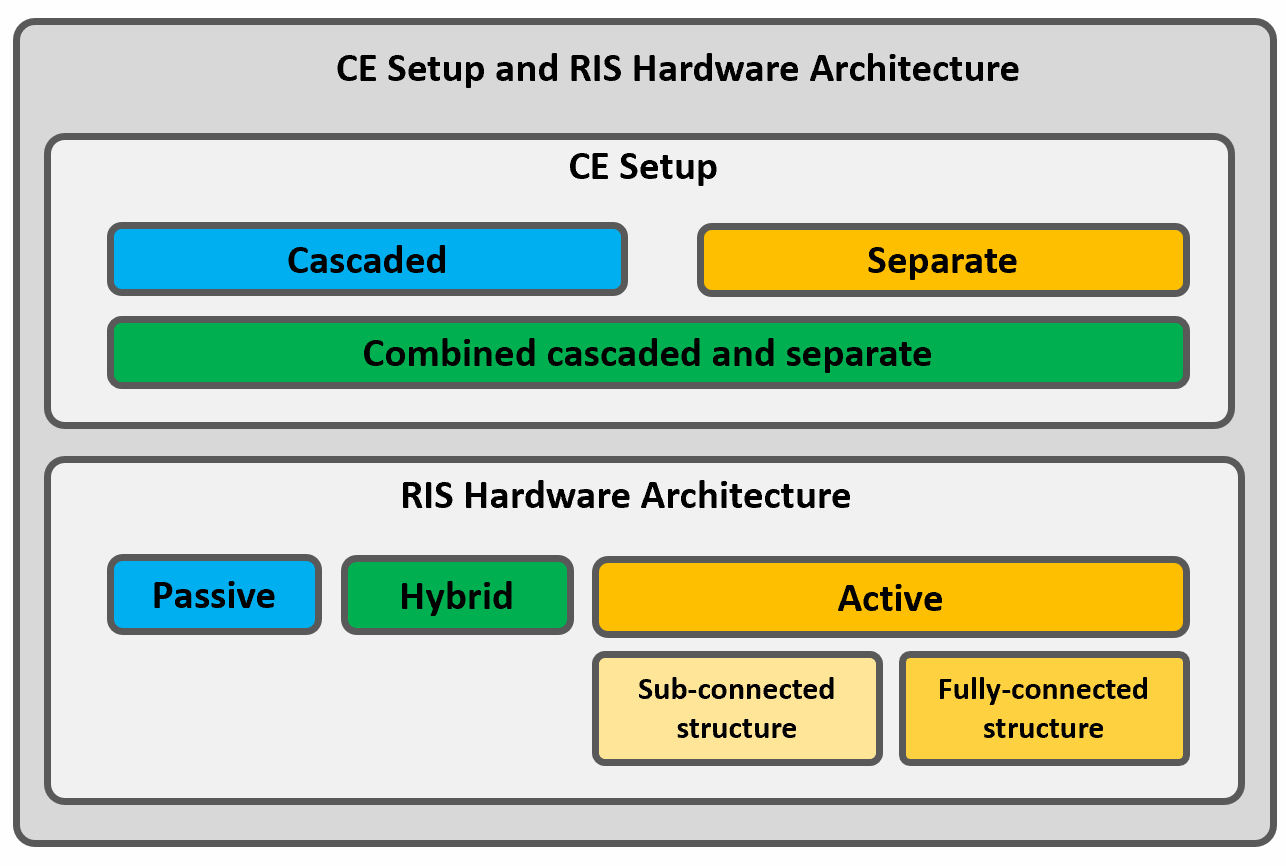}
\caption{Categorization of channel estimation setup and RIS hardware architecture.}
\label{fig:cesetup}
\end{figure}

This subsection provides CE categorization based on the setup (cascaded and separate CE) and the hardware architecture (passive, active, and hybrid RIS) as shown in Fig. \ref{fig:cesetup}. However, since CE for passive RIS has been comprehensively reviewed in \cite{zheng_survey_2022}, this subsection will focus more on active and hybrid RIS.

\subsubsection{Cascaded and Separate CE} 
\begin{figure}[tb]
\centering
\includegraphics[width=\linewidth]{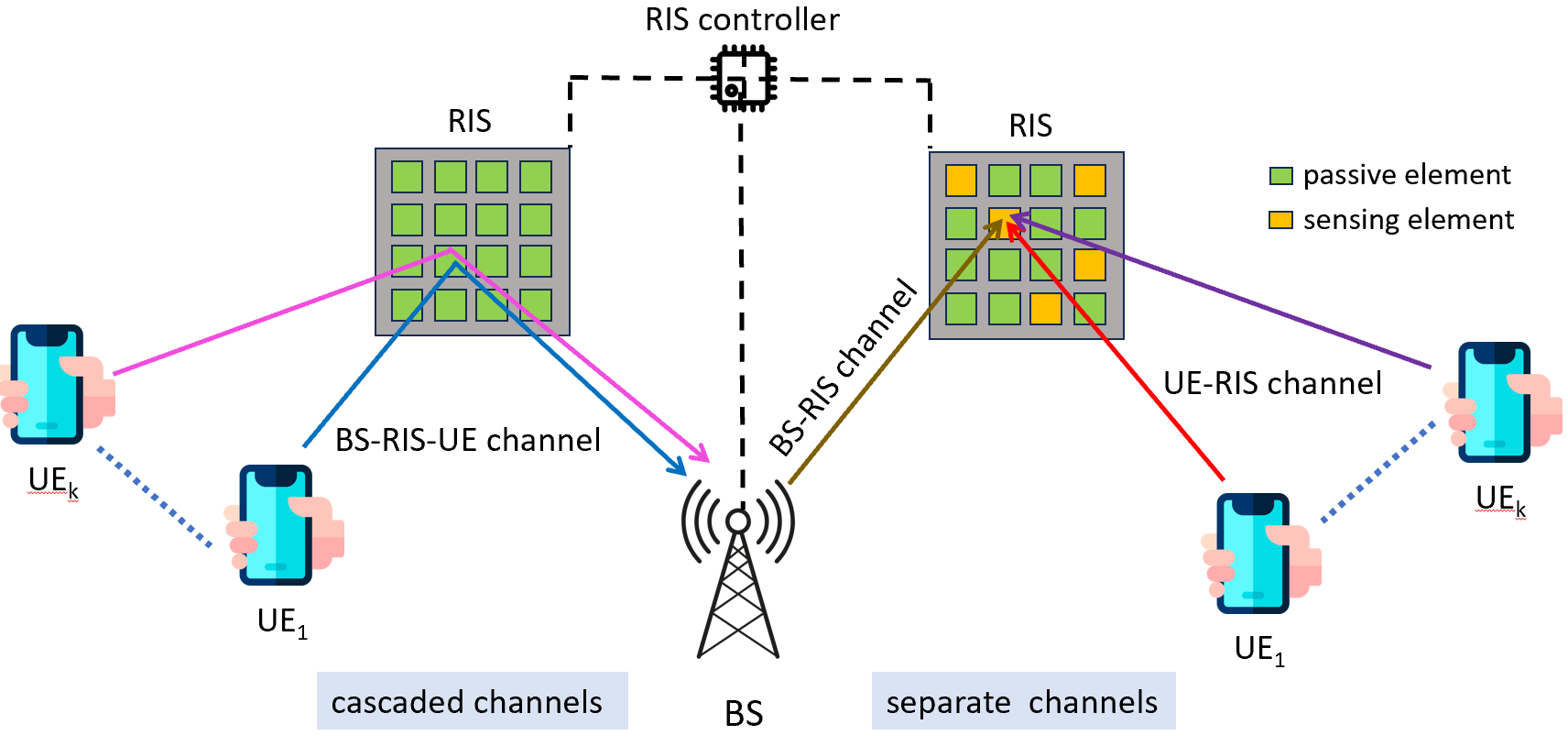}
\caption{Illustration of cascaded and separate channel estimation in RIS-assisted system.}
\label{fig:cascade}
\end{figure}

The CE setup in RIS-assisted systems can generally be categorized into two approaches: cascaded and separate  \cite{zheng_survey_2022}. This classification is based on how the BS–RIS–UE links are modeled and processed. In cascaded CE, the end-to-end channel encompassing the BS, the RIS, and the UE is estimated jointly, treating it as a unified channel. In contrast, separate CE involves estimating the BS–RIS and RIS–UE channels independently. Fig.\ref{fig:cascade} illustrates
the difference between cascaded and separate CE.

For simplicity, consider a channel estimation setup in which a BS is equipped with $M_\text{BS}$ antennas, $K$ users each equipped with $M_\text{UE}$ antennas, and an RIS with $N$ reflecting elements. Let $\boldsymbol{H}_{\text{BR}} \in \mathbb{C}^{M_\text{BS}\times N}$ denotes the BS–RIS channels, $\boldsymbol{H}_{\text{k}} \in \mathbb{C}^{N\times M_\text{UE}}$ denotes the $k$-th UE-RIS channels, and $\boldsymbol{\Theta}\in \mathbb{C}^{N\times N}$ denotes the diagonal reflection matrix of the RIS. A time division duplexing (TDD) setting with channel reciprocity is assumed, where the channel remains approximately constant over a short duration. Consequently, the uplink channel estimation can be reused for downlink transmission.

\paragraph{Cascaded CE}
The cascaded channel is typically estimated at the BS, which possesses sufficient computational resources. During the CE process, each UE transmits a known pilot symbol to the BS through the RIS. Assuming no direct link between the UE and the BS, the received symbol vector at the BS is given by 

\begin{equation}
\boldsymbol{y}_{\text{BS}} = \sum_{k=1}^{K} \, \sqrt{P_\text{k}}\,\underbrace{\boldsymbol{H}_{\text{BR}} \, \boldsymbol{\Theta} \, \boldsymbol{H}_{\text{k}}}_{\boldsymbol{H}_\text{cascaded}} \,\boldsymbol{x}_{\text{k}} + \boldsymbol{v}_\text{BS},
\label{eq:cascade}
\end{equation}
where $P_\text{k}$ is the $k$-th UE transmit power, $\boldsymbol{x}_{\text{k}}\in \mathbb{C}^{M_\text{UE}\times 1}$ represents the pilot symbol vector transmitted by the $k$-th UE, and $\boldsymbol{v}_\text{BS}\in \mathbb{C}^{M_\text{BS}\times 1}$ denotes the additive white Gaussian noise (AWGN) vector at the BS with $\boldsymbol{v}_{\text{BS}} \sim \mathcal{CN}(0, \sigma_b^2 \boldsymbol{I}_{M_{\text{BS}}})$. From (\ref{eq:cascade}), it can be seen that the effective channel observed by the BS is the $k$-th UE's cascaded channel: $\boldsymbol{H}_\text{cascaded} \in \mathbb{C}^{M_\text{BS} \times M_\text{UE}}$. Using the Kronecker product, (\ref{eq:cascade}) can be reformulated as

\begin{equation}
\boldsymbol{y}_\text{BS} = \sum_{k=1}^{K} \sqrt{P_\text{k}} \left( \left( \boldsymbol{x}_{\text{k}} \right)^T \otimes \boldsymbol{I}_{M_\text{BS}} \right) \text{vec}\,({\boldsymbol{H}}_{\text{cascaded}}) \boldsymbol  + \boldsymbol{v}_\text{BS},
\label{eq:cascade1}
\end{equation}
where $\otimes$ represents the Kronecker product, and vec() denotes the vectorization operation. 

Considerable research has been devoted to cascaded CE. As prior contributions, particularly that published before 2022, have been thoroughly reviewed in \cite{zheng_survey_2022}, our attention shifts to highlighting more recent advancements. In the following, we briefly discuss a few representative studies to illustrate the fundamental design principles, while a more comprehensive overview, including additional related works, is provided in the subsequent subsections. The first approach is by exploiting the spatial correlation of the cascaded channels with Karhunen-Loève transformation based linear minimal mean square error (KL-LMMSE) to reduce pilot overhead \cite{li_low-overhead_2024}. The second one exploits the sparsity of the channels. The methods, such as sparse Bayesian Learning (SBL) \cite{yang_regularized_2024} and message passing SBL \cite{li_exploiting_2024}, have demonstrated significant reductions in pilot overhead for uplink multi-user single-antenna scenarios when compared to least squares (LS) method. The study in
\cite{tang_joint_2024} proposes a hybrid low-rank and sparsity model with two-stage process. Evaluated in a multi-user single-antenna scenario, this method achieves more accurate channel estimation than the LS method. In \cite{chung_location-aware_2024}, an atomic norm minimization (ANM)-based estimation takes advantage from multi-dimensional ANM to represent the cascaded channel as a sum of steering vectors, further reducing training overhead, especially with location information. Another approach exploits tensor signal modeling and decomposition. Studies in \cite{fernandes_joint_2024} and \cite{fernandes_tensor_2024} arrange pilot measurements as  tensors and employ CANDECOM / PARAFAC (CP) to simultaneously estimate the cascaded and direct channels. This method achieves more accurate estimation without switching the RIS off, resulting in lower training overhead compared to ON–OFF-based method.

It is worth noting that cascaded CE is more suitable for purely passive RIS since they cannot sense the channel by themselves, and therefore, this task is conducted by the BS, which corresponds better to the fully and partially controlled RIS configuration that we discussed earlier. However, the main challenge of this estimation lies in jointly designing the pilot sequences at the UEs, the phase shift at the RIS, and the signal processing at the BS, while minimizing the overhead.

\paragraph{Separate CE}
In this approach, the objective is to estimate $\boldsymbol{H}_{\text{BR}}$ and $\boldsymbol{H}_{\text{k}}$ separately. It is usually more convenient to assume that separate CE is performed by the RIS, and for that, some sensing elements are incorporated into it as illustrated in Fig.~\ref{fig:cascade}. In this setup, the BS and the UEs  simultaneously transmit $N_\text{X}$ pilot symbols to the RIS. Considering $N_\text{SE}$ sensing elements at the RIS, the received symbol matrix at the sensing elements are given by 
\begin{align}
\boldsymbol{Y}_\text{SE}= & \, \mathbb{Q} \left( \sqrt{P_\text{BS}} \tilde{\boldsymbol{H}}_\text{BR} \boldsymbol{X}_\text{BS} + \sum_{k=1}^{K}\sqrt{P_\text{k}} \tilde{\boldsymbol{H}}_\text{k} \boldsymbol{X}_\text{k}+ \boldsymbol{V}_\text{SE} \right), 
 \label{eq:sepa}
\end{align}
where $\boldsymbol{X}_\text{BS}\in \mathbb{C}^{M_\text{BS} \times N_\text{X}}$,  $\boldsymbol{X}_\text{k}\in \mathbb{C}^{M_\text{UE} \times N_\text{X}}$, and $\boldsymbol{V}_\text{SE}\in \mathbb{C}^{N_\text{SE} \times N_\text{X}}$ denote pilot symbols transmitted by the BS, pilot symbols transmitted by the $k$-th UE, and the AWGN at the sensing elements, respectively. $\mathbb{Q}(.)$ denotes the quantization function and $P_\text{BS}$ is the BS transmit power. 

The integration of sensing elements at the RIS introduces additional hardware cost and energy consumption compared to a purely passive RIS. To reduce such hardware and energy costs, only a relatively small number of low-cost sensing devices, such as low-resolution analog-to-digital converters (ADCs), are used. Another significant challenge in this context lies in accurately reconstructing the full channel state information (CSI) using a limited number of low-cost sensing devices.

\subsubsection{Channel Estimation with Active RIS}
Based on the implementation design of the active RIS element, we distinguish two types of active RIS design : the one which incorporates power amplifiers together with active channel sensors and the one which incorporates only power amplifiers \cite{kang_active-irs-aided_2024}. The presence of the sensor enables the active RIS for direct and more accurate channel estimation, leading to better performance. In contrast, an active RIS without sensors must rely on  indirect methods of channel estimation, which can be less accurate but simpler and cheaper to implement. The first type of active RIS is exploited in \cite{jin_channel_2021}, but it is utilized in combination with passive RIS, making it a hybrid RIS rather than a purely active one. 

\begin{figure}[tb]
\centering
\includegraphics[width=\linewidth]{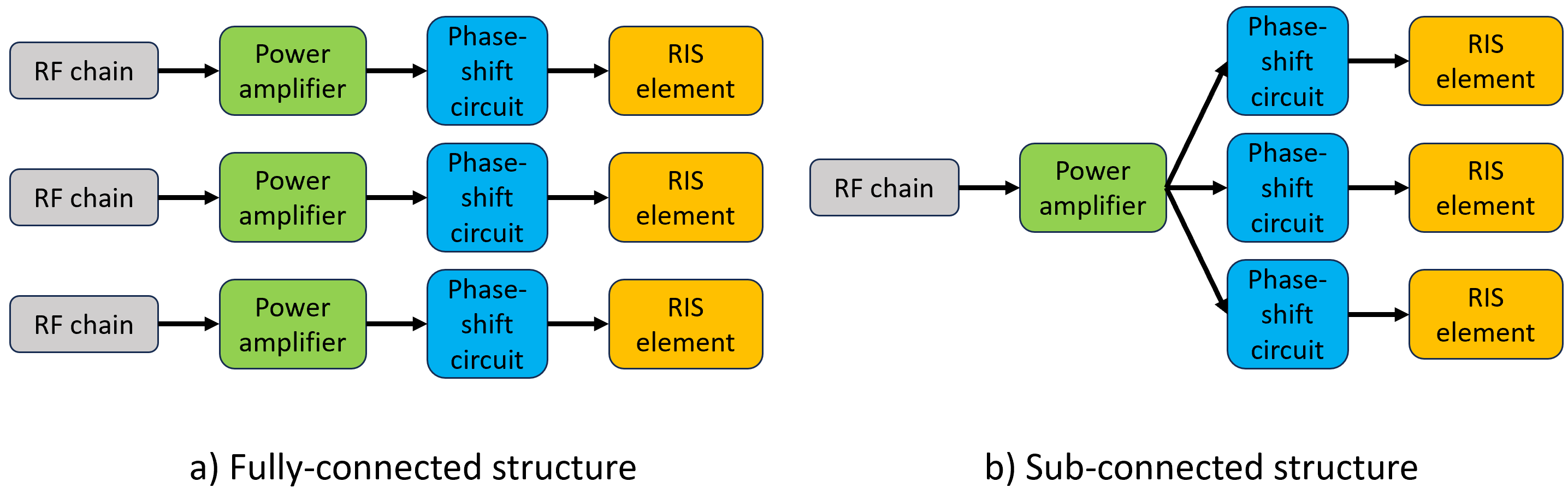}
\caption{Two types of active RIS structures\cite{yang_active_2023}.}
\label{fig:structure}
\end{figure}

Based on the connection structure between the RF chain, the power amplifier, and the RIS element, active RIS can be categorized into the two types depicted by Fig.\ref{fig:structure}: fully-connected and sub-connected structures, which was introduced later in \cite{liu_active_2022}. The fully-connected structure consumes more power due to the integration of dedicated power amplifiers for each element. Meanwhile, a sub-connected structure shares a single power amplifier over multiple  elements while controlling their phase shifts independently. This reduces the number of power amplifiers, even  improving energy efficiency up to 22\% \cite{liu_active_2022}.

If the system has a requirement for the power of each element to be controlled separately to conduct channel estimation, a fully-connected structure is implemented, such as in \cite{chen_channel_active_2023}. Otherwise, sub-connected type, like the one used in \cite{yang_active_2023}, is preferred.

\begin{table*}[tb]
\centering
\caption{Research in Channel Estimation using Purely Active RIS}
\begin{tabular}{|c|c|c|c|c|}
\hline
\textbf{Year} & \textbf{Ref} & \textbf{Structure} & \textbf{Sensing Capability} & \textbf{Channel Estimation Technique} \\ \hline
2023 & \cite{chen_channel_active_2023} & Fully-connected & No & LS \\ \hline
2023 & \cite{yang_active_2023} & Sub-connected & No & Bayesian CS \\ \hline
2024 & \cite{kang_active-irs-aided_2024} & With sensor; No sensor& Yes; No & Separate CE; Combined cascaded and separate CE \\ \hline
\end{tabular}
\label{tab:activ}
\end{table*}

\begin{figure}[tb]
\centering
\includegraphics[width=\linewidth]{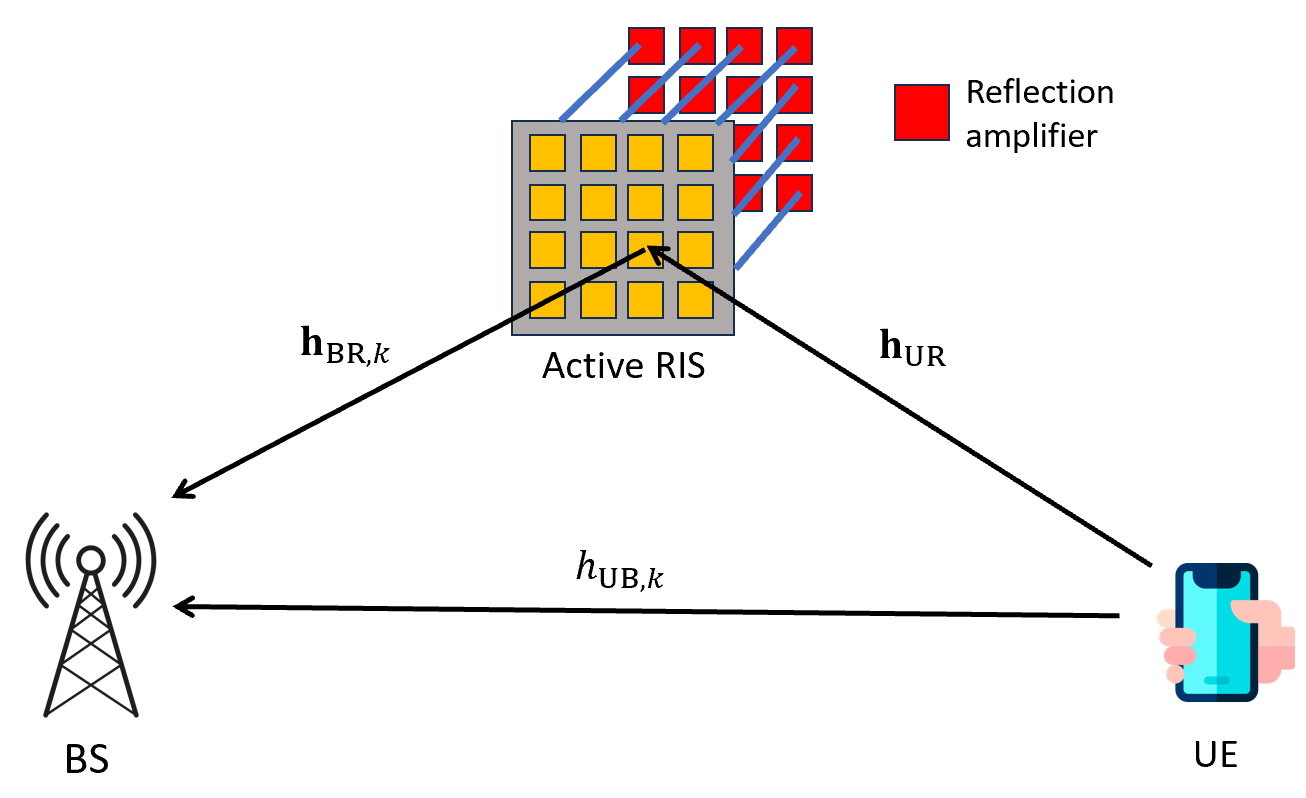}
\caption{Schematic illustration for considered case in \cite{chen_channel_active_2023}.}
\label{fig:activ2}
\end{figure}

\begin{figure}[tb]
\centering
\includegraphics[width=\linewidth]{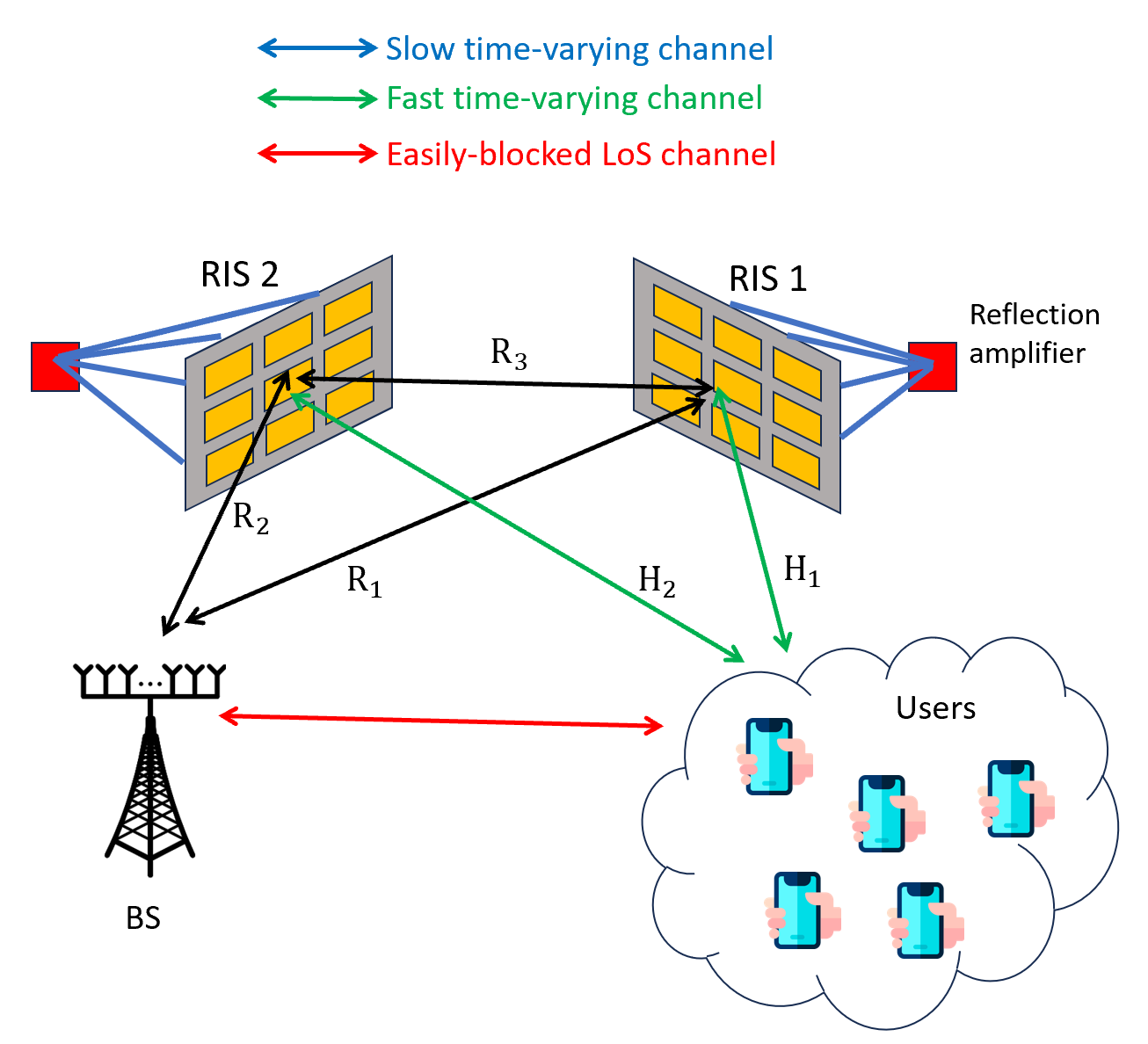}
\caption{Active double-RIS multiuser MIMO setup in \cite{yang_active_2023}.}
\label{fig:activ3}
\end{figure}

In \cite{chen_channel_active_2023}, the authors investigate an  active RIS-assisted system with a fully connected structure as illustrated in Fig.\ref{fig:activ2}. 
An uplink transmission with $K$ antennas at BS and a single-user SIMO system is considered. $\boldsymbol{H}_{\text{BR}}$ and $\boldsymbol{H}_{\text{k}}$ are estimated using combined cascaded and separate CE involving the active RIS and the BS. The study in \cite{yang_active_2023} considers a sub-connected structure and combined cascaded and separate CE in a double-RIS multi-user MIMO configuration shown in Fig. \ref{fig:activ3}. In a more recent study \cite{kang_active-irs-aided_2024}, sensors are explored in the context of active RIS deployment. The authors investigate both approaches for CE in active RIS: with and without sensor. Sensors enable the active RIS to estimate the BS-RIS and RIS-UE channels separately. Without the sensor, one possible way is combining separate and cascaded CE techniques. Hence, RIS-mounted sensors enable direct channel estimation, whereas in their absence, indirect methods must be employed. The literature review of this subsection is summarized in Table \ref{tab:activ}. 

\subsubsection{Channel Estimation with Hybrid RIS}
\label{hybrid}
Hybrid RIS combines the passive reflection capabilities of traditional RIS elements with the amplification and sensing functionalities of active elements. CE in hybrid RIS typically leverages the incident signals captured by the active elements to be processed. The passive elements just reflect the incoming signal without further processing. In \cite{koutsonas_deep_2023}, the authors consider CE for a downlink transmission. The RIS active elements receive the signal and then the CE is performed by an RIS microcontroller. The work in \cite{iren_simple_2023} investigates downlink CE in a UAV-assisted communication system with only one active element, aiming to reduce system complexity. The authors in \cite{haider_sparse_2022} also consider a downlink transmission and the CE in both BS-RIS and UE-RIS channels are performed by the RIS. The active elements are strategically placed in an L-shaped sparse array configuration, and they help resolve ambiguities in the sparse channel by minimizing mutual coupling and interference among the elements.

\begin{figure*}[tb]
\centering
\includegraphics[width=\linewidth]{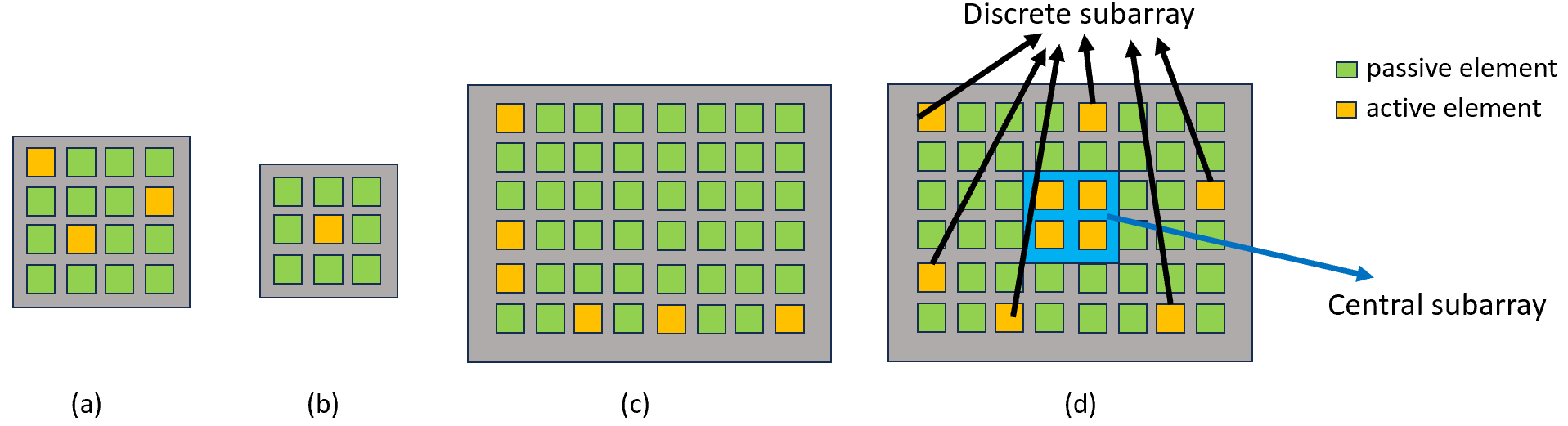}
\caption{Various pattern of the active elements placement in hybrid RIS: (a) random-like \cite{schroeder_two-stage_2022}, (b) single \cite{iren_simple_2023}, (c) L-shape \cite{haider_sparse_2022}, (d) subarray \cite{lu_low-overhead_2024}.}
\label{fig:hyb}
\end{figure*}

In \cite{schroeder_two-stage_2022, choi_joint_2024, yang_separate_2023}, the authors consider an uplink transmission scenario. The active elements at the RIS receive the signals and perform channel estimation for the UE–RIS link. The CE for the RIS–BS link, however, is carried out at the BS, resulting in a two-step estimation process. In \cite{nguyen_decision-directed_2024, li_joint_2022}, a two-stage protocol is adopted, consisting of a CE stage followed by a data transmission stage. During the CE stage, the RIS performs data detection as the UE transmits both pilot and data symbols. To minimize power consumption during the data transmission stage, the sensing functionality of the RIS's active elements is turned off, and the incoming signal is entirely reflected. In \cite{lu_low-overhead_2024}, the hybrid XL-RIS architecture employs active elements to assist in CE by processing signals received from two distinct subarrays: the central and discrete subarrays, as illustrated in Fig. \ref{fig:hyb}. The central subarray, located at the center of the XL-RIS and operating in active mode, is used to estimate the AoA. Meanwhile, the discrete subarray, consisting of strategically selected, non-adjacent elements, is activated to estimate the path distances of multipath components. By leveraging both subarrays, the active elements enable a decoupled channel estimation strategy, separating AoA estimation and distance estimation into two coordinated tasks. 

\subsection{Channel Estimation  Techniques}
\begin{figure}[tb]
\centering
\includegraphics[width=\linewidth]{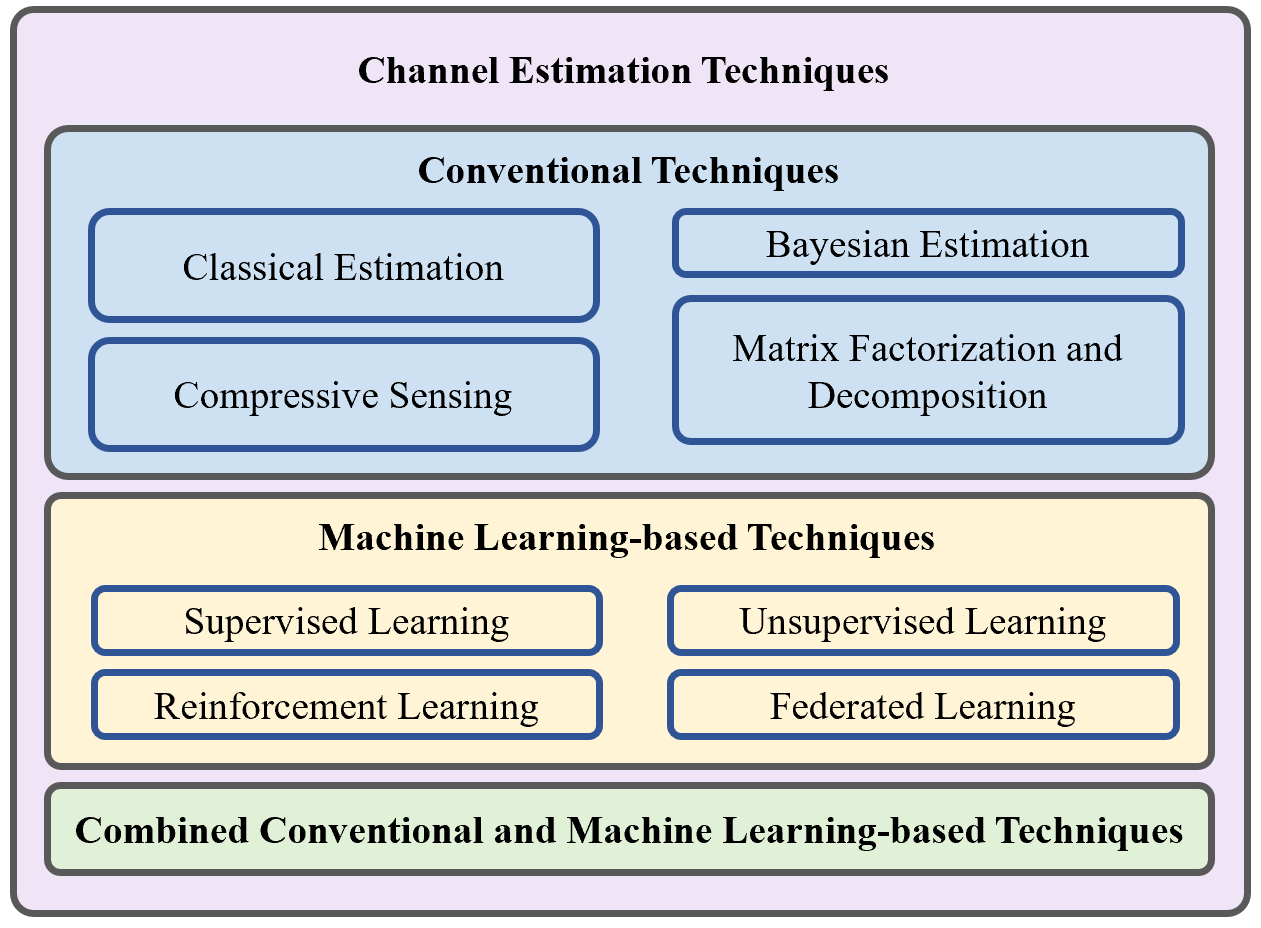}
\caption{Categorization of channel estimation techniques in RIS-assisted communication systems.}
\label{fig:estimation}
\end{figure}

Channel estimation techniques in RIS can be classified into two broad categories: conventional techniques, which rely on a certain model, and machine learning-based techniques, which leverage data-driven models to improve estimation performance. The categorization of the channel estimation techniques is shown in Fig. \ref{fig:estimation}.

\subsubsection{Conventional Techniques}
These techniques depends on mathematical models and require prior knowledge of the underlying system structure. In this survey, we classify them into 5 categories: classical estimation, Bayesian estimation, compressive sensing (CS), matrix factorization and decomposition, and other techniques. The representative studies about conventional CE are summarized in Table \ref{tab:conven}.

\begin{table*}[tb]
\caption{Summary of representative studies employing conventional channel estimation technique on RIS}
    \centering
    \begin{tabular}{|l|l|l|l|l|l|}
        \hline
        \textbf{Ref} & \textbf{Estimation method} & \textbf{Setup} & \textbf{Transmission} & \textbf{Channel Model} & \textbf{RIS Type} \\
          \hline
        \cite{jeong_joint_2023} & LS & MU-MIMO & Uplink & Frequency-selective  & Single-Passive \\
        \hline
         \cite{su_channel_2024} & RSBS-OMP  & MU-MIMO & Downlink  & Rician  & Single-Passive \\
        \hline
        \cite{syed_design_2023} & LS & SU-MISO & Downlink & Kronicker & Single-Passive \\
        \hline

         \cite{naamani_three-stage_2023} & OMP  & MU-MIMO & Downlink  & Static and dynamic  & Single-Passive \\
        \hline\cite{yang_regularized_2024} & Sparse Bayesian Learning  & MU-SIMO & Uplink  & Saleh-Valenzuela  & Single-Passive \\
        \hline
        \cite{eddine_zegrar_reconfigurable_2021} & EVD & MU-MIMO & Downlink  & Keyhole MIMO  & Single-Passive \\
        \hline
        
       \cite{gomes_channel_2023} & PARAFAC & MU-MIMO & Downlink  & Rayleigh  & Single-Passive \\
        \hline
       \cite{choi_joint_2024} & Generalized LS & MU-SIMO & Uplink & Rayleigh & Single-Hybrid \\
        \hline
        \cite{fernandes_channel_2023} & LS & SU-SIMO & Uplink & Rayleigh & Single-Active \\
      
        \hline
        \cite{long_channel_2023} & RS-LS & SU-SIMO & Uplink & Block fading & Single-Passive \\
        \hline
        
        \cite{wang_channel_2020} & LMMSE & MU-SIMO & Uplink  & Rayleigh & Single-Passive \\
        \hline
        \cite{gorty_channel_2022} & LMMSE  & SU-SISO & Uplink  & Quasi-static block fading  & Double-Passive \\
        \hline
         \cite{gurgunoglu_combating_2024} & Maximum Likelihood & MU-SISO (2 BSs) & Uplink  &  Rayleigh   & Double-Passive \\
        \hline
        \cite{zhu_estimating_2023} & GD & MU-SIMO & Uplink  & Rayleigh & Single-Hybrid \\
        \hline
        
        \cite{ma_robust_2022} & Variational Bayesian & MU-SIMO & Uplink  & Rayleigh & Single-Passive \\
        \hline
        
        \cite{li_kalman-filter_2024} & Kalman filtering & MU-SIMO & Uplink  & Block fading & Single-Passive \\
        \hline
        \cite{yu_kalman_2024} & Kalman filtering & MU-SIMO & Uplink  & Rician & Single-Passive \\
        \hline

    \end{tabular}
    \label{tab:conven}
\end{table*}

\paragraph{Classical Estimation Techniques}

This category encompasses traditional statistical signal processing methods, which can be further divided into linear and nonlinear estimation approaches.  Least squares (LS) and linear minimum mean square error (LMMSE) fall under the linear estimation category, whereas maximum likelihood falls under the nonlinear estimation.

In LS estimation, the goal is to minimize the sum of squared errors between received signals and the estimated channel matrix. To simplify the the basic idea of LS, let us consider the model in (\ref{eq:cascade}) for a UE with $P_\text{UE}$ transmit power, so that the received signal is
\begin{equation}
    \boldsymbol{y}_\text{BS} = \sqrt{P_\text{UE}}\,\boldsymbol{H}_\text{cascaded}\,\boldsymbol{x} + \boldsymbol{v}_\text{BS}.
\end{equation}
Hence, the LS problem is formulated as:

\begin{align}
    \hat{\boldsymbol{H}}_\text{LS} 
    &= \arg\min_{\boldsymbol{H}_\text{cascaded}} \left\| \boldsymbol{y}_\text{BS} - \sqrt{P_\text{UE}}\,\boldsymbol{H}_\text{cascaded}\,\boldsymbol{x} \right\|_2^2 \notag \\
    &= \frac{1}{\sqrt{P_\text{UE}}}\boldsymbol{y}_\text{BS}\, \boldsymbol{x}^H \left( \boldsymbol{x} \boldsymbol{x}^H \right)^{-1}.
\end{align}

A number of works have reported their successful implementation of LS to estimate the channel in RIS-assisted systems, i.e., \cite{azizzadeh_pilot-aided_2023, ding_two-phase_2024,fernandes_channel_2023,syed_design_2023,chen_channel_active_2023,jeong_joint_2023,long_channel_2023,yang_novel_2022,zheng_progressive_2023,zheng_intelligent_2022,zheng_efficient_2021,you_double,zheng_uplink_2021, demir_efficient_2025}. 

Unlike LS estimation, which operates without any prior statistical knowledge, LMMSE estimation seeks to minimize the overall mean squared error (MSE) by leveraging the second-order statistics of both the channel and the noise, as follows:
\begin{equation}
\hat{\boldsymbol{h}}_{\text{LMMSE}} = \boldsymbol{W}_{\text{LMMSE}}\, \boldsymbol{y}_\text{BS},
\label{eq:lm}
\end{equation}
with
\begin{equation}
\boldsymbol{W}_{\text{LMMSE}} = \arg\min_{\boldsymbol{W}} \, \mathbb{E} \left\{ \left\| \boldsymbol{W} \boldsymbol{y}_\text{BS} - \vec{\boldsymbol{h}} \right\|^2 \right\},
\end{equation}
where $\boldsymbol{W} \boldsymbol{y}_\text{BS}$ denotes a linear estimate of the true channel $\vec{\boldsymbol{h}}$, obtained by applying the estimator matrix $\boldsymbol{W}$ to the $\boldsymbol{y}_\text{BS}$. Studies in \cite{demir_efficient_2025} and \cite{long_mmse_2024} applied LMMSE at the BS to estimate the cascaded UE–RIS–BS channel, enhancing performance by incorporating the full spatial correlation matrix prior to estimation, which improved accuracy and reduced pilot overhead. The study in \cite{demir_efficient_2025} addresses cascaded channel estimation in RIS-assisted systems under electromagnetic interference (EMI). The received signal at the BS includes both the pilot signal reflected by the RIS and the EMI scattered by it.  The authors of \cite{demir_efficient_2025} utilized the equation from \cite{kay_fundamentals_1993}, so that the closed form solution of \eqref{eq:lm} for the cascaded channel is given by
\begin{subequations}
\begin{align}
\widehat{\boldsymbol{h}}_{\text{LMMSE}} &= \sqrt{P_\text{UE}} \, \boldsymbol{R}_{\boldsymbol{h}} \boldsymbol{\Phi}^{\text{H}} 
\left( P_\text{UE} \boldsymbol{\Phi} \boldsymbol{R}_{\boldsymbol{h}} \boldsymbol{\Phi}^{\text{H}} + \boldsymbol{C} \right)^{-1} 
\boldsymbol{y}_\textbf{BS}, \label{eq:lmmse_a} \\
\boldsymbol{C} &= \sigma_w^2 \boldsymbol{\Phi} \boldsymbol{R}_{\boldsymbol{w}} \boldsymbol{\Phi}^{\text{H}} + \sigma_v^2 \boldsymbol{I}_{M_\text{BS}}, \label{eq:lmmse_b}
\end{align}
\end{subequations}
where $
\boldsymbol{R}_{\boldsymbol{h}} = \mathbb{E} \left\{ \boldsymbol{h} \boldsymbol{h}^{\text{H}} \right\} \in \mathbb{C}^{M_\text{BS}N \times M_\text{BS}N}
$ and $
\boldsymbol{R}_{\boldsymbol{w}} = \mathbb{E} \left\{ \boldsymbol{w} \boldsymbol{w}^{\text{H}} \right\} \in \mathbb{C}^{M_\text{BS}N \times M_\text{BS}N}
$ denote the full
spatial correlation channel matrix and the EMI autocorrelation matrix, respectively. $\boldsymbol{\Phi}\in \mathbb{C}^{M_\text{BS}\tau_p \times M_\text{BS}N}$ is the RIS phase matrix with $\tau_p$ corresponds to the number of pilot symbols in each coherence block. $ \sigma_w^2$ and $ \sigma_v^2$ represent the EMI and noise variances at the BS, respectively. Authors of \cite{he_joint_2024} employed LMMSE at the BS to estimate both the cascaded UE–RIS–BS channel and the direct UE–BS channel. Their estimation procedure was integrated into an iterative optimization framework, in which the BS jointly optimized its training signal design and the RIS reflection pattern. This joint design effectively reduced estimation errors, as the accuracy depends not only on how the BS processes the received training signals but also on how the RIS modifies the signals before they reach the BS.

Maximum likelihood estimation (MLE) is another channel estimation technique that obtains the channel estimate by maximizing the likelihood of the received signal, given the known pilot signals and RIS configurations. When the noise is modeled as AWGN,  MLE reduces to LS, since maximizing the likelihood becomes equivalent to minimizing the squared error in this case\cite{fernandes_channel_2023}. \cite{huang_semi-blind_2022} proposed a semi-blind MLE-based method to estimate the cascaded BS–RIS–UE channel at the BS by jointly utilizing all available data symbols and a reduced number of pilot symbols, thereby lowering the pilot overhead compared to traditional full-pilot schemes. In \cite{haghshenas_parametric_2024}, MLE was applied to estimate the BS–UE and RIS–UE channels under the assumption of a static and known BS–RIS channel in a narrowband system, which was later extended to the wideband case in \cite{kosasih_parametric_2024}. Studies \cite{bjornson_maximum_2022} and \cite{haghshenas_efficient_2023} combined MLE with adaptive RIS configuration to reduce the number of required pilot symbols for estimating the UE–RIS channel, assuming the BS–RIS channel is known.

\paragraph{Bayesian Estimation}
Bayesian estimation is a statistical approach that uses Bayes' theorem to estimate unknown parameters by incorporating prior knowledge and updating this knowledge with observed data. Several works 
\cite{ma_robust_2022,li_variational_2024,nakul_variational_2024} used variational Bayesian based channel estimation for RIS-assisted systems. The authors of \cite{yang_regularized_2024} used Sparse Bayesian Learning (SBL), enhanced with a regularization technique. The method starts by vectorizing the channel matrix and applying a regularization technique that imposes penalties on the hyperparameters, which control the sparsity of the estimated channels. Study in \cite{kim_bayesian_2023} uses variational inference-sparse Bayesian learning (VI-SBL). Kalman filtering, as one type of Bayesian estimation, is implemented in \cite{li_kalman-filter_2024, yu_kalman_2024}.

\paragraph{Compressive Sensing}
Compressive sensing is a signal processing technique that allows for the reconstruction of sparse or compressible signals using far fewer samples than traditionally required by the Shannon-Nyquist sampling theorem \cite{candes_introduction_2008}. Several CS-based CE algorithms have been proposed in the literature of RIS aided systems: orthogonal matching pursuit (OMP) \cite{naamani_three-stage_2023}, regularized sensing beam split-based orthogonal matching pursuit (RSBS-OMP) \cite{su_channel_2024}, matching pursuit with phase rotation (MP-PR) \cite{guerra_channel_2024},  polar-domain double sparsity orthogonal matching pursuit (PDS-OMP)  \cite{jin_near-field_2023}, Subspace Pursuit (SP) OMP \cite{oyerinde_iterative_2024}, Bayesian learning-based compressive sensing \cite{yang_active_2023}, Atomic Norm Minimization (ANM) \cite{schroeder_channel_2022, schroeder_two-stage_2022}, ANM and OMP \cite{zhou_individual_2024}, Sparse Kalman Filter (SKF) \cite{zhang_sparse_2023}, Multi-user Triple-Structured-Compressive-Sensing simultaneous orthogonal matching pursuit (MTSCS-SOMP) \cite{shi_triple-structured_2022}, and double-layer bilinear generalized
approximate message passing (DL-BiGAMP) \cite{xiong_separate_2024}.

\paragraph{Matrix Factorization and Decomposition}
Matrix factorization and decomposition are mathematical techniques used to break down a matrix into simpler components, often to uncover underlying structure, reduce dimensionality, or make computations more efficient. Such techniques have been proposed in several studies for channel estimation in RIS-assisted systems, where they aim to estimate the cascaded BS–RIS–UE channels by exploiting structural properties, for example, parallel factor (PARAFAC) was used in \cite{beldi_parafac_2023,benicio_tensor-based_2024,gomes_channel_2023,gherekhloo_nested_2023}, singular value decomposition (SVD) was applied in \cite{zhang_low-complexity_2023},  eigenspace projection (EP) was employed in \cite{li_channel_2024}, canonical polyadic decomposition (CPD) was utilized in \cite{li_parametric_2024},  structured matrix completion was adopted in \cite{haider_sparse_2022}, and higher-dimensional rank-one approximations (HDR) was implemented in \cite{asim_structured_2025}.

\subsubsection{Machine Learning-based Techniques}
In this subsection, ML-based channel estimation techniques are categorized into four types based on the applied learning process: supervised learning, unsupervised learning, reinforcement learning, and federated learning, as summarized in Table \ref{tab:ml}.

\begin{table*}[tb]
\caption{Summary of the studies employing machine learning-based channel estimation on RIS}
    \centering
    \begin{tabular}{|l|l|l|l|l|l|}
        \hline
        \textbf{Ref} & \textbf{Learning Type} & \textbf{Learning Model} & \textbf{Setup} & \textbf{Transmission} & \textbf{Channel Model} \\
        \hline
        \cite{elbir_deep_2020} & Supervised Learning & Twin CNN & MU-MISO & Downlink & Saleh-Valenzuela \\
        \hline
        \cite{kim_deep_2023} & Reinforcement Learning & DQN & SU-SISO & Downlink & Wideband Geometric \\
        \hline
        \cite{haider_channel_2024} & Supervised Learning & U-LSTM-RNN & SU-MISO & Downlink & Time-Varying Multipath Fading \\
        \hline
        \cite{elbir_federated_2022} & Federated Learning & CNN & MU-MIMO & Downlink & Geometric \\
        \hline

        \cite{zhang_deep_2021} & Supervised Learning & CNN & SU-SISO & Downlink & Frequency-Selective Fading \\
        \hline
         \cite{yang_joint_2024} & Supervised Learning & Dual CNN & SU-SISO & Uplink & Rician \\
        \hline
        \cite{dampahalage_supervised_2022} & Supervised Learning & CNN & SU-SIMO & Uplink & Sparse Angular Domain \\
        \hline
        \cite{fredj_variational_2023} & Unsupervised Learning & VI-based DNN & SU-SIMO & Uplink & Rayleigh and Sparse Angular Domain \\
        \hline
        
        \cite{xiao_multi-scale_2023} & Supervised Learning & LapWRes & SU-SIMO & Uplink & Clustered Statistical MIMO \\
        \hline
        \cite{chen_attention-based_2024} & Supervised Learning & Attention-based Denoising NN & SU-SIMO & Uplink & Saleh-Valenzuela \\
        \hline
       
     \cite{tong_diffusion_2024} & Supervised Learning & U-Net & SU-SIMO & Uplink & Saleh-Valenzuela \\
        \hline 
         \cite{xu_ordinary_2021} & Supervised Learning & ODE-CNN & SU-SIMO & Downlink & OFDM \\
        \hline
        \cite{feng_deep_2024} & Supervised Learning & JDCNet & SU-MISO & Downlink & Quasi-Static Block Fading \\
        \hline
        \cite{li_channel_2020} & Supervised Learning & Two Stage NN & SU-MIMO & Downlink & Geometric with Sparse \\
        \hline
        \cite{gao_deep_2021} & Supervised Learning & Three-Stage DNN & SU-MIMO & Uplink & mmWave Geometric \\
        \hline
        \cite{qi_channel_2024} & Supervised Learning & Attention-based DNN & SU-MIMO & Downlink & Time-Varying Geometric \\
        \hline
        \cite{rahman_deep_2023} & Supervised Learning & Bi-LSTM-NN & MU-MISO & Downlink & 3D Saleh-Valenzuela \\
        \hline
        \cite{zhang_self-supervised_2023} & Supervised Learning & DNN & SU-MISO & Downlink & Rician \\
        \hline
        \cite{seo_dbpn-based_2023} & Supervised Learning & DBPN & MU-MISO & Downlink & Quasi-Static Block Fading \\
        \hline        \cite{ahmadinejad_performance_2024} & Supervised Learning & ANN & MU-MIMO  & Downlink & Rician \\
        \hline
        \cite{shen_deep_2023} & Supervised Learning & SRDnNet & MU-SIMO & Uplink & Rician \\
        \hline
        \cite{xiao_u-mlp-based_2023} & Supervised Learning & U-MLP & MU-SIMO & Uplink & Hybrid-Field \\  \hline\cite{shen_federated_2022} & Federated Learning & FDReLNet & MU-SISO & Downlink & Rician \\
        \hline
        \cite{nguyen_channel_2023} & Supervised Learning & CNN+LSTM & MU-MIMO & Downlink & Rayleigh \\
        \hline
        \cite{jin_channel_2021} & Supervised Learning & EDSR + MDSR & MU-MIMO & Uplink & mmWave Geometric \\
        \hline
        \cite{marques_guerra_ris-aided_2022} & Supervised Learning & MLP-NN & MU-SISO & Uplink & Narrowband Flat-Fading \\
        \hline
        \cite{zhai_cnn_2023} & Supervised Learning & CNN & MU-MIMO & Uplink & Classical Narrowband Ray-Based \\
        \hline
    \end{tabular}
    \label{tab:ml}
\end{table*}

\paragraph{Unsupervised Learning}
Unsupervised learning is a type of machine learning where the algorithm learns from unlabeled data to uncover hidden patterns.  One example applied in channel estimation is variational inference (VI), which approximates complex-intractable probability distributions with simpler-tractable ones to estimate the marginal distributions of hidden variables\cite{tzikas_variational_2008,blei_variational_2017}. The authors of \cite{fredj_channel_2024} and \cite{fredj_variational_2023} employed VI to jointly estimate $\boldsymbol{H}_\text{UR}$ and $\boldsymbol{H}_\text{BR}$ using uplink training signals in a passive RIS setting. The process begins with the received pilot signal  $\boldsymbol{y}_\text{BS}$, whose conditional probabilty $p(\boldsymbol{y}_\text{BS}|\boldsymbol{H}_\text{UR}, \boldsymbol{H}_\text{BR}  )$ is the input to two neural encoders. Each encoder maps $\boldsymbol{y}_\text{BS}$ to a set of variational parameters, denoted by $\boldsymbol{\mu}_1$ and  $\boldsymbol{\mu}_2$, that define the respective auxiliary distributions $q_{\boldsymbol{\mu}_1}(\boldsymbol{H}_\text{UR}|\boldsymbol{y}_\text{BS})$ and $q_{\boldsymbol{\mu}_2}(\boldsymbol{H}_\text{UR}|\boldsymbol{y}_\text{BS})$. These distributions serve as tractable approximations to the intractable true posteriors of the latent variables $\boldsymbol{H}_\text{UR}$ and $\boldsymbol{H}_\text{BR}$. The objective function in VI, known as evidence lower bound (ELBO), is defined as the negative of the loss function, i.e., $\log p(\boldsymbol{y}_\text{BS}) 
\triangleq -\mathcal{L}(\boldsymbol{y}_\text{BS}; \boldsymbol{\mu})$  \cite{zhang_advances_2019}. Since the neural network is trained to maximize the ELBO, this is equivalent to minimizing the corresponding loss function, which is expressed as
\begin{align}
&\mathcal{L}(\boldsymbol{y}_\text{BS};\bm{\mu}_1, \bm{\mu}_2) \notag\\
&= \underbrace{\mathbb{E}_{\boldsymbol{H}_\text{UR}^{\text{DFT}} \sim q_{\bm{\mu}_1}(\boldsymbol{H}_\text{UR}^{\text{DFT}} | \boldsymbol{y}_\text{BS})} 
\left[ \log \frac{q_{\bm{\mu}_1}(\boldsymbol{H}_\text{UR}^{\text{DFT}} | \boldsymbol{y}_\text{BS})}{p(\boldsymbol{H}_\text{UR}^{\text{DFT}})} \right]}_{\mathcal{L}_1} \notag\\
&\quad + \underbrace{\mathbb{E}_{\boldsymbol{H}_\text{BR}^{\text{DFT}} \sim q_{\bm{\mu}_2}(\boldsymbol{H}_\text{BR}^{\text{DFT}} | \boldsymbol{y}_\text{BS})}
\left[ \log \frac{q_{\bm{\mu}_2}(\boldsymbol{H}_\text{BR}^{\text{DFT}} | \boldsymbol{y}_\text{BS})}{p(\boldsymbol{H}_\text{BR}^{\text{DFT}})} \right]}_{\mathcal{L}_2} \notag\\
&\quad - \underbrace{\mathbb{E}_{\boldsymbol{H}_\text{UR}^{\text{DFT}}, \boldsymbol{H}_\text{BR}^{\text{DFT}} \sim q_{\bm{\mu}}(\boldsymbol{H}_\text{UR}^{\text{DFT}}, \boldsymbol{H}_\text{BR}^{\text{DFT}} | \boldsymbol{y}_\text{BS})}
\left[ \log p(\boldsymbol{y}_\text{BS} | \boldsymbol{H}_\text{UR}^{\text{DFT}}, \boldsymbol{H}_\text{BR}^{\text{DFT}}) \right]}_{\mathcal{L}_3},
\end{align}
where $\boldsymbol{H}_\text{UR}^{\text{DFT}}$ and $ \boldsymbol{H}_\text{BR}^{\text{DFT}}$ denote the UE-RIS and BS-RIS channels represented in the angular domain via DFT, respectively. Loss functions $\mathcal{L}_1$ and $\mathcal{L}_2$ corresponds to the Kullback-Leibler (KL)-divergence between the
auxiliary distribution and the prior of $\boldsymbol{H}_\text{UR}^{\text{DFT}}$ and $\boldsymbol{H}_\text{BR}^{\text{DFT}}$, respectively. Loss function $\mathcal{L}_3$ is the reconstruction error of the estimated pilot signal $\hat{\boldsymbol{y}}_\text{BS}$. In final step, the maximum a posteriori (MAP) estimates  $\hat{\boldsymbol{H}}_\text{UR}$ and  $\hat{\boldsymbol{H}}_\text{BR}$ are obtained by selecting the most likely latent variables under each auxiliary distribution.

\paragraph{Supervised Learning}
Different from unsupervised learning, supervised learning relies on ground-truth labels to find hidden relationships between inputs and their corresponding outputs. The goal is to learn a mapping from inputs to outputs by minimizing the difference between its predictions and the true labeled values. Once trained, the model can then predict outputs for new, unseen inputs. Since depending on labels, supervised learning requires a high-quality dataset. All the referenced studies in this context employ datasets generated using simulation tools. Datasets obtained from practical RIS-aided deployments are still limited in availability. The performance of supervised learning models is influenced not only by data quality but also by the chosen network architecture. While increasing the number of hidden layers can enhance performance, it also introduces greater computational complexity and longer training time. Therefore, the selection of neural network architectures must balance accuracy with training cost. Neural network architectures are divided into three main categories:  convolutional neural network (CNN), deep neural network (DNN), and recurrent neural network (RNN). CNNs are usually chosen for their ability to extract local spatial features using convolutional filters, like in \cite{yang_joint_2024}, U-Net \cite{tong_diffusion_2024}, and SRDnNet, which is a unification of super-resolution CNN (SRCNN) and denoising CNN (DnCNN) \cite{shen_deep_2023}. On the other hand, DNNs are selected for their general-purpose approximation capability and flexibility. They are well-suited for structured input data, such as statistical features or flattened vectors, and are commonly applied when the input-output relationship is non-sequential and lacks strong spatial or temporal structure. Attention-based DNN has been implemented for channel estimation in SIMO \cite{chen_attention-based_2024}, MISO \cite{fan_spatial-attention-based_2024}, and MIMO \cite{qi_channel_2025,xiao_multi-scale_2024,guo_parallel_2024} setups. Other implementation of DNN includes joint detection and classification network (JDCNet) \cite{feng_deep_2024}, U-shaped network dedicated multilayer perceptron (U-MLP) \cite{xiao_u-mlp-based_2023}, deep back-projection network (DBPN)\cite{seo_dbpn-based_2023}, DNN-based self-supervised learning (SSL) \cite{zhang_self-supervised_2023}, channel estimation neural network architecture search (CENAS) \cite{shi_automatic_2024}, and accurate-and-deep denoising convolutional neural network (ADDnet) \cite{wang_data_2025}.
Lastly, RNNs are chosen for tasks involving temporal or sequential data, as they maintain internal memory to capture time-dependent patterns. Estimating channel in RIS using RNN has been applied in long short-term memory (LSTM)  \cite{haider_channel_2024} and
bidirectional long short-term memory (Bi-LSTM) \cite{rahman_deep_2023}. Several performance metrics have been used as loss functions in recent studies, including MSE \cite{seo_dbpn-based_2023}, normalized mean squared error (NMSE) \cite{zhang_self-supervised_2023}, and root-mean-square error (RMSE)\cite{haider_channel_2024}.

\paragraph{Reinforcement Learning}
In reinforcement learning (RL), a learning agent interacts with an environment to make sequential decisions with the objective of maximizing cumulative rewards. The agent observes the current state of the environment, selects an action, and receives rewards. The state represents the observable condition of the environment, the action is the agent’s behavior that can change the state, and the reward means feedback received after taking an action, indicating the immediate performance of that action. Unlike supervised learning, RL does not rely on explicit input–output labels. Instead, the agent gradually learns a policy, which defines how actions are chosen based on observed states. To the best of authors' knowledge, only \cite{kim_deep_2023} explicitly considers the application of reinforcement learning for channel estimation in RIS. In \cite{kim_deep_2023}, the agent is deep Q-network (DQN), which is deployed at the BS. The state includes the RIS phase shift configuration, the UE’s location, and the UE’s SNR, representing the current environment for channel estimation. The action specifies where and how frequently to place pilot signals in time and frequency domains to support channel estimation. The reward encourages the agent to achieve low estimation error with fewer pilots by assigning higher values to efficient pilot configurations. The system used a masked autoencoder (MAE) channel estimator then introduced deep RL agent to learn
pilot allocation policies through MAE. In \cite{kim_deep_2023}, the state space is defined by a tuple that captures the current system configuration at time slot $t$, including the RIS phase shift matrix $\Phi(t)$, the UE location $L_\text{UE}(t)$, and the UE’s average signal-to-noise ratio (SNR) $SNR_\text{UE}(t)$. Accordingly, the state at time $t$ is given by $s(t) = \{\Phi(t), L_\text{UE}(t), SNR_\text{UE}(t)\}$. On the other hand, the action space defines how the agent allocates $X_{\Delta}$ pilots on the resource grid. Specifically, the agent selects the inter-pilot intervals $\Delta_f$ from the first subcarrier and $\Delta_t$ from the $\Delta_i(t)$-th symbol on the frequency and the time axis, respectively.
So, the action is represented as $a(t) = \{\Delta_i(t), \Delta_f(t), \Delta_t(t)\}$. The reward is higher when the pilot interval is wider and the corresponding MSE is smaller. Accordingly, the reward is expressed as
\begin{equation}
R(s_t, a_t) =
\begin{cases}
\displaystyle \frac{X_\Delta}{\alpha \cdot \mathrm{MSE}}, & \text{if } X_\Delta \leq \Delta_{\max} \\
-\alpha \cdot \mathrm{MSE}, & \text{otherwise},
\end{cases}
\end{equation}
where $\alpha$ serves as a scaling factor to adjust for the smaller MSE compared to $\Delta_f+\Delta_t$. Since the authors of \cite{kim_deep_2023} used  DQN, the
state-action values are updated by
\begin{equation}
Q^{\pi}(s, a) = \hat{R}(s, a) + \beta \sum_{s' \in \mathcal{S}} P^a_{s, s'} \sum_{a' \in \mathcal{A}} \pi(a' \mid s') Q^{\pi}(s', a'),
\end{equation}
where $\hat{R}(s, a)$ denotes the future reward, $\pi$ represents the pilot allocation policy, $\beta$ is the discount factor, and $P^a_{s, s'}$ represents the probability of transitioning from state $s$ to state $s'$ according to the action. The loss function is determined by the MSE:
\begin{equation}
\mathcal{L} = \frac{1}{J} \sum_{j=1}^{J} \left\| \widehat{H} - H \right\|_F^2,
\end{equation}
where $J$ denotes the total of channel samples, generated from DeepMIMO dataset\cite{alkhateeb_deepmimo_2019}. $\widehat{H}$ and $H$ are the estimated and true channel matrices, respectively.  $\|\cdot\|_F$ represents the Frobenius norm.

\paragraph{Federated Learning}
Unlike traditional centralized learning, federated learning (FL) for channel estimation is a decentralized ML paradigm in which UEs collaboratively train a shared model under the coordination of a central aggregator, such as a BS. FL enables each UE to train a local neural network (e.g., CNN \cite{elbir_federated_2022}, residual network\cite{qiu_federated_2024}, or hierarchical architecture \cite{dai_distributed_2022}) on its own observations, such as received pilot signals. Instead of transmitting raw measurements, UEs periodically upload only model parameters updates to the BS. The BS then performs model aggregation to produce an updated global model. This model is subsequently redistributed to the UEs for the next training round. To enhance efficiency and scalability, some FL frameworks introduce adaptive participation strategies, such as user clustering \cite{asaad_cheema_channel_2025} or hierarchical regional coordination\cite{qiu_federated_2024}, allowing selective or grouped UE participation across training rounds. Overall, this decentralized approach significantly reduces communication overhead and preserves data privacy by keeping raw UE data localized.

\subsubsection{Combined Conventional and Machine Learning-based Techniques}
There are works combining machine learning technique and conventional technique together to conduct channel estimation in RIS-assisted system. In \cite{abdallah_ris-aided_2023}, the authors integrate compressive sensing and CNN-based supervised  learning for cascaded channel estimation. While a DnCNN-based model is employed when sufficient training data and computational resources are available, the proposed double-structured orthogonal matching pursuit for frequency-selective channel estimation (DSOMP-FS-CE) serves as an alternative in scenarios where offline training and real-time deployment of data-driven methods are infeasible. In \cite{gao_two-stage_2023}, a two-stage approach is proposed for mmWave communication systems employing a hybrid RIS. In the first stage, a CNN maps the angular-domain sparsity of the channel to estimate its amplitude. This intermediate result is then used in an LS estimator to recover the full channel. In contrast, studies \cite{liu_deep_2023, liu_deep_2022,feng_mmwave_2023} first apply an LS estimator to obtain a rough channel estimate, which is then refined using neural networks. Specifically, \cite{feng_mmwave_2023} uses 
global attention residual network (GARN), \cite{liu_deep_2022} utilizes a CNN-based deep residual network (CDRN), and \cite{liu_deep_2023} employs a skip-connection attention (SC-attention) network to refine the channel estimation using supervised learning with MSE as the loss function. Other works specifically combine compressive sensing and deep learning: denoising and attention mechanism assisted residual learned approximate message passing (DA-RLAMP)\cite{zheng_hybrid_2024}, row compression two stage learned AMP (RCTS-LAMP)\cite{tsai_deep_2024}, hypernetwork-assisted LAMP (HN-LAMP)\cite{tsai_low-complexity_2022}, and ResU-Net\cite{xie_deep_2022}.

\subsection{Channel Estimation in High Mobility Scenario} 

\begin{table*}[ht]
\centering
\caption{Summary of the references of channel estimation with RIS in high mobility scenarios, arranged in chronological order}
\label{tab:high}
\resizebox{\textwidth}{!}{
\begin{tabular}{|c|c|c|c|c|c|c|}
\hline
\textbf{Year} & \textbf{Ref.} & \textbf{User} & \textbf{Maximum User Speed} & \textbf{User Setting} & \textbf{Antenna Setup} & \textbf{Channel Estimation Technique} \\ \hline
2022&\cite{zheng_intelligent_2022} & LEO satellite & 756 m/s & Single User & MIMO & LS + distributed estimation scheme \\ \hline
2022 & \cite{xu_reconfigurable_2022} & UAV & 671 mph ($\approx$ 300 m/s) & Single User & SISO & MMSE \\ \hline
2022&\cite{li_double_2022} & High speed train & 150 m/s & Multi User & MIMO & PARAFAC decomposition + Kalman filter \\ \hline
2022&\cite{xu_deep_2022} & General & 100 km/h ($\approx$ 27.8 m/s) & Single User & SISO & Neural network \\ \hline
2023&\cite{xu_channel_2023} & Unmanned aircraft systems & 90 mph ($\approx$ 40.2 m/s) & Single User & MISO & MMSE \\ \hline
2023&\cite{huang_roadside_2023} & Car (roadside RIS only) & 50 m/s & Single User & MISO & Off-line and
online training scheme  \\ \hline
2023&\cite{huang_intelligent_2024} & Car (also vehicle-side RIS)& 50 m/s & Single User & MISO & Off-line and
online training scheme  \\ \hline
2024&\cite{qi_channel_2024} & General & 400 km/h ($\approx$ 111.1 m/s) & Single User & MIMO & MHA \\ \hline
\end{tabular}%
}
\end{table*}

High mobility, in the context of wireless communication systems, refers to scenarios where one of the communicating entities are moving at high speeds (e.g., drones, vehicles, satellites). This rapid movement significantly affects the communication channel, leading to fast-varying channel conditions, short coherence times, and substantial Doppler shifts. These factors present significant challenges in ensuring reliable and efficient communication, particularly in systems that demand precise and real-time channel estimation. High mobility typically occurs in environments where traditional static or low-mobility communication models fall short, necessitating the use of advanced techniques to handle the dynamic variations in the wireless channel. All of the contribution papers about CE in high mobility scenarios are summarized in Table \ref{tab:high}.

\begin{figure}[tb]
\centering
\includegraphics[width=\linewidth]{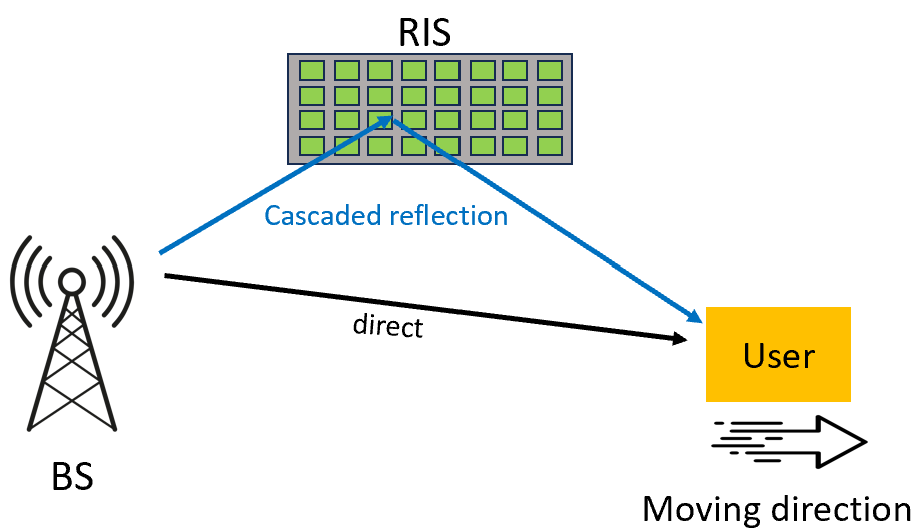}
\caption{Schematic illustration for case in \cite{xu_reconfigurable_2022,xu_channel_2023}.}
\label{fig:high1}
\end{figure}

Most of the channel estimation studies on high-mobility scenarios are based on conventional methods. The research in \cite{xu_reconfigurable_2022,xu_channel_2023} considered conventional method applied in passive RIS for single-user. The authors assumed a moving UE with the presence of the direct-link between the UE and the BS
as depicted in Fig.\ref{fig:high1}. Both studies utilized MMSE estimation, by formulating a statistical model that incorporates the known characteristics of the channel, such as the Doppler shift and the multipath effects. By accounting for both direct and reflected links, the MMSE estimator optimizes the channel estimates, enabling the RIS to be configured in a way that maximizes signal quality at the receiver, even under rapidly changing conditions. 

Continuing the exploration for single-user in high mobility scenarios, the study in \cite{qi_channel_2025} proposes a supervised learning-based channel estimation scheme using a hybrid RIS architecture. The method consists of two neural network models: an RIS Element Selection Model (RESM) and a Full Channel Estimation Model (FCEM). The input to RESM is the preliminary CSI estimated via the LS method for a subset of RIS elements, treated as real-valued tensors. RESM selects the most informative subset of RIS elements using a convolutional scorer and a differentiable Top-N operation. The scorer is implemented as a CNN that assigns relevance scores to each RIS element based on its contribution to the channel reconstruction. The differentiable Top-N operation adds random noise to the scoring values and uses a smoothing technique to approximate the selection of the top-ranked RIS elements. This allows the selection step to be compatible with gradient-based learning during neural network training. FCEM then reconstructs the full channel using a deep neural network that integrates residual blocks and an improved transformer equipped with multi-head attention mechanism (MHA), which is used to capture long-range dependencies and mitigate rapid channel fluctuations. The entire model is trained offline using supervised learning with the Huber loss function. Specifically, the Huber loss uses a squared residual formulation for small residuals and switches to a linear form for large residuals. 

\begin{figure}[tb]
\centering
\includegraphics[width=\linewidth]{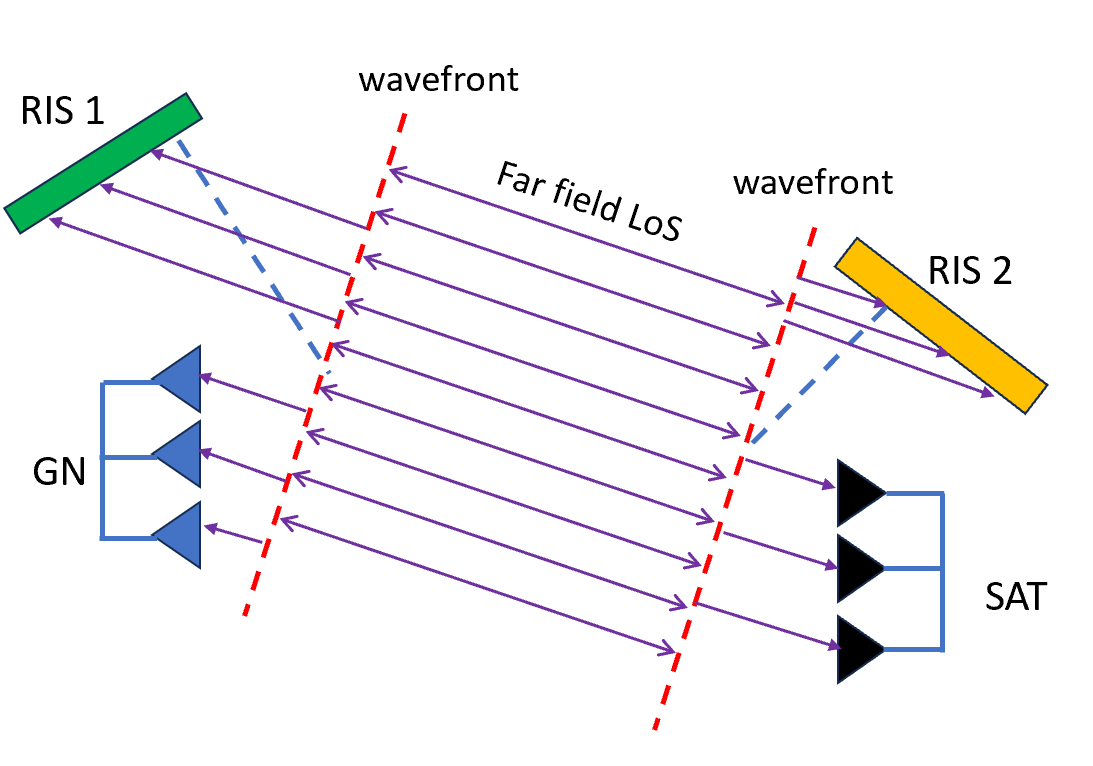}
\caption{Schematic illustration for satellite-ground communication in \cite{zheng_intelligent_2022}.}
\label{fig:sat}
\end{figure}

A study that specifically targets the channel estimation for low-earth orbit (LEO) satellite and ground node (GN) communication is presented in
\cite{zheng_intelligent_2022}, the considered system model is illustrated in Fig. \ref{fig:sat}. This study uses single-user MIMO setup. The authors propose the use of two RISs: one is mounted on the satellite (SAT) and another one is deployed on earth near the GN. The channel estimation is conducted at both SAT and GN with LS method to avoid the high-complexity estimation. Each side conducts its own local estimation using downlink and uplink pilots. This method enables each side to handle its portion of the CE autonomously, without the need for centralized control or information exchange between SAT and GN during the estimation process. 

\begin{figure}[tb]
\centering
\includegraphics[width=\linewidth]{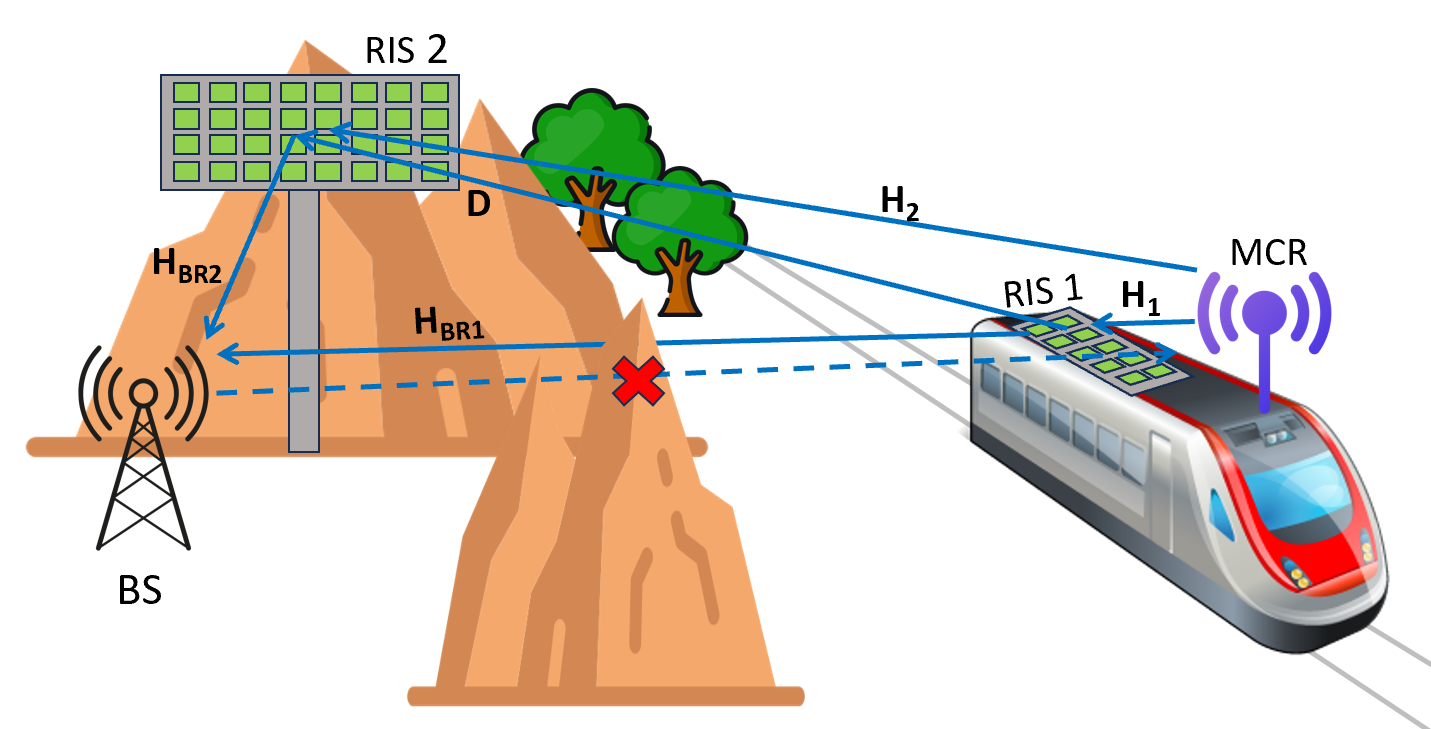}
\caption{RIS-assisted high-speed railway communication in \cite{li_double_2022}.}
\label{fig:train}
\end{figure}

A channel estimation framework for RIS-assisted high-speed railway communications is proposed in \cite{li_double_2022}. This study exploits two RISs, as illustrated in Fig. \ref{fig:train}. Specifically, one RIS (RIS 1) is  mounted on the train roof near the mobile communication relay (MCR), which communicates with users onboard, and the second one (RIS 2) is positioned near the BS. Due to the deployment geometry, the RIS 1 – MCR link benefits from a strong LoS component. The system adopts a TDD protocol where uplink pilots are transmitted by the MCR and reflected via both RISs toward the BS. MCR is equipped with $M_\text{MCR}$ antennas. RIS 1 and RIS 2 have $N_\text{1}$ and $N_\text{2}$ elements, respectively. Channel estimation is conducted exclusively at the BS and targets the individual components of the double-reflection and single-reflection links. In uplink transmission, the equivalent channel matrix $\bar{\boldsymbol{H}}\in \mathbb{C}^{M_\text{BS} \times M_\text{MCR}}$ can be expressed as
\begin{equation}
\bar{\boldsymbol{H}} = \boldsymbol{H}_{\text{BR}_\text{2}} \boldsymbol{\Theta}_2 \boldsymbol{D} \boldsymbol{\Theta}_1 \boldsymbol{H}_1 + \boldsymbol{H}_{\text{BR}_\text{2}} \boldsymbol{\Theta}_2 \boldsymbol{H}_2 + \boldsymbol{H}_{\text{BR}_\text{1}} \boldsymbol{\Theta}_1 \boldsymbol{H}_1,
\label{eq:mcr}
\end{equation}
where  $\boldsymbol{H}_{\text{BR}_\text{1}} \in \mathbb{C}^{M_\text{BS} \times N_1}$ and $\boldsymbol{H}_{\text{BR}_\text{2}} \in \mathbb{C}^{M_{BS} \times N_2}$ are the channel matrices of BS - RIS 1 and BS - RIS 2, respectively. $\boldsymbol{\Theta}_1 \in \mathbb{C}^{N_\text{1} \times N_\text{1}}$ and $\boldsymbol{\Theta}_2 \in \mathbb{C}^{N_\text{2} \times N_\text{2}}$ denote the diagonal reflection matrix of RIS 1 and RIS 2, respectively.  $\boldsymbol{D} \in \mathbb{C}^{N_2 \times N_1}$ represents the RIS 2 - RIS 1 channel matrix. $\boldsymbol{H}_1 \in \mathbb{C}^{N_1 \times M_\text{MCR}}$ and $\boldsymbol{H}_2 \in \mathbb{C}^{N_2 \times M_\text{MCR}}$ are the channel matrices of RIS 1 - MCR and RIS 2 - MCR, respectively. Leveraging channel reciprocity, the downlink transmission reuses the channel acquired during the uplink phase. The channel estimation is divided into two phases: large-timescale estimation to capture quasi-static channels (e.g., $\boldsymbol{H}_{\text{BR}_\text{2}}$ and $\boldsymbol{H}_1$), and small-timescale estimation for fast-varying channels (e.g., $\boldsymbol{D}$). In the large-timescale phase, the quasi-static channels are estimated using a dual-link pilot scheme and refined via coordinate descent. In the small-timescale phase, PARAFAC decomposition is used to separate the contributions of the three reflection paths in (\ref{eq:mcr}). Kalman filtering and Doppler compensation are further employed to enhance estimation robustness under mobility constraints. 

\begin{figure}[tb]
\centering
\includegraphics[width=\linewidth]{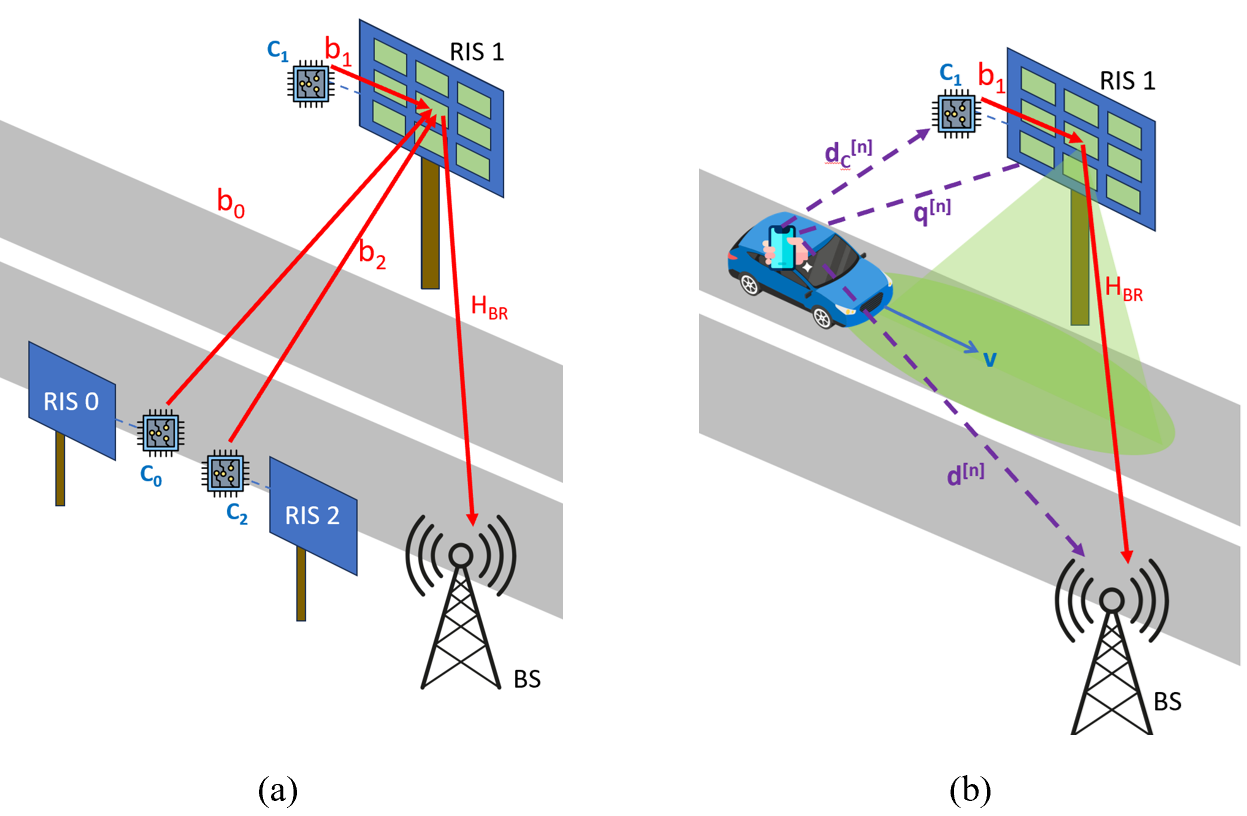}
\caption{Roadside RIS-assisted vehicular communication in \cite{huang_roadside_2023} for (a) offline estimation (b) online estimation.}
\label{fig:road}
\end{figure}

The study in \cite{huang_roadside_2023} investigates a vehicular communication scenario where a mobile UE communicates with a BS assisted by a roadside RIS. To address the challenges posed by high mobility, a two-stage channel estimation scheme is proposed, comprising offline and online stages, as illustrated in Fig.~\ref{fig:road}. The authors of \cite{huang_roadside_2023} consider a central serving RIS (RIS 1) accompanied by two neighboring RISs (RIS 0 and RIS 2), each associated with a controller \( C_k \), where the  index \( k \in \{0,1,2\} \)  corresponds to its respective RIS. Each controller is equipped with RF chains, capable of receiving pilot signals, and connected to the BS via a dedicated and reliable wireless control channel. The static $C_k $ - RIS 1 channel is denoted by $\boldsymbol{b}_k$. The cascaded $C_k $ - RIS 1 - BS channel is represented by $\boldsymbol{R}_k=\boldsymbol{H}_\text{BR} \operatorname{diag} (\boldsymbol{b}_k)$. RIS 0 and RIS 2 are symmetrically deployed across from RIS 1. In the offline stage, only the static channels are estimated prior to the vehicle entering the coverage area. Since RIS is passive, estimating $\boldsymbol{H}_\text{BR}$ is easier by leveraging the cascaded channels observed at the BS when nearby controllers transmit pilots through RIS 1. However using a single controller would lead to ambiguity, as the observed signal contains not just the desired $\boldsymbol{H}_\text{BR}$, but also $\boldsymbol{b}_k$, which contains direction parameters scaled by an unknown complex gain. This factors challenge the accurate extraction of spatial structure of $\boldsymbol{H}_\text{BR}$ needed for beamforming of RIS 1. This motivated the authors of \cite{huang_roadside_2023} to involve a second controller. However, this controller should be placed on the opposite side of RIS 1 to allows the BS to combine the cascaded observations. Their symmetrical deployment allows the cancellation of the unknown gain algebraically and facilitates the extraction of the angular characteristic of $\boldsymbol{H}_\text{BR}$. Offline stage refers to the initial phase conducted before the vehicle enters the coverage area. In this stage, $C_0$ and $C_2$ transmit pilots to the BS via RIS 1. This enable the BS to estimate  $\boldsymbol{H}_\text{BR}$, which in turn is used to calculate the initial passive beamforming vector of RIS 1. This vector is then sent to $C_1$, which subsequently transmits pilots to the BS via RIS 1. The BS estimates $\boldsymbol{b}_1$ and sent the results back to $C_1$. In the online stage, $C_1$ is responsible for estimating and predicting the angular parameters of time-varying UE-$C_1$ channel. Before RIS 1 begins to serve the UE (i.e., for block $n \leq 0$), $C_1$ passively receives the pilot signals transmitted by the UE. It estimates the azimuth and elevation angles $\left(\vartheta^{[n]}, \psi^{[n]}\right)$ using $\boldsymbol{b}_1$ in offline stage, based on the received signal model as follows:
\begin{equation}
    \boldsymbol{y}^{[n]} = \bar{a}^{[n]} \boldsymbol{\Phi} \operatorname{diag}\left(\bar{\boldsymbol{b}}_1\right) \boldsymbol{u}\left(\vartheta^{[n]}, \psi^{[n]}\right) + \boldsymbol{d}_C^{[n]} \boldsymbol{1}_X + \boldsymbol{v}^{[n]},
\end{equation}
where \( \bar{a}^{[n]} \) denotes the effective path gain, \( \boldsymbol{\Phi} \in \mathbb{C}^{X \times N_1} \) is the training reflection matrix of RIS 1, and \( X \) represents the number of pilot symbols used per block. \( \bar{\boldsymbol{b}}_1 \in \mathbb{C}^{N_1 \times 1} \) denotes the scaled $C_1$ - RIS 1 channel vector. \( \boldsymbol{u}\left(\vartheta^{[n]}, \psi^{[n]}\right) \in \mathbb{C}^{N_1 \times 1} \) is the RIS 1 array 
phases along the x- and y-axis directions.  \( \boldsymbol{d}_C^{[n]} \in \mathbb{C} \) represents the time-varying UE-$C_1$ channel, \( \boldsymbol{1}_X \) is an all-ones vector of length \( X \), and \( \boldsymbol{v}^{[n]} \in \mathbb{C}^{X \times 1} \) is the AWGN vector. In the serving phase of the online stage (i.e., for $n \geq 0$), the RIS begins actively assisting the uplink transmission by dynamically adjusting its passive beamforming based on predicted angle information. At each block, $C_1$ predicts the UE’s angular parameters $\{\hat{\vartheta}[n], \hat{\psi}[n]\}$ based on previously estimated values obtained during the pre-serving phase. These predicted angles are used to construct the RIS reflection vector $\boldsymbol{\theta}[n]$  through the RIS array response vector $\boldsymbol{u}(\hat{\vartheta}[n], \hat{\psi}[n])$, combined with the initial beamforming vector obtained during the offline stage. Meanwhile, the UE continues to transmit uplink pilot signals, allowing the BS estimate the overall uplink channel at block $n$, modeled as
\begin{equation}
\boldsymbol{h}[n] = \mathbf{Q}[n] \boldsymbol{\theta}[n] + \boldsymbol{d}[n]
\label{eq:h_online}
\end{equation}
where $\boldsymbol{Q}[n]$ and $\boldsymbol{d}[n]$ denots the cascaded UE - RIS 1 - BS channel and the direct UE-BS channel, respectively. This process is repeated for each block until the end of the RIS coverage area. The authors of \cite{huang_roadside_2023} stated that this strategy yields significantly higher achievable rates compared to those in \cite{hu_two-timescale_2021} and \cite{wang_channel_2020}, which were set up for low-mobility scenario. The work in \cite{huang_intelligent_2024}, which extends \cite{huang_roadside_2023}, provides a comprehensive comparison between roadside and vehicle-side RIS deployment strategies for high-mobility communications. The authors examine key aspects such as channel variation dynamics, handover complexity, and deployment cost. To address the temporal characteristics of each setup, they propose distinct two-timescale beam alignment frameworks. Simulation results demonstrate that the vehicle-side RIS achieves a higher and more stable effective channel gain compared to the roadside RIS, primarily due to the static RIS–UE link and uninterrupted coverage. These results highlight practical considerations for RIS deployment in high-mobility scenarios.

In a high-mobility SISO scenario, authors of \cite{xu_deep_2022} developed a deep learning-based channel estimation framework for RIS-assisted systems, where the user speed reaches 100 km/h and the carrier frequency is 28 GHz. The proposed method leverages a cascaded neural architecture consisting of two components: a neural ordinary differential equation-enhanced RNN (ODE-RNN) to interpolate the temporal evolution of the cascaded channel, and a Runge-Kutta-modified DNN to extrapolate over the antenna domain. The framework is trained using subsampled pilot data along both time and spatial dimensions. Simulation results show that the model maintains a low NMSE across various sampling ratios and SNR levels.

There are studies other than those in Table \ref{tab:high} that have addressed channel estimation in RIS-assisted systems under high-mobility scenarios, such as \cite{ chen_reconfigurable_2023,huang_transforming_2021,sun_channel_2021,benicio_tensor-based_2024,sahoo_mobility-aware_2022}. However, these studies do not explicitly mention the user speeds at which their methods are applicable. For instance, in \cite{sun_channel_2021}, the proposed technique uses Doppler Shift Adjustment (DSA) that compensates for the phase shifts and distortions introduced by the Doppler effect. In \cite{sahoo_mobility-aware_2022}, the authors propose a hybrid channel estimation scheme using LS, which reduces pilot overhead while considering the mobility of the BS, RIS, and UE channels. The mobility factor determines what channel estimation framework to use. If it is greater than two (fast movement), the two-timescale scheme is used. If it is equal to one (zero-to-slow movement), a simpler scheme is employed. The work in \cite{benicio_tensor-based_2024} proposes tensor-based channel estimation approach with PARAFAC to  estimate the channel and improve the RIS-assisted MIMO in terms of NMSE and BER. In \cite{li_kalman-filter_2024}, Kalman filtering is employed to iteratively estimate and track the channel gain over time. Another work, \cite{huang_transforming_2021}, proposes efficient two-stage transmission protocol. First, the cascaded BS-RIS-UE channel is estimated with compensation for the Doppler effect. Next, the fast-fading BS-UE channel is estimated, and the RIS reflection is continuously adjusted to align with the cascaded channel. This approach ensures a high passive beamforming gain as well as converting the overall channel from fast to slow fading, making it easier to estimate and track the channel.

In practical scenarios, acquiring CSI feedback from the receiver, especially in high-mobility environments with rapidly changing channels, can incur a significant time cost.  The time consumed by the channel estimation process reduces the available time for data transmission, creating a trade-off between channel estimation and throughput. Therefore, determining how frequently channel estimation should be performed is crucial for balancing this trade-off. The study in \cite{zhang_how_2024} suggests that it is beneficial to perform channel estimation less often when the channels demonstrate a stronger correlation over time.

Overall, in RIS-assisted communication, high mobility complicates the process because the system must adapt quickly to the rapidly changing environment to maintain effective signal reflection and beamforming, making real-time channel estimation a critical component of the system's design and operation. This complication could be a contributing factor to the relatively low number of studies conducted on this topic. 

\subsection{Channel Estimation in Multi-RIS Deployment}
\label{sec:multi}

The deployment of multiple RIS is a promising approach to further enhance coverage, spectral efficiency, and link reliability especially where a single RIS suffers from coverage constraints and path loss, particularly in large-scale networks or highly obstructed environments. Multi-RIS deployments overcome these limitations by extending signal reach and creating alternative reflection paths to bypass blockages. However, this multi-hop reflection mechanism significantly increases the complexity of channel estimation, as it requires accurately characterizing multiple cascaded channels, including inter-RIS interactions and RIS-user links. Unlike single-RIS scenarios, where only direct and single-reflected paths are considered, multi-RIS systems introduce additional variables that can lead to severe pilot overhead, high-dimensional channel matrices, and complex signal correlations. Thus, efficient channel estimation techniques that can mitigate these challenges while maintaining scalability and accuracy are essential. 

Several studies have been conducted on the channel estimation problem in multi-RIS scenarios, which can be classified based on the type of RIS employed: passive RIS and active RIS. The studies involving passive RIS can be further divided into single-user case \cite{gorty_channel_2022,you_double,zheng_intelligent_2022,ardah_double-ris_2022} and multi-user case \cite{ding_two-phase_2024,zheng_efficient_2021,zheng_uplink_2021,wu_common_2023,gherekhloo_nested_2023,li_double_2022}. All of these studies focus on double-RIS deployment.

The first study on the single-user case with passive RIS is presented in \cite{you_double} which estimates one cascaded channel, UE → RIS-1 → RIS-2 → BS, and applies an LS estimator for channel estimation, also assuming no LoS path from the UE to BS. In contrast, the work in \cite{zheng_intelligent_2022} 
considers three cascaded channels in a double-RIS-assisted high-speed satellite-ground node communication system: SAT → RIS-1 → GN, SAT → RIS-2 → GN, and SAT → RIS-1 → RIS-2 → GN, while also accounting for the presence of a direct link (LoS) SAT → GN. It introduces an LS estimator combined with a distributed estimation scheme to perform CE. The authors in \cite{gorty_channel_2022} also consider three cascaded channels to be estimated: UE → RIS-1 → BS, UE → RIS-2 → BS, and UE → RIS-1 → RIS-2 → BS. The study employs an LMMSE estimator for all three cascaded links, assuming no LoS between the UE and BS. Lastly, the study in \cite{ardah_double-ris_2022} compares double-RIS and single-RIS configurations using an ALS method, considering cascaded channel estimation at the UE. In the double-RIS case, the cascaded channel consists of three links: Tx → RIS-1, RIS-1 → RIS-2, and RIS-2 → Rx. Because directly solving for each channel simultaneously is complicated (non-convex), the solution uses ALS, an iterative approach that updates one channel estimate while keeping the others two channels fixed, iteratively updating each estimate until convergence. Hence, the ALS method yields separate estimates for each channel. The findings suggest that, with an optimized distribution of RIS elements, double-RIS deployment achieves lower training overhead and requires fewer channel coefficients to be estimated compared to the single-RIS case.

Unlike the previously discussed single-user studies, the works in \cite{ding_two-phase_2024,
zheng_efficient_2021,
zheng_uplink_2021,
wu_common_2023,
gherekhloo_nested_2023,
li_double_2022}
focus on the multi-user case.  The study in \cite{wu_common_2023} applied common structured sparsity based orthogonal matching pursuit (CSS-OMP), where the channel estimation problem is reformulated as a sparse signal recovery problem, enabling efficient channel reconstruction. In \cite{ding_two-phase_2024}, the LS method is employed for channel estimation, with a specific operating mode where RIS-1 remains fixed, while RIS-2 dynamically adjusts based on the time slot. The studies in \cite{zheng_efficient_2021,zheng_uplink_2021} also adopt the LS method but with an always-ON mode for both RISs throughout the entire channel training period to maximize the reflected signal power.
Additionally, \cite{gherekhloo_nested_2023} employs an ALS with PARAFAC decomposition for channel estimation, utilizing tensor-based methods to enhance estimation accuracy while maintaining computational efficiency. Similarly, \cite{li_double_2022} extends the use of PARAFAC decomposition by integrating it with a Kalman filter, providing a robust adaptive estimation framework for tracking dynamic multi-user channels.

While all of the studies above use passive RIS, \cite{yang_active_2023} uses active RIS. Its channel estimation is based on Bayesian learning. The active RIS architecture, with its ability to receive signals through a single RF chain, supports this process by facilitating the acquisition of slow and fast time-varying channels separately, allowing for a reduction in pilot overhead.

However, the aforementioned studies use conventional CE techniques. The authors in \cite{liu_deep_2023} consider a machine learning-based approach called Skip-Connection Attention (SC-attention) to further improve performance from the LS estimator in estimating the channel.


An important insight from the simulation results presented in \cite{al-nahhas_performance_2024} demonstrates that multi-RIS deployments outperform single-RIS configurations in terms of system performance. Specifically, the findings reveal that downlink throughput improves as the number of RISs increases, but only up to an optimal point. Beyond this point, further increments in the number of RISs lead to performance degradation due to rising interference levels. Such behavior is expected, as in practical wireless systems, the continuous addition of network elements does not always guarantee performance improvements.

\begin{figure*}[tb]
\centering
\includegraphics[width=\linewidth]{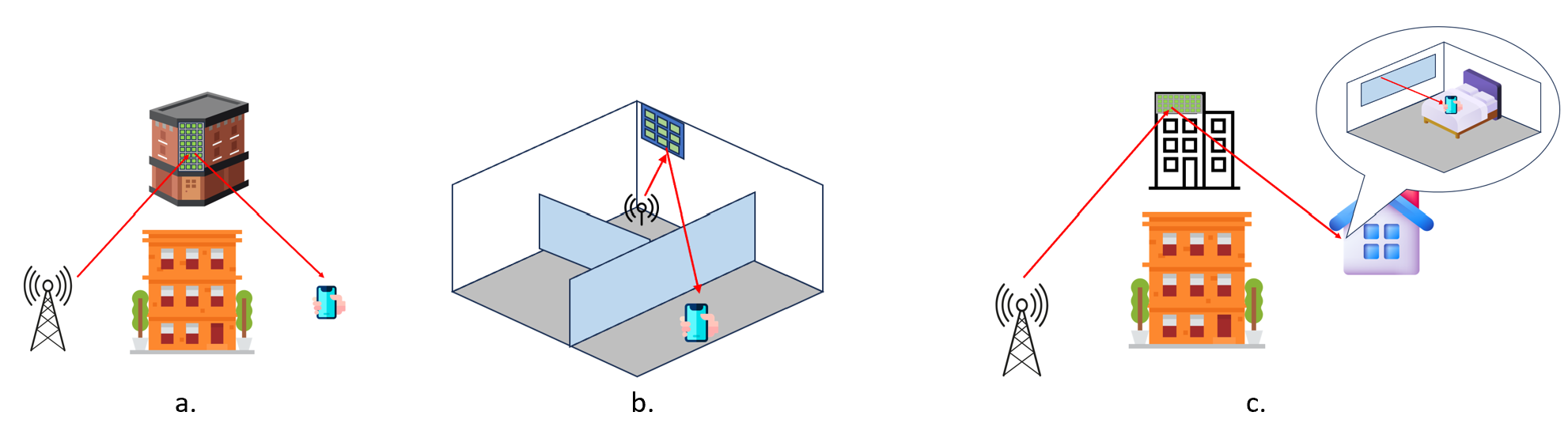}
\caption{Illustration of 3 types of RIS deployment scenario based on GR RIS 001: (a) outdoor (b) indoor (c) outdoor-to-indoor.}
\label{fig:deployment}
\end{figure*}

\section{RIS from Standardization and Industrial Perspectives}
\label{sec:sec5}

This section discusses how standardization and industry efforts are contributing to the integration of RIS into modern communication systems. It begins by outlining key standardization activities led by major organizations, followed by an overview of industrial initiatives aimed at RIS commercialization 

\subsection{Standardization}
RIS is emerging as a transformative technology in the telecommunications industry, offering significant enhancements in signal propagation and network performance. From a standardization perspective, leading international organizations, including the ETSI and 3GPP, are conducting research project to investigate the potential integration of RIS into current and next-generation wireless communication standards. These standardization efforts are crucial for ensuring interoperability, defining technical specifications, and establishing rigorous testing protocols to facilitate the deployment of RIS-enabled systems.

\subsubsection{RIS Standardization Efforts in ETSI}

In September 2021, ETSI established the Industry Specification Group (ISG) on RIS, marking the first dedicated ETSI initiative aimed at investigating, advancing, and standardizing RIS technology in preparation for future 6G networks. The ISG publishes its findings through Group Reports (GRs) and operates in three phases: 2022–2023 (Phase 1), 2024–2025 (Phase 2), and 2026–2027 (Phase 3) \cite{liu_sustainable_2024}. During the first phase alone, three GRs were published: GR RIS 001 \cite{etsi01}, GR RIS 002 \cite{etsi02}, and GR RIS 003 \cite{etsi03}.

GR RIS 001 outlines the relevant use cases for RIS, deployment scenarios, and operational requirements\cite{etsi01}. A comprehensive discussion on RIS use cases, including those outlined in GR RIS 001 by ETSI, has been presented in Section \ref{sec:sec2}. Regarding the deployment scenario, this report analyzes three primary deployment scenarios: indoor, outdoor, and hybrid, which is illustrated in Fig. \ref{fig:deployment}. In complex indoor environments like hallways, corners, and stairwells, RIS can be mounted on walls or furniture to enhance signal strength and data rates by establishing new transmission paths. For outdoor settings, RIS can be strategically placed on building facades or near BSs to mitigate signal loss caused by buildings and trees, significantly benefiting mobile devices and UE, especially those on high-speed trains traveling up to 350 km/h, through dynamic beam tracking. In hybrid scenarios, where indoor and outdoor environments intersect, such as outdoor-to-indoor transitions, RIS can be integrated into building facades or used as transparent surfaces like windows to efficiently redirect or focus signals, offering a cost-effective solution compared to deploying micro-BSs indoors. Next, the report divides the deployments into 2 types: static and nomadic. For static RIS where the RIS is installed on stationary structures like building walls, BS-RIS channel is often static, but the RIS-UE may suffer from fast fading because of the user's mobility. For nomadic RIS where RIS is installed on mobile structures like vehicles, both BS-RIS and RIS-UE channels might suffer from fast fading because of the moving user and the moving RIS. The document also emphasizes the need for RIS to be cost-efficient, ensuring that it is more affordable than conventional solutions such as small base stations or relays, considering factors like manufacturing, assembly, and component costs. Additionally, RIS needs to be easy to install and maintain, taking into account aspects such as backhaul integration, power supply options, and mobility. Effective signal power boosting is crucial to improve coverage and spectral efficiency, which depends on the size and number of RIS elements. The technology should also be flexible, allowing for both dynamic and semi-static reconfiguration to optimize performance for different user scenarios, with considerations for reconfiguration speed, resolution, and scope. RIS compatibility with various network operators and users is essential, requiring it to operate seamlessly across different frequencies, bandwidths, and modes. Finally, RIS must meet regulatory standards, particularly concerning electromagnetic field exposure and maximum allowable power limits in unlicensed frequency bands. All of these requirements are summarized in Table \ref{tab:require}. The second phase of GR RIS 001 is planned to be published at 2025 \cite{etsi01-3}.

\begin{table*}[h!]
\caption{RIS Requirements based on GR RIS 001 by ETSI \cite{etsi01}.}
\label{tab:require}
\centering
\renewcommand{\arraystretch}{1.5} 
\setlength{\tabcolsep}{6pt} 
\begin{tabular}{|c|p{3.5cm}|p{6cm}|p{6.2cm}|}
\hline
\textbf{No} & \textbf{Requirement} & \textbf{Values and configurations} & \textbf{Comments} \\ \hline

\multirow{3}{*}{1} & \multirow{3}{4cm}{\textbf{Hardware Cost}} 
    & Device cost: Should be lower compared to relays and repeaters 
    & Includes the production cost, assembly, and component costs \\ \cline{3-4}
 & & Signal processing: Both analog and digital processing are required for advanced signal processing and channel estimation 
    & Indicates whether the RIS needs to receive and demodulate signals \\ \cline{3-4}
 & & RIS type: Active RIS, passive RIS, hybrid RIS 
    & The hardware cost varies depending on the type of RIS \\ \hline

\multirow{3}{*}{2} & \multirow{3}{4cm}{\textbf{Deployment and maintenance requirements}} 
    & RIS controller: Should support cellular/WiFi/Bluetooth/RAT 
    & The backhaul and control interface of RIS could impact the ease of deployment and maintenance \\ \cline{3-4}
 & & RIS power supply: Should support passive mode, power cables, batteries, or energy harvesting 
    & Considerations for maintaining and providing power supply \\ \cline{3-4}
 & & Mobility: Should support deployment on vehicles, trains, or UAVs 
    & The mobility and mounting locations of the RIS can also influence the deployment and maintenance requirements \\ \hline

\multirow{3}{*}{3} & \multirow{3}{4cm}{\textbf{Power boosting}} 
    & RIS dimensioning: The RIS path is expected to provide comparable gains to the path without RIS 
    & The size of RIS and the number of elements affect its power boosting capability \\ \cline{3-4}
 & & Re-radiation power amplification: Active amplification is required if the signal needs significant boosting 
    & Whether the use of power amplifiers is active or not impacts the power boosting capacity of RIS \\ \cline{3-4}
 & & Efficiency: -2 dB 
    & Represents the RIS hardware power loss between the input and output signals \\ \hline

\multirow{5}{*}{4} & \multirow{5}{4cm}{\textbf{Reconfigurability}} 
    & Configuration flexibility: 1-3 bits 
    & Represents the granularity of the RIS phase shifters \\ \cline{3-4}
 & & Polarization: Single or dual polarization 
    & The RIS could be polarization-sensitive or insensitive, or even able to transform polarization \\ \cline{3-4}
 & & Reconfiguration rate: Varies from quasi-static configurations to symbol-level reconfiguring rates 
    & Defines the rate at which RIS configurations can change \\ \cline{3-4}
 & & Reconfiguration latency: Must be compatible with the scheduler 
    & Indicates the delay in applying the RIS configuration \\ \cline{3-4}
 & & Number of supported configurations/codebooks: Example codebook size is 16 (4 bits) 
    & This corresponds to the bits supported by the communication bus \\ \cline{3-4}
 & & Number of RIS elements configured simultaneously: The total number of RIS elements can range from tens to over a thousand 
    & Defines the number of independently configurable RIS elements, based on the total number of elements and RIS partitioning (e.g., column-wise configuration) \\ \hline

\multirow{6}{*}{5} & \multirow{6}{4cm}{\textbf{Interoperability}} 
    & Carrier frequency: FR1/FR2/Sub-THz/THz 
    & \\ \cline{3-4}
 & & Bandwidth: Ranges from hundreds of MHz to GHz 
    & \\ \cline{3-4}
 & & Operating mode: Supports reflection/transmission modes 
    & \\ \cline{3-4}
 & & Duplex mode: Supports full-duplex operation 
    & \\ \cline{3-4}
 & & Unlicensed operation: Determines whether it is supported or not 
    & \\ \cline{3-4}
 & & Reciprocity: Supports uplink and downlink reciprocity 
    & \\ \hline

\multirow{2}{*}{6} & \multirow{2}{4cm}{\textbf{Regulatory}} 
    & Subject to applicable standards 
    & \\ \cline{3-4}
 & & Maximum EIRP in unlicensed bands: Required if unlicensed operation is supported 
    & In unlicensed bands, separate/reduced EIRP limits might be needed depending on regulatory and standards requirements \\ \hline

\end{tabular}
\end{table*}

Next, GR RIS 002\cite{etsi02} reported on operation mode, technological aspects of RIS, its network architecture, and its physical layer aspect. It details the operational modes of RIS, including control types, capabilities, and complexity.  The capability aspects of RIS, however, depend on the controlling type of RIS. For example, in network-controlled RIS, the control information is entirely determined by the network, meaning RIS only needs to tune its elements according to received commands, while the RIS controller handles communication with the network. It is different with network-assisted RIS, which has similar panel capabilities but includes limited processing power in its RIS controller to collect data and configure the RIS based on assistance from the network. If RIS supports a multi-functional mode of operation, it can further report specific modes, such as reflection, refraction, and absorption mode. Complexity in RIS correlates with its capabilities; a more capable RIS requires a more complex controller. The report also describes technological aspects of RIS, such as functional modules and internal architecture. RIS can be modeled as a combination of at least an RIS controller and an RIS panel. The internal architectures include single-connected impedance structures, where elements are isolated from one another, and multi-connected architectures, where elements are interconnected. Various fabrication techniques for RIS panels, such as RFMEMS, PN diodes, varactor diodes, MOSFETs, photo-conductive materials, ferroelectric materials, and liquid crystals, are examined, each with distinct implications for frequency operation, power consumption, cost, and performance metrics. The document also explores different network topologies for integrating RIS for ISAC and localization. These topologies depend on the RIS capabilities and specific use cases, such as extending coverage or improving positioning accuracy. Also, in the radio and physical layers, significant efforts are required particularly in beamforming, channel estimation, control signaling, and interference management. Additionally, RIS-assisted transmissions require precise time alignment to mitigate latency and synchronization issues, particularly in NLoS scenarios. In frequency-division duplex (FDD) systems, RIS selection strategies differ for downlink and uplink due to non-reciprocal channels, requiring independent optimization. 

Furthermore, GS RIS 003 \cite{etsi03} outlines ETSI  perspectives on communication models, channel modeling, channel
estimation methods, and evaluation methodology for RIS-integrated communications, localization, and sensing. RIS should be able to perform, not just functions, but also refraction, scattering, and modulation. Hence the types of RIS includes reflecting, refracting, joint reflecting-refracting, absorbing, and sensing surfaces. Each type has specific functionalities suited to different applications. The document presents different models for representing RIS behavior in wireless communication systems. Three primary models are discussed. First, the locally periodic discrete model, which assumes that RIS elements are periodically arranged and characterized by complex reflection coefficients. Second, the mutually coupled antenna model, which considers the electromagnetic interactions between closely spaced RIS elements and models them similarly to MIMO systems. Third, the inhomogeneous sheets of surface impedance model, which treats RIS as a continuous metasurface rather than a discrete structure, allowing for the derivation of macroscopic surface parameters such as electric and magnetic surface impedances to define the RIS response to incident electromagnetic waves. 

GS RIS 003 also brings up that  channel modelling in RIS needs to balance electromagnetic accuracy and computational efficiency for practical performance evaluation. The document distinguishes between deterministic and statistical models. Additionally, the report details empirical channel modeling, using real-world measurements to refine propagation assumptions and introduce RIS-specific components such as multipath reflections, re-radiation, and scattering effects. The multipath models can be divided into rich scattering environments (e.g., sub-6 GHz channels) and structured multipath scenarios (e.g., millimeter-wave and terahertz bands). Multi-mode reradiation models are introduced to analyze RIS-induced secondary radiation, ensuring that RIS does not inadvertently introduce harmful interference. The interference is categorized into types, including irregular reflections, discontinuous time-varying emissions, and unwanted electromagnetic noise. Security concerns such as eavesdropping vulnerabilities and unintended radiation leakage are also considered, highlighting the importance of RIS control in preventing information leaks. Finally, the document explores polarized RIS element modeling, considering electromagnetic field polarization effects, which are particularly relevant for high-frequency bands where polarization mismatches can significantly impact signal reception. 

The document also discusses scenarios in which RIS can or cannot perform on-board channel estimation, with the capability being attributed to the installment of active elements. It also discusses categorization of whether the estimation is for individual channels (BS-RIS and RIS-UE) or end-to-end channel (BS-RIS-UE). Another categorization is based on the availability of CSI: instantaneous, two-timescale, or fully long-term CSI.

Currently, there are 4 other ongoing works which are GR RIS 004, 005, 006, and 007. GR RIS 004, initiated in late January 2023, is being developed to provide insights into implementation aspects and practical considerations of RIS \cite{etsi04}. This study analyzes implementation factors and addressing practical challenges associated with RIS across various frequency ranges and deployment contexts, while also presenting viable solutions and prototype validation results. As of now, the final draft has been officially registered. 

Initiated in May 2023, GR RIS 005 aims to advance the understanding of diversity and multiplexing in RIS-assisted communications. This work encompasses various aspects, including the identification of relevant use cases, feasibility assessment across different RIS hardware architectures and operating modes, evaluation of transmitter-receiver complexity across multiplexing scenarios, characterization of RIS-assisted channel properties, analysis of achievable performance gains, and examination of the impact on channel estimation. At this stage, a draft has been received.

Simultaneously with the development of GR RIS 005, ETSI has also initiated GR RIS 006, which focuses on the modeling, optimization, and operation of multi-functional (MF) RIS. GR RIS 006 addresses key technological challenges and explores technical solutions that integrate multiple functionalities, including transmission, reflection, sensing, and computation. Additionally, it investigates critical aspects such as channel modeling, coefficient optimization, deployment strategies, and resource allocation for MF-RIS. Furthermore, the study aims to propose practical deployment approaches for MF-RIS in real-world scenarios while assessing its potential performance enhancements across various application contexts. 
 
The most recent initiative, GR RIS 007, began in March 2024 and focuses on near-field channel modeling and the fundamental mechanics of RIS.  This work encompasses several critical aspects, including the identification of near-field use cases and deployment scenarios, the development of channel models specific to the near-field region, and the characterization of cascaded channel modeling. Additionally, it explores novel technical solutions tailored for near-field RIS applications and evaluates the potential impact on future specifications. However, currently, it is in the early draft stage.

\subsubsection{RIS Standardization Efforts in 3GPP}
RIS has not yet been standardized by 3GPP, but insights from previous 3GPP efforts offer valuable guidance\cite{yuan_reconfigurable_2022}. The integration of RIS in 3GPP standardization can draw on lessons from past technologies like LTE Relay in 3GPP Release 10 (Rel-10) and Full-Dimension MIMO (FD-MIMO) in 3GPP Release 13-15 (Rel-13 - Rel-15). LTE Relay was originally introduced to improve network coverage and system performance, but it failed to see widespread deployment due to over-engineered designs, complex implementation, and a lack of practical use cases. The rushed standardization process, which included multiple competing relay types, resulted in a specification that was too burdensome for operators. In contrast, FD-MIMO in 3GPP Rel-13 was successfully deployed due to its gradual and structured standardization process. This structured evolution from research to commercial deployment serves as a model for RIS, ensuring that overly ambitious designs are avoided.

\begin{figure}[tb]
\centering
\includegraphics[width=\linewidth]{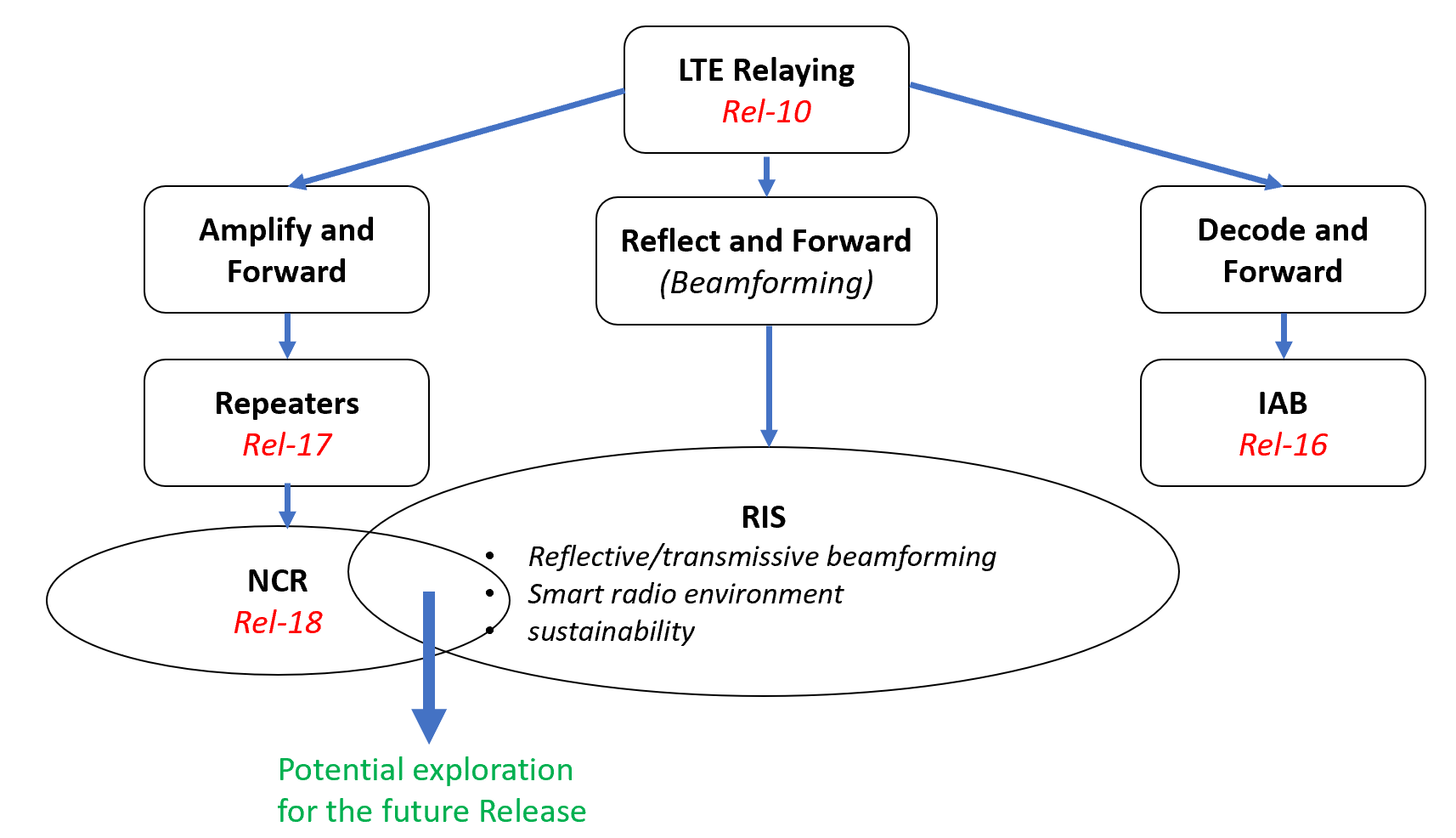}
\caption{The possibility of RIS for the next standardization of 3GPP \cite{garcia_nokia}.}
\label{fig:3gpp}
\end{figure}

To facilitate the smooth adoption of RIS, the author in \cite{yuan_reconfigurable_2022} proposes a multi-stage standardization approach across 3GPP Releases 18, 19, and 20 (Rel-18, Rel-19, and Rel-20, respectively). Rel-18  (2022–2023) introduced Network-Controlled Repeaters (NCRs), which can direct radio beams. The basic mechanism and signaling structure defined for NCRs in Rel-18 can serve as a basis for RIS relay control channels, making NCR a relevant reference point for the standardization of RIS in future 3GPP releases \cite{yuan_reconfigurable_2022, 3gpp_3rd_2025}. Moreover, \cite{su_reconfigurable_2024} and \cite{wen_shaping_2024} present a technical comparison between NCR and RIS, through which several structural and functional similarities are identified. Rel-19 (2023-2025) is focusing on advanced network topology \cite{wen_shaping_2024}. As illustrated in Fig.~\ref{fig:3gpp}, 3GPP currently envisions RIS typically as a relay technology \cite{garcia_nokia}. Following the focus of amplify-and-forward relay type on repeaters in Release 17 (Rel-17), Rel-18 shifted attention to control aspects and introduced NCR. At the Rel-19 workshop, several companies and research institutions proposed initiating a feasibility study of RIS. For instance, NTT DoCoMo expressed interest in RIS but conceptualized it as a reflect-and-forward (beamforming) relay type, highlighting that RIS can be seen as a special case of NCR. However, RIS has been excluded from 3GPP Rel-19 and is unlikely to be incorporated into 5G-Advanced. Rel-20, planned for the 2025–2026 timeframe, will mark the beginning of the 6G standardization phase through a study item, although it will not yet involve normative work. Rel-20, however, may serve as the final opportunity for RIS to be considered within the standardization trajectory for 6G, as 3GPP Release 21 (Rel-21) will initiate the formal specification process for 6G, leveraging insights derived from Rel-20. This gradual approach prevents RIS from facing the same pitfalls as LTE Relay in Rel-10 and allows for progressive hardware improvements, control refinements, and better-defined deployment strategies.

\subsubsection{RIS Standardization Efforts in ITU}
The International Telecommunication Union (ITU) has also been advancing efforts to standardize RIS. During the ITU-R meeting in 2020, held for the development of the IMT Future Technology Trends, Chinese companies submitted a proposal for a preliminary draft report highlighting RIS as a promising technology for the physical layer of the next-generation wireless networks \cite{liu_path_2022, ahmed_survey_2023}. Subsequently, companies from other countries also incorporated RIS in their draft proposals in later meetings, reflecting its growing recognition in global standardization discussions. Between 2022 and 2023, the ITU Journal published 5 articles addressing various aspects of RIS. These studies explored AI-based architecture and algorithm design for RIS\cite{lu_artificial_2023}, RIS-assisted millimeter-wave beamforming\cite{torkzaban_ris-aided_2023}, resource slicing in RIS-assisted systems\cite{liaskos_scheduling_2023}, non-coherent modulation techniques in RIS-assisted MIMO systems \cite{hu_non-coherent_2022}, and RIS performance analysis considering spatial correlation\cite{lu_performance_2023}.

\subsubsection{RIS Standardization Efforts in Other Institutions}
Beyond international efforts, standardization initiatives also extend to the national level \cite{liu_path_2022}. For instance, in October 2020, the China Communications Standards Association (CCSA) approved a proposal to initiate a study item (SI) dedicated to RIS. Subsequently, in December 2020, another Chinese standardization body, The Future Forum, established a working group to explore the integration of RIS into wireless networks.

Outside standardization bodies, academia and enterprises collaborations have also contributed to paving the way for RIS standardization. For instance, during 2020–2021, institutions such as the University of Oulu, Nokia Bell Labs, Southeast University, Guangdong Communications and Networks Institute, UNISOC, and Lenovo published white papers dedicated to RIS \cite{jian_standardization_2021}. Additionally, several other studies have explored the vision for RIS standardization, with a primary focus on channel modeling, control mechanisms, and interference management \cite{liu_how_2022,li_considerations_2022,tohidi_ris-assisted_2024,lou_standardization_2024,su_reconfigurable_2024}.

\subsection{Industrial Perspectives}
Other than standardization efforts, it is essential to consider the industrial perspective on RIS, where practical implementation challenges and opportunities for commercial deployment come to the forefront. For instance, in October 2021, NTT Docomo, the biggest Japanese mobile phone operator, published a special article on their work about RIS \cite{kitayama_research_2021}. They explore RIS as a key enabler for the intelligent radio environment (IRE) in next-generation wireless networks. The study focuses on transparent RIS using a metasurface lens, developed in collaboration with AGC Inc., which allows for adaptive reflection and transmission of signals while maintaining visual transparency, making them suitable for urban deployment. In the same year, they also released the results of the first use of their user-tracking metasurface reflector \cite{ntt_docomo_press_2021}. The test result showed the possibility of employing RIS to enable high-speed millimeter-wave indoor communications. 

One of the European telecommunication operators based in France, Orange, is exploring RIS to provide wireless connectivity with minimum environmental consequences and controlled electromagnetic field (EMF) exposure \cite{Orange2022}. From their perspective, RIS offers promising solutions for improving coverage and data rates in both indoor and outdoor environments without emitting extra radio waves. Furthermore, RIS has the potential to reduce costs and energy consumption compared to installing additional access points. A research conducted by Orange demonstrated that RIS can enhance antenna beamforming, increasing data rates to target devices while reducing EMF exposure \cite{awarkeh_electro-magnetic_2021}. They have also created an RIS prototype featuring continuous control solutions for reflection phases, marking a significant advancement over current discrete control solutions. However, Orange acknowledges that RIS technology is still in its early stages and faces challenges related to cost, control, energy efficiency, reflection quality, and coexistence between operators. Their ongoing investigations and field trials are part of the European project RISE-6G. Additionally, while RIS holds significant promise for mmWave communications, its commercial viability remains uncertain as of now \cite{dinh_orange}.

In \cite{huawei_communication_2022}, Huawei believes that RIS is the key to mMIMO in 5G evolving to support extreme spectrum efficiency in B5G/6G. They found that RIS over MIMO can theoretically enhance the spectrum efficiency by up to 24 bit/s/Hz at 30 dB SNR. Additionally, it boosts energy efficiency by up to 5 bit/kJ at the same SNR. Huawei's research also demonstrated that both dynamic and static RIS at C-band in a real-world indoor setting enhanced the signal level at the receiver side by at least 15 dB and 19 dB, respectively. These findings suggest the potential for recovering lost video connections due to blockages and maintaining connection stability despite mobility.

It is reported in \cite{zte_zte_2024} that ZTE's research on RIS technology has led to the development of Dynamic RIS (D-RIS) 2.0, a breakthrough towards 5G-Advanced (5G-A) networks. Unlike conventional passive RIS, which relies on static reflection properties, D-RIS enables intelligent coordination with base stations, dynamically adjusting the propagation of electromagnetic waves to optimize network performance. This capability enables D-RIS to mitigate the challenges posed by high-frequency bands, such as millimeter-wave (mmWave) and terahertz (THz) communications, where signal attenuation and blockage are more pronounced. ZTE’s D-RIS 2.0 introduces several key enhancements, including high gain, low power consumption (as low as 30W), and simplified deployment, making it an attractive solution for widespread network deployment. By integrating D-RIS 2.0 with 5G-A base stations, network coverage can be extended by 30\%, with signal gains of up to 30 dB, improving connectivity, spectral efficiency, and user experience. This enhancement is particularly beneficial in urban environments with dense infrastructure, indoor coverage scenarios, and remote areas where traditional base station deployment is costly and inefficient.
Beyond improving coverage, D-RIS 2.0 also contributes to network sustainability by reducing power consumption and minimizing the need for additional base stations, aligning with global efforts toward green and energy-efficient wireless communication technologies.

Furthermore, \cite{liu_path_2022} addresses the challenges from an industrial point of view. First is the electromagnetic consistency of the channel models. Developing accurate channel models for RIS is more complex than traditional models because RIS elements need to be precisely modeled, including the electromagnetic characteristics and inter-element coupling effects. Existing models often oversimplify these aspects, particularly in NLoS scenarios. A lot of channel measurements to
characterize RISs accurately in sub-6GHz and mmWave bands are still required. Secondly, while RISs offer similar environmental shaping capabilities as relays, but with passive elements, the trade-offs between RISs and traditional relays involve considerations of cost, power, and control overhead. The comparison in this paper reveals that RISs need significant optimization to match the performance benefits provided by relays. Thirdly, as nearly-passive devices, RISs face challenges in channel estimation and dynamic scheduling. Efficient algorithms are necessary to estimate the cascaded channels accurately, and control signaling needs to be designed for low power consumption and complexity. Moreover, RISs lack the capability to process signals at fine granularity based on frequency bands, leading to potential interference issues in adjacent frequency bands and limitations in frequency-selective scheduling. The large-scale deployment of RISs requires affordable manufacturing processes, especially since each RIS needs hundreds of elements. Additionally, the materials used in RISs must be durable and maintain their electromagnetic properties under unfriendly environmental conditions.

\subsection{Challenges and Future Directions}
This subsection presents research challenges and possible
future directions of RIS.
\subsubsection{RIS with Near Field Setup}
Previous works focus on small to moderately sized RIS, where the far-field assumption generally holds and neither the BS nor the UE falls within the near-field region of the RIS. However, as the size of the RIS increases, the near-field distance also grows. According to the near-field distance criterion, both the BS and the UE may lie within the near-field region of a large RIS panel, making it necessary to account for near-field propagation characteristics in modeling and system design. Consequently, for example, far-field codebooks become suboptimal, and new codebook designs for RIS phase shifts and channel state information (CSI) estimation are required to accommodate near-field characteristics. Some recent works have started to investigate this aspect, for instance through theoretical analysis and simulations \cite{chen_near-field_2024,pan_ris-aided_2023,lv_ris-aided_2024,emenonye_ris-aided_2023,zhang_hybrid_2023,tian_near-field_2024,shen_multi-beam_2023}, and implementation-oriented studies \cite{ramzan_reconfigurable_2023, tang_path_2022,mei_study_2022}.
\subsubsection{RIS in Wideband Commmunication}
Although many studies have shown that RIS beamforming can improve communication performance, most of them have focused on narrowband systems. However, the increasing demand for higher data rates and enhanced sensing capabilities has led to a growing interest in higher carrier frequencies and wider bandwidths. The wider bandwidth and the larger number of reflective elements cause the beams associated with different subcarriers to be reflected into different physical directions, which is known as beam split effect. Since RIS elements are passive with frequency-independent phase shifts, this effect introduces significant gain fluctuations across the bandwidth and distorts the reflected signal in the intended direction. As a result, the effective bandwidth becomes limited, reducing the overall communication performance. The beams formed by frequency-flat phase shifters can only be well-aligned with the target user over a narrow frequency region around the central frequency. Consequently, only subcarriers near the center frequency achieve high array gain, while the others experience severe degradation. Therefore, the beam split effect leads to a significant drop in achievable rate, offsetting the potential gains from wider bandwidth and degrading the distance resolution. Hence, RIS-aided wideband communication requires new approaches \cite{zhang_wideband_2023,wang_wideband_2024,su_wideband_2023,zhou_wideband_2024,mo_reconfigurable_2024,demir_wideband_2024,qian_wideband_2025}.

\subsubsection{Phase-Shift Design}
Despite being a core function of RIS operation, phase-shift design remains a significant challenge, particularly in practical and dynamic wireless environments. Most current approaches assume ideal or near-ideal conditions, where phase shifts can be perfectly configured to maximize beamforming gain. However, in reality, phase shifts are subject to hardware constraints such as limited quantization levels and non-linear responses, which reduce design flexibility and impact overall performance. Moreover, optimal phase-shift design often relies on accurate CSI, which may be difficult to obtain in scenarios involving user mobility, wideband fading, or low-latency requirements. These limitations are further compounded in systems with multi RIS configuration or hybrid architectures. As a result, designing robust, low-complexity, and adaptive phase-shift strategies remains an open research problem, although some recent studies have begun to explore this direction \cite{yang_how_2023,subhash_optimal_2023,vu_performance_2023,soumana_hamadou_grey_2024,jiang_joint_2023}.
\subsubsection{Beyond Diagonal RIS}
Most existing models of RIS assume a diagonal reflection matrix, where each RIS element independently applies a phase shift to the incident signal. While this simplification facilitates tractable analysis and design, it neglects potential inter-unit-cell coupling, mutual scattering, and surface wave interactions that may arise, especially in dense or large RIS deployments. These physical phenomena result in a more general non-diagonal reflection behavior, which cannot be accurately captured by traditional models. Modeling RIS as a full matrix beyond the diagonal approximation becomes increasingly relevant. However, this introduces new challenges in reflection characterization, phase-shift optimization, and channel estimation, as the system behavior becomes more complex and less separable. Investigating the theoretical limits, practical design, and signal processing frameworks for non-diagonal RIS is a promising direction for future research. Research interest in this area is growing \cite{sousa_de_sena_beyond_2025,zhou_optimizing_2024,li_beyond_2024,li_reconfigurable_2024,li_beyond_2025,sun_new_2024,bjornson_capacity_2025,nerini_dual-polarized_2025,fang_low-complexity_2024}. However, the majority of these works remain theoretical, with practical implementations still scarce due to the considerable challenges associated with realizing a fully functional BD-RIS.

\subsubsection{Real-World RIS Deployment}
RIS practical deployment are still under active investigation. One major factor influencing real-world deployment is the lack of extensive channel sounding studies in outdoor environments. Most existing experimental campaigns focus on controlled indoor or O2I (outdoor-to-indoor) settings, where propagation conditions are more manageable and stable. In contrast, outdoor scenarios involve more complex and dynamic channel characteristics, such as large-scale fading, mobility-induced variation, and environmental blockage, all of which complicate RIS placement, orientation, and phase control. These uncertainties make it difficult to generalize indoor findings to broader deployment contexts. Furthermore, site-specific constraints such as building geometry, available surfaces for RIS mounting, energy supply, and integration with existing infrastructure must be carefully considered. Given the frequency-selective and spatially non-stationary nature of outdoor RIS-assisted channels, practical deployment requires not only refined channel models but also environment-aware planning. Therefore, bridging the gap between controlled experimentation and scalable outdoor deployment remains a critical direction for future research \cite{hwang_environment-adaptive_2023, yang_beyond_2023, yuan_field_2024, encinas-lago_cost-effective_2024, sang_coverage_2024}.
\subsubsection{Green Communication}
RIS is often regarded as a promising enabler for green communication due to its potential to improve spectral and energy efficiency with minimal power consumption. Passive RISs can manipulate wireless signals without requiring significant power, making them attractive for low-energy network designs. However, the broader sustainability of RIS deployment also depends on factors beyond operational efficiency. This includes the energy and material costs associated with manufacturing RIS panels, particularly those involving large-scale use of tunable elements such as PIN diodes or varactors. Additionally, the increasing interest in deploying dense or large-surface RISs raises concerns about electronic waste, component recyclability, and lifecycle energy consumption. As RIS technology moves toward practical deployment, a comprehensive assessment of its environmental impact will be essential to ensure that its contribution to green communication is both effective and sustainable \cite{ding_intelligent_2024, wang_reconfigurable_2024, abualhayjaa_how_2025}.

\section{Conclusion}
\label{sec:sec7}

In this survey, we provided a comprehensive overview of the development of RIS, covering the theoretical foundations, recent advancements, and its prospects for real-world deployment.
To establish a solid basis for the subsequent sections, this survey begins by outlining the fundamental principles of RIS and categorizing their types: passive, active, and hybrid.
Then, we discuss the use cases of RIS and observe that it offers a wide range of promising applications across various domains. As standardization efforts continue, these use cases are expected to evolve in parallel with ongoing technological developments.
We also provided a detailed discussion on the control mechanisms of RIS. Controlling an RIS involves more than selecting suitable materials or tuning the phase of individual unit cells. It also extends to higher-layer operations such as inter-unit-cell communication managed by the RIS controller and inter-RIS coordination facilitated by the RISO. RIS control mechanisms can be categorized into three types: fully controlled, partially controlled, and fully autonomous. Furthermore, the control signaling can be implemented either with or without a dedicated communication channel.

This survey also reviews channel sounding efforts with RIS. Measurements are typically conducted in indoor, outdoor, or O2I environments, with most setups employing horn antennas for both transmission and reception. Most of these experiments use VNAs as the measurement peripheral. There is still a limited number of outdoor measurement studies, which may be attributed to the increased complexity of channel conditions in such environments. Additionally, the number of studies in RIS-assisted channel sounding at mmWave is less than in sub-6GHz. These gaps open opportunities for future research, as more real-world measurement data, particularly in outdoor and mmWave scenarios, are important to support the development and standardization of RIS technology. Beyond empirical measurements, the literature characterizes RIS-assisted communication channels using key parameters, including path loss, scattering gain loss, delay spread, small-scale fading, and fast fading.

The next section continues with the discussion about the channel estimation which can be either cascaded or separate. Most existing works on channel estimation focus on passive RIS under conventional techniques, especially in single-RIS scenarios. However, it seems that there is growing interest in ML-based estimation, particularly for MIMO and hybrid RIS configurations. By contrast, the study of channel estimation in high-mobility environments and multi-RIS deployments remains relatively limited, despite their critical importance for realizing RIS-assisted systems in dynamic and large-scale scenarios. For multi-RIS systems, existing works on channel estimation are mostly limited to double-RIS setups, while configurations involving more than two RISs remain significantly underexplored, opening potential avenues for future investigation.

We summarize the standardization efforts by major global institutions such as ETSI, 3GPP, and ITU, which cover various aspects including deployment scenarios, use cases, reconfigurability, hardware cost, regulatory issues, interoperability, power consumption, maintenance, control strategies, channel estimation, channel modeling, and more. From an industrial perspective, RIS appears to be a promising technology for commercialization. Mobile operators and telecommunications vendors are actively conducting studies, trials, and prototyping RIS systems. While substantial progress has been made, RIS is still in the early stages.

In conclusion, this survey consolidates the current state of RIS technology by reviewing its use cases, control mechanisms, channel sounding, channel estimation, and ongoing standardization and industrial efforts, providing a unified perspective that complements existing surveys. These components collectively form the foundation for bridging theoretical insights and practical deployment considerations. As RIS technology continues to evolve, we hope that the insights provided in this survey will serve as a useful reference to guide future efforts across both academic and industrial domains.


\bibliographystyle{IEEEtran_mod}
\bibliography{ref}

\newpage

\vspace{11pt}


\vfill

\end{document}